**Universidade de Aveiro** Departamento de Física
**Ano 2019**

**André Filipe Castanheira Horta**

# Otimização da síntese de ferrites para aplicações em fluidos magnéticos

# Synthesis optimization of Zn-Mn ferrites for magnetic fluid aplications

**Universidade de Aveiro** Departamento de Física
**2019**

**André Filipe Castanheira Horta**

# Otimização da síntese de ferrites para aplicações em fluidos magnéticos

# Synthesis optimization of Zn-Mn ferrite for magnetic fluids aplications

Dissertação apresentada à Universidade de Aveiro para cumprimento dos requisitos necessários à obtenção do grau de Mestre em Engenharia Física, realizada sob a orientação científica do Dr. João Horta, investigador júnior do IFIMUP na Faculdade de Ciências da Universidade do Porto, e co-orientação do Dr. João Amaral, investigador auxiliar do CICECO na Universidade de Aveiro.

*"In a world of magnets and miracles"*
*-Pink Floyd in "High Hopes"*

**o júri**

| | |
|---|---|
| **presidente** | Professor Auxiliar Dr. António Ferreira da Cunha<br>Professor Auxiliar do Departamento de Física na Universidade de Aveiro |
| **orientador** | Investigador Júnior Dr. João Filipe Horta Belo da Silva<br>Investigador Júnior do IFIMUP na Faculdade de Ciências da Universidade do Porto |
| **co-orientador** | Investigador Auxiliar Dr. João Cunha de Sequeira Amaral<br>Investigador Auxiliar do CICECO na Universidade de Aveiro |
| **arguente** | Professor Auxiliar Dr. André Miguel Trindade Pereira<br>Professor Auxiliar do IFIMUP na Faculdade de Ciências da Universidade do Porto |

## Acknowledgements

The delivery of my master thesis is certainly one of the most important marks of my life. This achievement would have not been possible without the support of everyone who I have met in this path. To all of them I am eternally grateful.

To my supervisor, Dr. João Horta, who embraced this project with me. Thank you for the wise thesis proposal, thank you for all the support, motivation and opportunities during this year, thank for the scolding that made me better, thank you for all you taught me. During this year, you were not only my supervisor but also a teacher and a friend. Thank you.

To my co-supervisor, Dr. João Amaral. When I first met you, you said was “do not count with me, I’m very busy”, turns out that I always could count with you. I’m thankful for all the advices, corrections and opportunities that you provided.

To PhD student Farzin Mohseni and Dr. Carlos Amorim, with who I had the pleasure to work. Thank for your expertise advices, promptitude and time spent with me. To Prof. Nuno Silva, for the availability when using his magnetic induction heating setup and for the useful lessons. To Prof. Vitor Amaral for the interest in the project and wise suggestions. To Rosário Soares, XRD technician, for helping in XRD matters. To Marta Ferro, TEM and SEM technician, for these techniques support and expertise.

I also have to big thanks my family, my partner and my friends. You were my support and strength during this year, the previous and the next. Thank you.

**resumo**

Nanopartículas de ferrite Manganês-Zinco são cada vez mais investigadas pelas suas propriedades desejadas para uma vasta gama de aplicações. Essas propriedades incluem controlo nanométrico de tamanho de partícula, propriedades magnéticas ajustáveis, elevada magnetização de saturação e baixa toxicidade, providenciando estas ferrites com os requerimentos necessários para tratamento de cancro por hipertermia magnética. Durante esta investigação, foram sintetizados e caracterizados pós de ferrite de Mn-Zn, visando otimizar as suas propriedades estruturais e magnéticas para futura aplicação num ferrofluido.
Amostras de $Mn_{1-x}Zn_xFe_2O_4$ (x=0; 0.5; 0.8; 1) foram sintetizadas pelos métodos de autocombustão de sol-gel e pelo método hidrotermal. Os pós sintetizados foram caracterizados por XRD, SQUID, SEM, TEM e aquecimento por indução magnética. Os difratogramas de XRD das amostras produzidas por hidrotermal apresentam a estrutura cristalina de espinela com elevada percentagem de fase-única (>88%). O refinamento de Rietveld e a análise de Williamson-Hall revelam decréscimos no parâmetro de rede (8.50 até 8.46 Å) e no tamanho médio de cristalite (61 até 11 nm) com o aumento da razão Zn/Mn. As imagens de TEM revelam uma estreita distribuição de tamanhos e um decréscimo do tamanho médio de partícula (41 até 7 nm) com o aumento da razão Zn/Mn. Os resultados de SQUID mostram que o aumento de Zn resulta num decréscimo de magnetização de saturação (79 até 19 emu/g) e de magnetização remanente (5 até aproximadamente 0 emu/g). Notoriamente, as curvas M(T) revelaram um desvio na temperatura de ordenamento magnético para mais baixas temperaturas com o aumento de Zn, de ~556 (estimado) até ~284 K. A experiência de aquecimento por indução também revelou um decréscimo na taxa de aquecimento com o aumento de Zn na ferrite.
Nano-cristais de ferrite de Mn-Zn produzidos pelo método hidrotermal apresentam melhor cristalinidade e propriedades magnéticas que as amostras de autocombustão de sol-gel. As amostras sintetizadas pelo método hidrotermal revelam dependência das suas propriedades estruturais e magnéticas com a razão Zn/Mn. A temperatura de ordenamento magnético destas ferrites pode ser usada como um mecanismo de aquecimento autorregulado, elevando estas ferrites para uma classe de materiais inteligentes.

**Abstract**

Manganese-Zinc ferrite nanoparticles have been the subject of increasing research due to their desired properties for a wide range of applications. These properties include nanometer particle size control, tunable magnetic properties, high saturation magnetization and low toxicity, providing these ferrites with the necessary requirements for cancer treatment via magnetic hyperthermia. During this research, powders of Mn-Zn ferrite were synthesized and characterized, aiming to optimize their structural and magnetic properties for further application in a ferrofluid.

Samples of $Mn_{1-x}Zn_xFe_2O_4$ (x=0; 0.5; 0.8; 1) were synthesized via the sol-gel auto-combustion and hydrothermal methods. Synthesized powders were characterized by XRD, SQUID, SEM, TEM and magnetic induction heating techniques. The XRD diffractograms of hydrothermally produced samples presented spinel crystal structure with high single-phase percentage (>88%). Rietveld refinement and Williamson-Hall analysis revealed a decrease of lattice constant (8.50 to 8.46 Å) and crystallite size (61 to 11 nm) with increase of Zn/Mn ratio. TEM images reveals narrow particle size distributions and decrease of the mean particle size (41 to 7 nm) with the Zn/Mn ratio increase. SQUID results showed that the increase of Zn results in a decrease of saturation magnetization (79 to 19 emu/g) and remnant magnetization (5 to approximately 0 emu/g). More noticeably, the M(T) curves present a shift in the samples magnetic ordering temperature towards lower temperatures with the increase of Zn content, from ~556 (estimated) to ~284 K. The magnetic induction heating experiment also unveiled a decrease in the heating rate with the increase of Zn in ferrite.

Nanocrystals of Mn-Zn ferrite produced by hydrothermal method present better crystallinity and magnetic properties than the sol-gel auto-combustion samples. The hydrothermally synthesized samples revealed dependence of its structural and magnetic properties with Mn/Zn ratio. The magnetic ordering temperature of these ferrites can be used as a self-controlled mechanism of heating, raising these ferrites to a class of smart materials.

# Content:

# List of Figures:

# List of Tables:

# List of Abbreviations:

cgs - Centimeter-Gram-Second

S.I. – International System
XRD – X-Ray diffractometer

TEM – Transmission electron microscope

SQUID – Superconducting Quantum Interference Device

SEM – Scanning electron microscope

FWHM – Full-width at Half Maximum

W-H – Williamson-Hall

FC – Field Cool

ZFC – Zero-Field Cool

p-p – peak to peak

# List of Nomenclatures:

$M$ - Magnetization

$M_S$ – Saturation Magnetization

$H_C$ – Coercive Field

$M_R$ – Remnant Magnetization

$T_C$ – Curie temperature

$T_B$ – Blocking temperature

$H$ – Applied magnetic field

$V$ - Volume

T - Temperature

$K_{Eff}$ – Anisotropy Constant

C – Curie Constant

$T_C$ – Curie Temperature

$T_N$ – Néel Temperature

$K_B$ – Boltzman Constant

$T_{Irr}$ – Irreversibility Temperature

$T_B$ – Block Temperature

U - Entropy

Q - Heat

W – Work

$D_{XRD}$ – Crystallite size

$d_{TEM}$ – Particle Size

$P_L$ – Power Loss

$H_A$ – Hysteresis Area

f - Frequency

# List of Symbols:

$\mu_e$ – Bohr Magneton

$m_e$ – Electron mass

e – Electron Charge

$h$ - Plank constant

ħ – Plank Constant/$2\pi$

$S$ – Spin Value

$L$ – Orbital Momentum Value

$J$ – Total Angular Momentum Value

$g_e$ – Landé Electron g-factor

μ - Atomic spin

χ – Susceptibility

$\tau_N$ – Néel Relaxation Time

$\tau_B$ – Brownian Relaxation Time

$\tau_{eff}$- Relaxation Time

$\delta$ – Inversion Parameter

$\lambda$ – Wavelength

$d$ – Interplanar Distance

$\beta_{TOTAL}$ - FWHM

$\beta_{SIZE}$ – $D_{XRD}$ contribution for FWHM

$\beta_{STRAIN}$ – Strain contribution for FWHM

$\eta$ - Strain

σ – Standard Deviation

μ - mean value

$\Theta_p$ – Paramagnetic Curie Temperature

$\mu_0$ – Vacuum permeability

# Chapter 1: INTRODUCTION

Magnetic materials attracted humanity's curiosity since ancient times. The characteristic distant interaction between a magnet and a magnetic field still nowadays amazes people who are not familiarized with magnetic phenomena. The first magnetic material discovered by Mankind was Lodestone in Greece, 800 BC. Lodestone contains magnetite, $Fe_3O_4$, which is a naturally magnetized mineral, that was used to create the first compasses, [1]. Today, in the new era of nano research, the magnetic materials are getting smaller and interesting magnetic properties are being studied and applied in novel applications such as medicine, higher efficiency devices or environmental aid.

Every material reacts somehow when a magnetic field is applied to it. Some materials can have a magnetic field of their own, those are magnets or ferromagnets. When a ferromagnet is heated, its magnetic properties weaken until it completely loses its magnetization. The temperature at which a ferromagnet loses its magnetization is the Curie temperature.

The main purpose of this master thesis is to tune the Curie temperature in nanoparticles of Mn-Zn ferrite ($Mn_{1-x}Zn_xFe_2O_4$) with the objective of developing self-regulated heating at the nanoscale.

The interest in the Curie temperature tuning in nanoparticles of this material, is the vast range of applications emerging from it. The Mn-Zn ferrite is frequently mentioned for hyperthermia treatments, due to its nano size, biocompatibility and Curie temperature close to room temperature, [2]. The self-regulated heating mechanism can also be used for self-pumping devices, energy harvesting, controlled melting, cookware, among many other applications, [3] [4] [5]. The advantage of preparing these systems at the nanoscale is the possibly of scaling them to a larger scale, as a bulk material, a ferrofluid, or even incorporated as a part of a composite.

## 1.1 MOTIVATION AND OBJECTIVES

The engineering of magnetic nanoparticles is a contemporary subject with applications in many areas of research. For this master thesis, the challenge is to synthesize magnetic nanoparticles capable of self-regulating their temperature in a magnetic induction heating experiment, via Curie temperature tuning. The material of election was Mn-Zn ferrite due to the dependence of the Curie temperature with the Zn/Mn ratio. In addition, the Curie temperature is expected to change from below room temperature until high temperature with increasing Mn content (~300ºC). Another valuable feature of Mn-Zn ferrite is its biocompatibility. In this point this ferrite outcomes other ferrites and hence signals the Mn-Zn ferrite as a potential candidate for biomedical treatments, for example, cancer treatment via hyperthermia, [6].

The main objective of this master thesis is to synthesize magnetic nanoparticles of Mn-Zn ferrite, $Mn_{1-x}Zn_xFe_2O_4$, capable of self-regulating their temperature for a further application in a ferrofluid. This family of ferrites had never been synthesized by this research group and, for this reason, the development of good quality samples, for further applications, is a priority. This master thesis objectives are the following:

- Synthesis optimization of $Mn_{1-x}Zn_xFe_2O_4$ nanoparticles with different compositions.
- Structural and magnetic characterization of the samples.
- Comprehension of the magnetic phenomena in this material and characterization techniques.
- Achievement of the self-regulated heating mechanism via Curie temperature.

## 1.2 THESIS LAYOUT

The present report is divided in 6 chapters. In “Chapter 1: Introduction”, the reader will find the objectives and motivations of this master thesis work. In “Chapter 2: Magnetic behavior”, an introduction of the magnetic phenomena related with the synthesized samples is discussed. Due to the nano-size nature of the samples, this chapter starts with an increasing in scale until macroscale is achieved and then drops again to the nanoscale magnetic behaviors. “Chapter 3: Mn-Zn ferrite” is about this remarkable compound, it starts with the synthesis methods, passing to the crystallographic structure and ends in magnetic structure. In “Chapter 4: Experimental Procedure”, the synthesis methods (sol-gel auto-combustion and hydrothermal) are detailed, then the physical principles of the characterization techniques are explored. “Chapter 5: Results and Discussion”, starts with the results and discussion of the samples synthesized by sol-gel auto-combustion method and ends in the results and discussion of the hydrothermally synthesized samples. This master thesis content ends in “Chapter 6: Conclusions”, in this chapter the reader will find some conclusions about the samples synthesized by both methods and suggestion for further work. After the conclusions the References are presented. In the last pages of this report a few Appendixes with complementary information are provided.

# Chapter 2: MAGNETIC BEHAVIOR

All materials present a magnetic behavior under the influence of a magnetic field, this behavior is explained by the presence of electrons (charges) in all materials. Magnetic fields are created by moving charges, like magnetic fields created by solenoids, or, as an intrinsic property of elementary particles, such as electrons, protons or neutrons - this quantum property is called spin. In ferromagnetic materials, electrons are responsible for the magnetic field. For protons and neutrons the magnetic moment is several orders of magnitude lower than for electrons, 2000 times lower, [1]. The magnetic moment of an electron, $\mu_e$, is 1.0012 $\mu_B$, [7], and is related to the Bohr magneton, $\mu_B$ =−9.284×10$^{-24}$ J/T (SI), by the relationship $\mu_e = g_e \frac{\mu_B}{\hbar} S$, where $g_e$ is the Landé g-factor ≈2 and S the spin value ($\frac{\hbar}{2}$).

Electrons surround the atomic nucleus in discrete energy levels, N, each energy level has 2N+1 orbital for electrons to occupy. Each of these orbitals is classified with an L value, related with the angular orbital momentum. The electrons' angular orbital momentum generates a magnetic field which interacts with the electrons' spin. Concluding, every atom magnetic moment is dependent of the total angular momentum, J, arising from the interaction between electrons spin and electrons orbit, calculated as |L+S|, …, |L-S|, for the ground state. The J value is proportional to the atomic magnetic moment, $\mu$, equation 2.1, where $g_J$ is the Landé g-factor, $\mu_B$ is the Bohr magneton, [8].

$$\mu = -g_J \frac{\mu_B}{\hbar} J \quad (2.1)$$

Temperature interferes with the electrons' distribution, accordingly with Fermi-Dirac distribution. Any temperature above 0K shifts the electrons' ground state for higher energies, probabilistically. The temperature plays a major role in the magnetic properties of all materials.

Every material has electrons, as consequence, every material is affected by electromagnetic fields. At the ground state, electrons group up accordingly with the Hund-rules which determines which electronic orbitals are first filled, the inner orbitals should be fully filed with paired electrons before outer-most orbitals. Paired electrons are found in almost all elements and compounds. These pairs of electrons are responsible for magnetic repulsion (diamagnetic behavior). Unpaired electrons, in the outer orbitals are responsible for magnetic attraction.

When dealing with a molecule, the electromagnetic interactions between atoms influence the spin value of each atom. The molecules magnetic properties are affected by the electromagnetic fields arising from the atomic arrangement (hyperfine structure), magnetic interactions (further explained), temperature, external magnetic fields and others.

Section 2.1: Magnetic Properties is an introduction to the magnetic properties commonly found in all materials. Section 2.2: Macro Magnetism is where the magnetic properties of the previous section are analyzed for different macroscale magnetic behaviors. Section 2.3: Nano Magnetism, is directly related to the synthesized samples and to ferromagnetic behavior at the nanoscale. Section 2.4 goes straight to the objective of this work, understanding the magnetic induction heating mechanisms and heat generation phenomena.

## 2.1 MAGNETIC PROPERTIES

This section intends to provide the reader with knowledge about fundamental magnetic properties. Magnetization and magnetic susceptibility, when an external field is present. Magnetism is a result of exchange interaction, superexchange interaction, magnetocrystalline anisotropy and dipolar interactions. The magnetic properties are temperature dependent, as will be detailed in section 2.2.

### *2.1.1 MAGNETIZATION*

The magnetization of a material is described as "the magnetic dipole moment regarding either unit of volume or unit of mass", [9]. The magnetization of a material is consequence of the alignment of the atomic dipolar moments of a material in the presence of a magnetic field. Considering a finite number of non-interacting magnetic moments, n, each with a magnetic moment, μ, (mis) aligned with the field direction by an angle θ, the total magnetization of the group is given by equation 2.1, [10].

$$\vec{M} = \int_0^n \mu_n \cos(\theta_n) dn \quad (2.1)$$

The more aligned moments, the higher the magnetization, assuming the same volume or mass. The magnetic order saturates for high fields when all magnetic moments are aligned. The magnetization units are, usually, normalized to the mass or volume of the sample, emu/g or emu/cm$^3$ (cgs).

### *2.1.2 Magnetic Susceptibility*

The magnetic susceptibility is a property of every material, since all materials react somehow when a magnetic field is applied to it. A material with unpaired electron is attracted towards a magnetic field because its atomic dipolar moments align in the field direction, experiencing attraction. Between two materials, the material with a higher magnetization for the same intensity of applied field, has a higher magnetic susceptibility.

Magnetic susceptibility, $\chi$, is defined as the amount of magnetization, $M$, induced in a material when a magnetic field, $H$, is applied to it, equation 2.2

$$\vec{M} = \chi \vec{H} \quad (2.2)$$

Magnetic susceptibility is measured with resource to magnetometers. Its measurement is done by changing the intensity of an applied magnetic field while measuring the materials magnetization, M(H) measurements. These measurements are performed at constant temperature. the magnetic susceptibility is temperature dependent. The temperature effect will be individually detailed for each magnetic behavior.

The magnetic alignment is field constant when considering non-interacting magnetic moments, like in paramagnetic behavior. When dealing with interacting magnetic moments, for example in the ferromagnetic state, the susceptibility varies with temperature.

### 2.1.3 Exchange and Superexchange Interaction

Electrons are indistinguishable particles, since all electrons present the same fundamental properties (mass, charge and spin), [9]. Electrons wave-function spread through space as time passes, [11]. In a molecule, covalent bonding causes the electrons' wave-function to overlap between atoms, which means that electrons have a non-zero probability of exchange its position with another electron of the molecule.

Exchange interaction is a purely quantum-mechanical phenomenon as it affects electrons in an atom or in close-neighbor atoms. When the exchange interaction affects close-neighbor atoms it is referred to as direct exchange. When the exchange interaction is mediated through a non-magnetic atom is called superexchange. The exchange interaction is the main cause of ferro- and antiferro-magnetism.

Figure 2.1 shows magnetic couplings in a ferromagnetic material, direct exchange and superexchange, both are mediated by the overlapping of the wave-functions, [1].

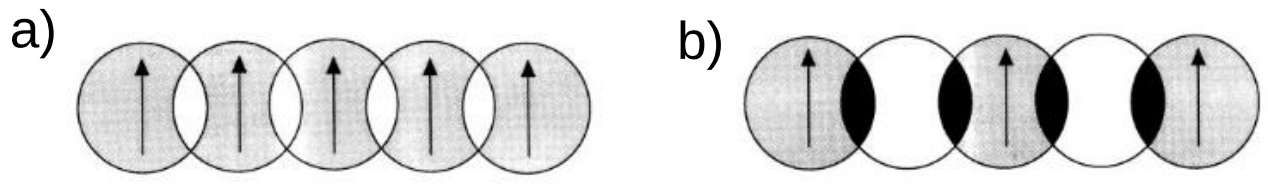

**Figure 2.1 –** Schematic representation exchange interactions in a ferromagnet: a) direct exchange. b) superexchange.

### *2.1.4 Magnetocrystalline Anisotropy*

"The magnetocrystalline anisotropy is manifested by locking magnetic moments in certain crystallographic directions", by V. Sechoveský [1]. The magnetocrystalline anisotropy is a consequence of magnetic alignment between neighbor atoms of a crystallographic plane. When the magnetic alignment occurs between close-neighbor atoms, this direction is called an easy axis. If the alignment occurs between further atoms is called a hard axis. The magnetocrystalline anisotropy causes the magnetic domains to be preferentially aligned in specific crystallographic directions.

The magnetocrystalline anisotropy is affected by the domain volume and the domain shape. Assuming a spherical shape domain, the larger the domain volume, the longer the chain of aligned magnetic moments, consequently, the larger the magnetocrystalline anisotropy, $K_a(\theta)$ given by equation 2.3, [12].

$$E_a(\theta) = K_{Eff}Vsen^2(\theta) \quad (2.3)$$

Where $\theta$ is the angle between the magnetization vector and an easy axis. The energy barrier between two easy axis correspond to the product of the material anisotropic constant and the domain volume, $KV$. The magnetocrystalline anisotropy has a maximum value when the magnetization is perpendicular to an easy direction, which corresponds to the highest potential energy of the system.

A polycrystalline material is composed of many crystals, or magnetic domains, each crystal has its own preferred direction for alignment. The interaction between magnetic domains is mainly dipolar.

### *2.1.5 Dipolar Interactions*

The dipolar interaction is the weakest of the interactions previously mentioned. This interaction is the classical magnetic force between a magnet (dipole) and a magnetic field. The dipolar interaction occurs between magnetic domains or magnetic particles. Its strength depends on the characteristics of the dipoles, such as magnetic moment, distance between dipoles and interaction angle.

The dipolar interaction is responsible for the directional magnetization of a ferromagnet. If a magnetic field is applied to a demagnetized ferromagnet, it aligns its domains in the field direction, through dipolar interaction. After the field is removed, the ferromagnet might retain its magnetization (remnant magnetization). In order to return the magnetization to zero, an opposite magnetic field can be applied (coercive field). Both coercive field and remnant magnetization are a direct consequence of the dipolar interaction between magnetic domains.

## 2.2 Macro magnetism

Macro magnetism section is dedicated to the different magnetic behaviors found in bulk materials. These behaviors are categorized as diamagnetic, paramagnetic, ferromagnetic, antiferromagnetic and ferrimagnetic. Other forms of magnetism will not be discussed, such as metamagnetism, spin glass and molecular magnetism.

### *2.2.1 Diamagnetism*

Diamagnet materials have the interesting property of repelling an external magnetic field, in opposition to all others magnetic behaviors. Some examples of diamagnetic compounds are water, graphite, silver, mercury and bismuth. The magnetic susceptibility in diamagnets have a negative value, the materials with higher diamagnetic behavior are pyrolytic carbon ($\chi$ = -23x10$^{-5}$ cm$^3$/g, [13]) and bismuth ($\chi$ = -16.6x10$^{-5}$), water has a susceptibility of -0,91x10$^{-5}$, [14]. The strongest diamagnetic materials are superconductors: they behave as ideal diamagnets, completely repelling magnetic fields. Superconductors have a susceptibility of -1, [1], which enables them to have a zero total magnetic field within their volume.

The characteristic repulsion of diamagnets for an applied magnetic field is responsibility of paired electrons. Due to compensated spins, the electrons pair has a null magnetic moment, thus, paired electrons do not align in the magnetic field. Instead, the movement of paired electrons create a magnetic field with opposing polarization, experiencing repulsion from a magnetic field, Lenz law. Every material has an intrinsic diamagnetic behavior, since most elements have paired electrons in their fully filled orbitals.

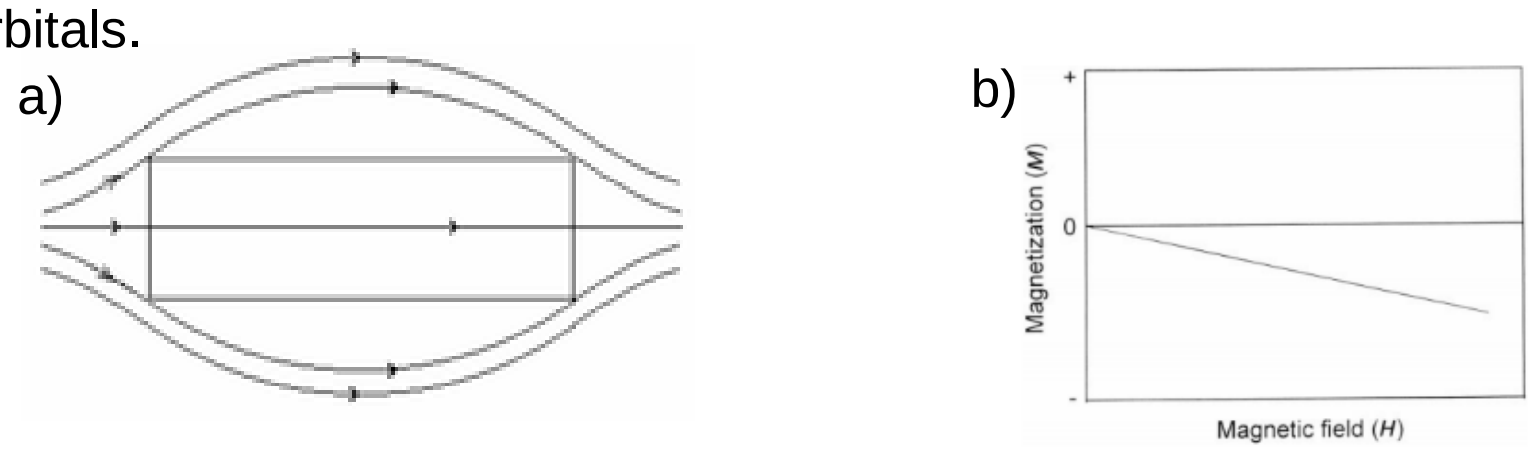

**Figure 2.2 -** Diamagnetism; a) Field lines in diamagnetic material; b) magnetization dependence of an external applied field.

### 2.2.2 PARAMAGNETISM

Paramagnetism is the most common type of magnetic behavior at room temperature. Some examples of paramagnetic materials are oxygen, aluminum and titanium. Paramagnets are weakly attracted to magnetic fields and do not retain any magnetization at zero magnetic field.

Paramagnetic materials have unpaired electrons. Without an external magnetic field, the electrons are randomly oriented, figure 2.3.a. The magnetization of a paramagnet, without an applied magnetic field, is 0 emu/g, as result of the randomly oriented atomic spins.

In the presence of an external magnetic field, the atomic spins tend to align parallelly to the magnetic field causing a weak attraction of the material towards the field source. Thus, in order to measure these materials magnetization and magnetic susceptibility an external magnetic field must applied. The magnetization follows a linear tendency with increasing magnetic field, which is consequence of a nearly field independent magnetic susceptibility, figure 2.3.b). The magnetization of paramagnets reaches a maximum, saturation magnetization, value when all the atomic spins are aligned in the field direction, although this phenomenon is only observable typically at high magnetic fields. When the external magnetic field is removed, the material returns to the zero-magnetization state, as the thermal agitation randomly orientates the atomic spins.

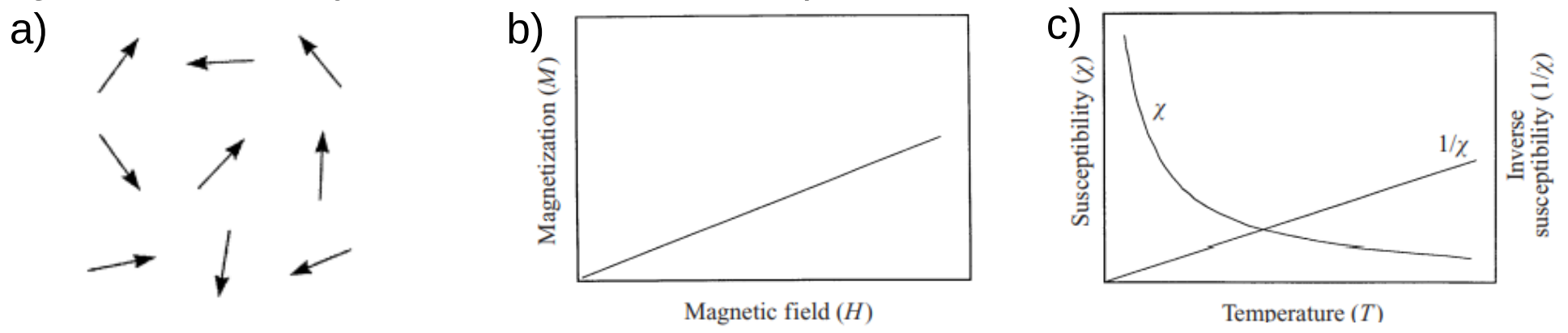

**Figure 2.3 –** Paramagnetism; a) Schematic representation of magnetic dipoles in a paramagnet; b) Magnetization dependence with the applied field; c) Susceptibility and inverse susceptibility dependence with temperature.

The magnetic susceptibility, is dependent of the applied magnetic field and temperature, as presented in section 2.1.2. This equation was formalized by Pierre Curie, [8], in the Curie law, equation 2.4.

$$\chi = \frac{M}{H} = \frac{C}{T} \tag{2.4}$$

Where $\chi$ is the susceptibility, C is the Curie constant of the material, H the magnetic field and T the temperature. By analyzing equation 2.4 it is noticeable that the susceptibility increases inversely proportional with the magnetic field and decreases with increasing temperature. Increasing temperature decreases magnetization, as it suppresses the magnetic interactions. This equation is not valid for high magnetic fields, due to the saturation of the magnetic domains or phase transitions caused by temperature.

Paramagnetic materials at room temperature might not be paramagnetic at lower temperatures. Paramagnetism is temperature dependent - the thermal energy is competing with the magnetic energy by changing the atomic spins orientation. At lower temperatures, paramagnets at room temperature might change its magnetic structure, this effect is referred as phase transition. The phase transition from paramagnetism to ferro-/ferri-magnetism is said to occur at the Curie temperature. For example, bulk Gadolinium is paramagnetic above 20ºC (293.15 K), while below this temperature it is paramagnetic. The phase transition from paramagnetism to antiferromagnetism occurs at the Néel temperature. Manganese oxide, $MnO_2$, is paramagnetic above -157ºC and becomes antiferromagnetic below this temperature. Paramagnets which are paramagnetic at 0K are called ground-state paramagnets.

### 2.2.3 FERROMAGNETISM

Ferromagnetism has a highlighted place between magnetic behaviors, due to the capacity of these materials to retain a ma gnetic field. Ferromagnetism is the most researched type of magnetic behavior and occurs at room temperature in elements such as Iron, Cobalt, Nickel and Rare-Earth elements. It can have several applications ranging from fridge magnets, electric motors, compasses, transformers and other devices, [8].

The remnant magnetization arises from ferromagnets as consequence of magnetic domains alignment inside the material. A magnetic domain is typically associated with the crystallite of the ferromagnet, which has its own preferred direction for magnetic alignment, as discussed previously in section 2.1.4. The atomic spins alignment within a magnetic domain is a result of preferred orientation between atoms of a crystallographic plane (magnetocrystalline anisotropy). Figure 2.4.a) is a representation of atomic spin arrangement in a ferromagnetic unit cell. Figure 2.4.b) is the representation of two domains, separated by domain-wall, [8]. Figure 2.4.c) is an image of magnetic domains in a ferromagnetic material, obtained using a Kerr-effect microscope, [15].

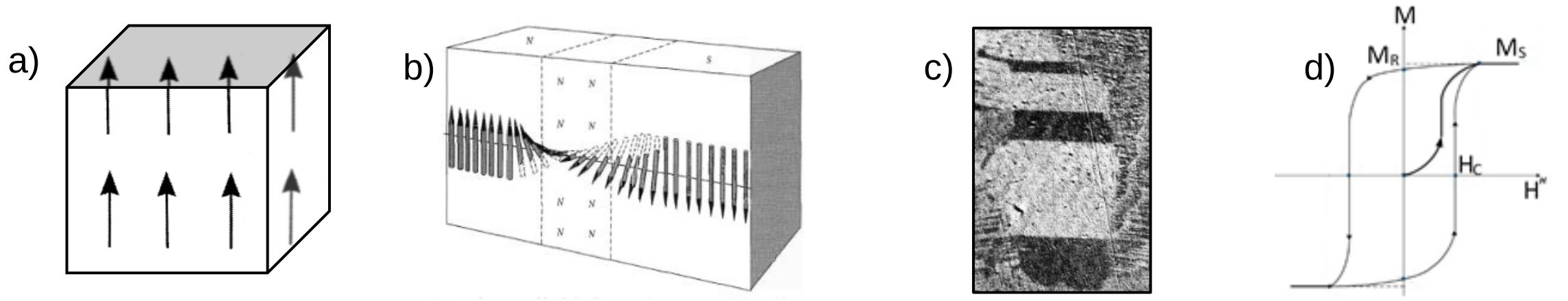

**Figure 2.4 –** a) Ferromagnetic unit cell. b) domain-wall dividing two antiparallel domains c) ferromagnetic domain observed by Kerr-effect microscope (author: C. V. Zureks). d) ferromagnetic M(H) curves: $M_S$ - saturation magnetization, $M_R$ - remant field, $H_C$ - coercive field.

#### 2.2.3.1 Hysteresis Loops

When a ferromagnet is demagnetized, its ferromagnetic domains are misaligned. If a magnetic field is applied to the ferromagnet, the magnetic domains tend to align parallelly with the applied magnetic field. When all the magnetic domains are pointed in the field direction the material reaches a point of magnetic saturation, $M_S$. When the applied field is removed, the aligned magnetic domains relax to more stable orientation, however, some magnetization is retained in the applied field direction, remnant magnetization, $M_R$. The remnant magnetization grants the ferromagnets with their own magnetic field. This remnant magnetization is only lost by heating the ferromagnet above the Curie temperature or by applying an opposite magnetic field with enough intensity for the magnetization to become zero, called coercive field, $H_C$. The dependence of the ferromagnet magnetization with an applied field is called a magnetic hysteresis loop, figure 2.4.d).

The hysteresis loops, obtained by M(H) measurements, are consequence of magnetic domain aligning their magnetic moments in the presence of a magnetic field. This behavior can be described by three characteristic points of a hysteretic curve: saturation magnetization, remnant magnetization and coercive field. Saturation magnetization occurs when all the sample magnetic domains are aligned in the field direction. Remnant magnetization is the sample remaining magnetization when the magnetic field is removed. Coercive field is the required field to take the sample magnetization to zero. All these points are marked in figure 2.4.d).

The inversion of magnetic domains, or domain walls movement, causes the crystal lattice to heat. The hysteretic loss can be used for heat generation via magnetic induction heating and will be detailed in section 2.4.

#### 2.2.3.2 Temperature Effect

The temperature effect on a material magnetization is usually observed in measurements of magnetization dependence of the temperature, M(T) measurements. In ferromagnets, the thermal energy (randomizing aligned spins) competes with the magnetic interactions (spins alignment). The temperature which causes a material to alter its magnetic behavior from the ferromagnetic state to the paramagnetic state is the Curie temperature, as referred in 2.2.2.

Below the Curie temperature, in the ferromagnetic state, the temperature is not sufficiently high to compete with the magnetic interactions mentioned in sections 2.1.3 to 2.1.5. While increasing the temperature, the overall magnetization weakens as the thermal energy gradually overcomes the magnetic exchange interaction energy, promoting the misalignment of atomic spins or magnetic domains. Above the Curie temperature, the thermal energy is higher that the magnetic exchange interactions and hence while temperature increases the magnetic domains are destroyed (paramagnetism).

The susceptibility, $\chi$, of a ferromagnet in the paramagnetic state was formalized by Pierre Currie and Pierre-Ernest Weiss in the Curie-Weiss law, equation 2.5.

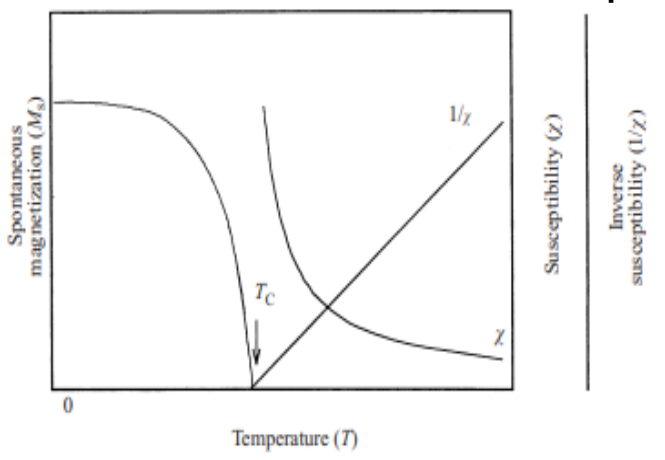

$$\chi = \frac{C}{T - \Theta_p} \qquad (2.5)$$

**Figure 2.5 –** Magnetic susceptibility and inverse susceptibility of a ferromagnetic material.

Where C is the material Curie constant, T is the temperature and $\Theta_p$ is the paramagnetic Curie temperature, [1]. Figure 2.5 shows the dependence of ferromagnetic magnetization with temperature, M(T) curve. Above the Curie temperature, the susceptibility decreases inversely proportionally with the increase of temperature. Measuring the inverse susceptibility, $\chi^{-1}$, a linear paramagnetic behavior is revealed, being possible to estimate $\Theta_p$, the Curie temperature of the ferromagnetic material by fitting its linear behavior as function of T. The $\Theta_p$ differs from the Curie temperature since it is a prediction in the paramagnetic state.

### *2.2.4 ANTIFERROMAGNETISM*

Antiferromagnetism is a magnetic ordering occurring in materials where crystallographic planes have anti-parallelly aligned atomic spins, figure 2.6.a). A group of antiparallel spins forms an antiferromagnetic domain, figure 2.6.b), [16]. The anti-parallel alignment is a consequence of exchange, or superexchange, interaction between atoms of a crystallite.

When a magnetic field is applied to the antiferromagnet, an increase in its magnetization is observed. The positive susceptibility verified is due to an unequal number of antiparallel spins in the domains, which causes a weak magnetization oriented in the field direction. The dependency of magnetic susceptibility with the magnetic field is linear until saturation, [1], resembling the paramagnets susceptibility, figure 2.6.c).

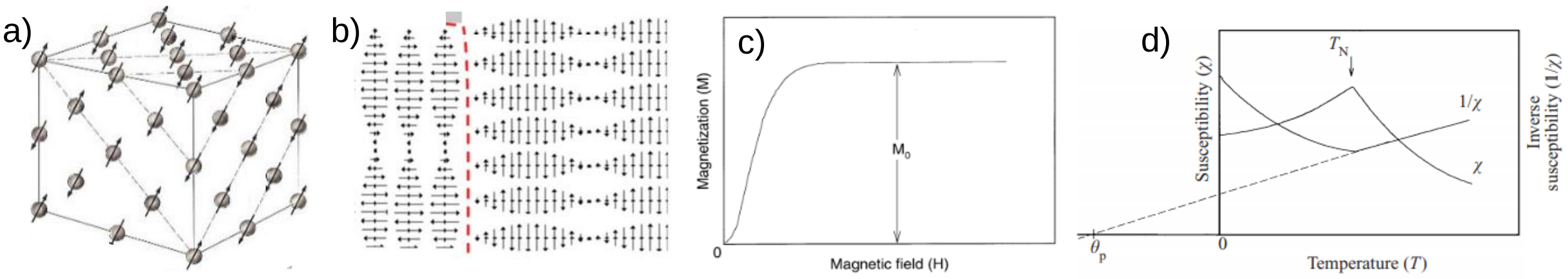

**Figure 2.6 –** a) Unit cell of antiferromagnetic moments. b) Antiferromagnetic domains. c) Magnetization increases with external field. d) Susceptibility and inverse susceptibility as function of temperature, the maximum value is the Néel temperature.

Figure 2.6.d), [1], shows the susceptibility dependence of temperature. The maximum value of $\frac{\partial}{\partial T}\chi(T)$ is the Néel temperature, $T_N$. At this point, temperature induces a magnetic phase transition: below $T_N$ the material is antiferromagnetic and above $T_N$ is paramagnetic.

Examples of antiferromagnetic material are hematite ($Fe_2O_3$), chromium and nickel oxide.

### *2.2.5 FERRIMAGNETISM*

Ferrimagnetism is the magnetic behavior present in most of the synthesized samples of this report. Ferrimagnetism was a term originally proposed by Néel to describe magnetic order of ferrites, [1]. Ferrites, and other ferrimagnets, are characterized for possessing ions with different magnetic moments in an antiferromagnetic arrangement. These materials present a spontaneous magnetization as consequence of the uncompensated magnetic moments. The dipolar interaction of the unequal antiparallel magnetic moments often induces canting effects in the weaker magnetic ion.

These materials are usually ceramic oxides, where the oxygen is the responsible for the antiferromagnetic alignment of different ions. Ferrimagnetic arrangement is a consequence of superexchange interaction. The ferrimagnetic materials resembles ferromagnets, once the uncompensated magnetic spins create magnetic domains. Ferrimagnetic domains possess a spontaneous magnetization.

When a magnetic field is applied to a ferrimagnet its magnetic domains align in the field direction. When the magnetic field is removed, they retain magnetic moment due to dipolar interaction between domains. The ferrimagnetic hysteresis loop is like the ferromagnetic hysteresis loop. The temperature has a similar effect between ferrimagnet and ferromagnet. When a ferrimagnet heats above the Curie temperature it becomes paramagnetic and loses its magnetic ordering. When cooled below the Curie temperature, the ferrimagnet is demagnetized due to the misalignment of the magnetic domain.

## 2.3 NANO MAGNETISM

Magnetic nanomaterials is an area that attracted many research efforts lately, due to its vast range of applications in the multiple fields, such as magnetic recording, [17], high-efficiency devices, biomedicine and other. During the last 60 years, improvements in synthesis methods and characterization techniques, allowed control over size, shape and magnetic properties of nanoparticles. Also, high resolution characterization techniques allowed researchers to understand nanoparticles' magnetic properties in detail.

Despite magnetic nanoparticles having the same magnetic ordering as the bulk material, the size effect has several implications in the nanoparticles magnetic behavior. The large surface/volume ratio, characteristic of nanoparticles, greatly impacts their structural and magnetic properties. In particular, $M_S$, $H_C$, $M_R$ and their temperature dependence is affected by their reduced size. Section 2.3.1, refers to ferromagnetic nanoparticles, due to their similarity with ferrimagnetic single domains.

a)

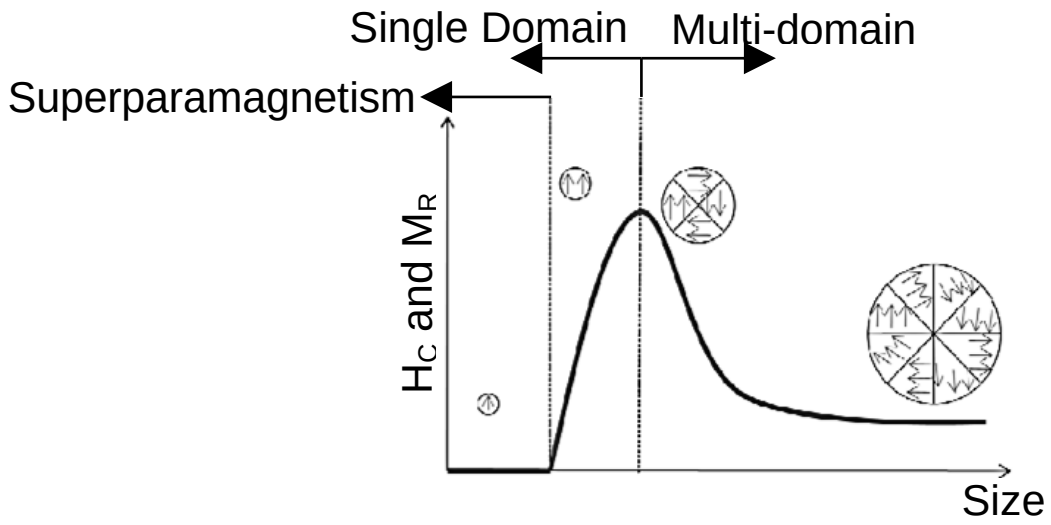

b)

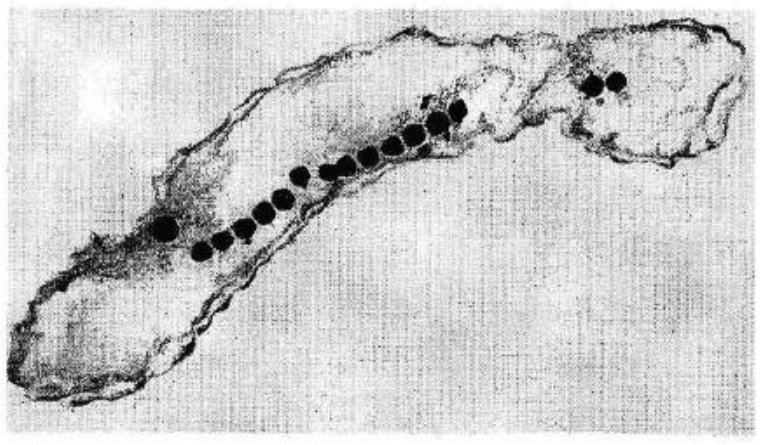

**Figure 2.7 –** a) $H_C$ and $M_R$ dependence with the particle size and single domain regimes. b) chain of magnetic nanoparticles of $Fe_3O_4$ found inside a magnetotactic bacteria (draw by Marta Puebla).

### 2.3.1 *SINGLE DOMAIN*

Ferromagnetic materials are composed of multiple domains, separated by domain walls. When the particle size is small, typically about 100nm, [18] , the formation of a domain wall is not energetically favorable. Thus, the entire particle becomes a magnetic domain, or, single domain, [19]. The ferromagnetic single domain has all atomic spins aligned in the same direction, which confers the particle a high value of spontaneous magnetization. Furthermore, a single domain has a high value of coercive field, since in order to rotate the particle magnetization the entire magnetic domain rotates. When a magnetic field is applied to a system of single domains, an increase in magnetization is expected, as result of single domains alignment in the field direction. Without an applied field the system present a low value of remnant magnetization, consequence of the low magnitude dipolar interactions between particles contributing to reduce the system magnetization.

#### 2.3.1.1 SIZE AND SHAPE EFFECTS

The size and shape greatly affect the magnetic properties of the single domain. The increase of the single domain volume increases the magnetic anisotropy, similarly with what occurs with the magnetocrystalline anisotropy. The shape anisotropy of the single domain can lead to an increase of the magnetic anisotropy, affecting preferred orientation for the magnetization.

The single domain is usually represented by the Stoner–Wohlfarth model, figure 2.8.a). In this model, the magnetic anisotropy is consequence of the particle magnetization rotating with a phase diference of the easy axis, φ-ϴ. The angle between magnetic field and particle easy-axis is ϴ and with magnetization is φ. For a system of Stoner-Wohlfarth particles, when a magnetic field is applied, the single domains rotate in the field direction until all magnetic moments are aligned, saturating the magnetization of the single domains system. However, it is frequently reported that systems composed of single domains do not saturate, even at high fields. The cause of this is that the atomic surface spins are misaligned with the spins of the single domain core. While the core spins are bounded by exchange interactions to the surrounding spins, the superficial spins are only bounded with the core, thus, they are relatively free to align in other direction. This phenomenon is known as canting and it is a common cause for the very high fields required to achieve complete saturation magnetization, figure 2.8.b) and c), [20].

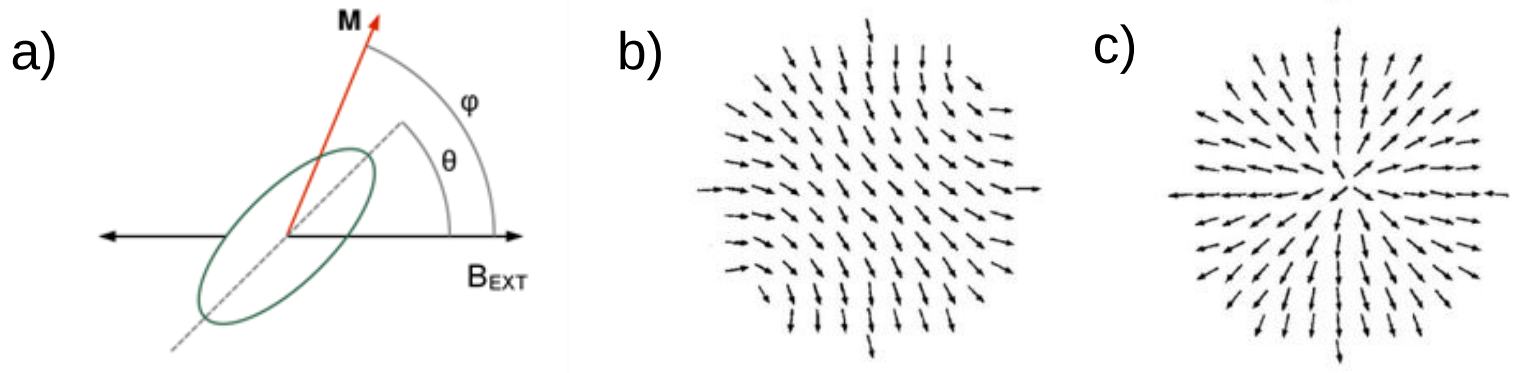

**Figure 2.8 –** Single domain representations. a) Stoner–Wohlfarth model. $B_{EXT}$ is the externally applied magnetic field direction. M is the magnetization vector rotated φ from the applied field. ϴ is the angle between an easy axis and $B_{EXT}$. b) and c) is the representation of atomic spin canting at the nanoparticle surface.

#### 2.3.1.2 TEMPERATURE EFFECTS

The effect of the temperature in a single domain has similarities with the bulk material, if the bulk material presents phase transitions, the single domain might also present the same phase transitions. The magnetic phase transition from ferromagnetic to paramagnetic, Curie temperature, also occurs for ferromagnetic single domains. However, due to finite-size effects, the temperature firstly affects the superficial spins, transiting them to the paramagnetic state. The paramagnetic layer formed at the surface of the single domain is known as dead layer, figure 2.19.a).

Temperature and time have curious effects in magnetic nanoparticles. As time passes the particles' magnetization can be flipped to a different easy axis. The time it takes for the magnetization to be thermally fluctuated is called Néel relaxation time, equation 2.6.

$$\tau_N = \tau_0 \exp\left(\frac{K\,V}{K_B\,T}\right) \tag{2.6}$$

The relaxation time is dependent of the energy barrier between easy axis (anisotropy), KV, and the thermal energy, $K_B\,T$. $\tau_0$ is approximately $10^{-9}$ s, depends on the chemical composition of the single domain, [20]. The larger the anisotropy, or the lower the temperature, the longer the relaxation time. This equation explains why the magnetization of a bulk sample is not spontaneously changed by temperature, due to its large volume, the anisotropy is orders of magnitude larger that the thermal energy. When the thermal energy is comparable with the anisotropy energy, the nanoparticle magnetization rapidly fluctuates between easy axis. When measuring a system of single domains, the time window of the measurement is relevant in order to determine the system magnetic behavior. If the Néel relaxation time is significantly longer that the measurement window, the magnetization appears to be static during the measurement - the single domain is in the blocked state. If the Néel relaxation time is shorter that the time window, the magnetization flips during the measurement and, consequently, the magnetometer will measure with equal probability a up (positive) or down (negative) magnetization - the single domain is in the superparamagnetic state. The temperature which separates both states is the blocking temperature, $T_B$. A system with a distribution of volumes will simultaneously present a distribution of blocking temperatures, accordingly with equation 2.6, $\tau_N$ depends on volume.

The thermal fluctuations in the superparamagnetic regime demote the system of any remnant magnetization and coercive field. However, each particle is still characterized by a high value of magnetization, which grants these systems a high susceptibility to the magnetic field, [18].

The blocking temperature distribution is usually identified in magnetization versus temperature measurements, M(T) measurements, 2.9.b). The M(T) measurements are performed under two protocols: cooling the sample with and without magnetic field, field cool (FC) and zero-field cool (ZFC), respectively. The FC imposes that the single domains are blocked with a preferred alignment in the field direction. ZFC, allows the nanoparticles to block freely oriented, thus, decreasing the total magnetization of the system. Both measurements are performed during sample warming and with an external field applied. When plotting both curves, figure 29.b), at the lowest measured temperature, typically the FC and ZFC have different values of magnetization, which is a consequence of the nanoparticles being blocked in different magnetic arrangements. Increasing the temperature will allow the ZFC single domains to be unblocked and align in the field direction, until a maximum value of magnetization is obtained. The magnetization drops due to the superparamagnetic behavior above the blocking temperature. The temperature at which both curves merge is known as irreversibility temperature, $T_{Irr}$, and is the temperature where the particle with higher blocking temperature is unblocked, [21]. For higher temperatures, both curves superimpose into a single curve, due to all the single domains being above the blocking temperature, in the superparamagnetic state. Further increasing temperature will contribute for the superparamagnetic nanoparticles to transform its magnetic phase from superparamagnetic to paramagnetic, at the Curie temperature.

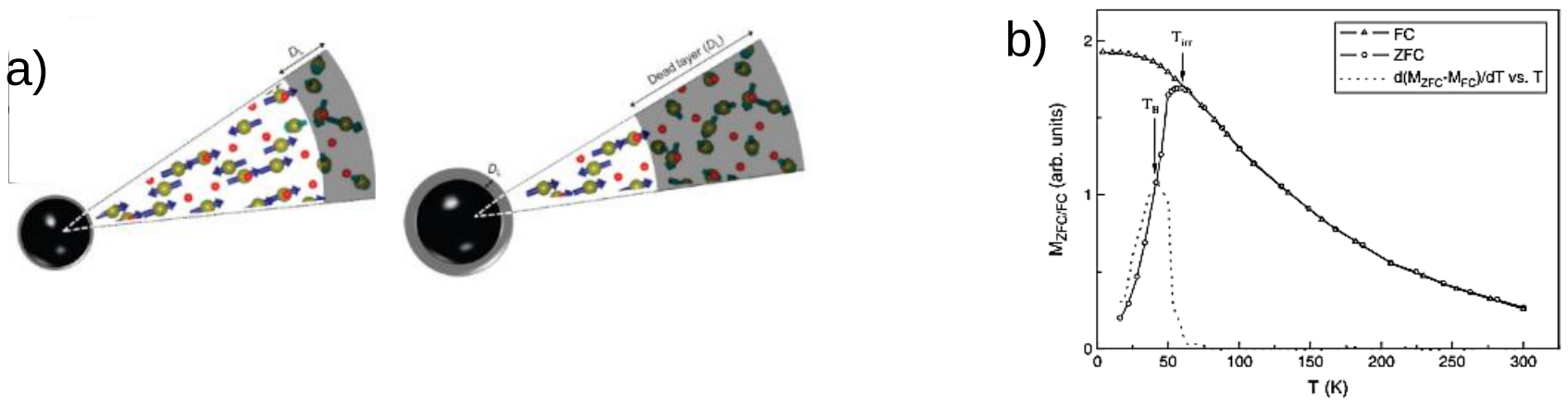

**Figure 2.9 –** a) Dead layer increase with temperature increase. b) M(T) measurement of single domains and $T_B$ distribution, $f(T_B)$ (point dashed). $\langle \mathbf{T_B} \rangle$ and $T_{Irr}$ are signed in the M(T) curve.

## 2.4 MAGNETIC INDUCTION HEATING MECHANISMS

The magnetic induction heating mechanisms section is dedicated to the heat generation mechanisms found in magnetic materials, or magnetic nanoparticles. Magnetic induction is a known phenomenon since Michael Faraday discovered it in 1831, [22]. Since then, magnetic induction heating has become more efficient and widely available as a heating mechanism. Nowadays, magnetic induction heating mechanisms are used in induction plates for cooking, biomedical treatments or in the metallurgical industry for metal melting, [23], [24] [25]. In some cases, magnetic induction heating is an undesired effect, for example, through the heat generated in a transformer core or in wireless battery chargers.

Magnetic induction heating is based on the electromotive force experienced by a conductive material upon the application of a magnetic field. In order to an electrical current to be induced in the material, the intensity of the magnetic field should vary in time. For this reason, most magnetic induction heating setups use oscillating magnetic fields with frequencies in the kHz or MHz ranges.

Heating a ferromagnet via magnetic induction is highly efficient due to the multiple heat generation mechanisms that these materials presents. Conductive materials generate heat only through Joule effect, ferromagnets not only generate heat via Joule effect but also by hysteretic loops. In nanoparticles form, ferromagnets generate heat through Néel and Brownian relaxation. The following sections are dedicated to the heat generation mechanism found in ferromagnetic materials.

### *2.4.1 EDDY CURRENTS*

Eddy currents are based on Faraday law of induction, which states that electrical charges are affected by the electromotive force (emf) induced by an alternating magnetic field. Furthermore, Lenz law states that the alternating magnetic field induces current loops whose corresponding magnetic field opposes the applied magnetic field changes - this behavior is similar with diamagnetic behavior. Since electrons move in the metal surface, they decrease the magnetic field penetrating the material. The electrons movement on the metal surface causes the metal to heat via Joule effect. Nowadays, Eddy currents are used in metallurgical industry for metal melting. This phenomenon is also applied in magnetic breaks and to separate metallic materials from non-metallic, [26] .

### *2.4.2 Hysteresis Power Loss*

A ferromagnet will generate heat at every hysteretic loop performed. The heat source is the magnetization-demagnetization process, [27]. Heat is released as the domains change their magnetic orientation and/or domain walls move along the ferromagnet.

The first law of thermodynamics states that the total energy, U, of an isolated system is neither created nor destroyed. It is usually presented as ΔU = Q – W. Where ΔU is the change of the system entropy, Q is the energy given to the system and W is the work performed by the system. For an adiabatic system, which does not transfer energy out of the system, W = 0. The increase of entropy can be performed by demagnetizing the system. The energy change per hysteretic cycle is equal to the hysteresis area, present equation 2.7, [28].

$$\Delta U = -\mu_0 \oint M \, dH \tag{2.7}$$

The total power loss in an induction heating experiment, considering only hysteretic losses is proportional to hysteresis area, $H_A$, and to the frequency, f, of the alternating magnetic field, $P_L = H_A.f$, [29].

### *2.4.3 Néel-Brown Relaxation*

Magnetic induction heating mechanisms occurring in single domain nanoparticles have a significant contribution of both Néel relaxation and Brownian motion. Although the heat generation occurs differently in both mechanisms, they occur parallelly. Once an external field is applied to the single domain, its magnetization tries to follow the magnetic field, with a time lag, [29]. It is not energetically favorable for the single domain to align an individual atomic moment, thus, for the magnetization to align with the applied field, either the entire nanoparticle rotates, Brown relaxation, or the materials' magnetization rotate, Néel relaxation.

Néel relaxation is the rotation of the single domain magnetization, between easy axis, leaving the nanoparticle orientation unaltered. The heat generation leads to an increase of the single domain temperature which is a consequence of the magnetization transposing the energy barrier that separates two, or more, easy axis.

The Brownian motion is a heating generation mechanism that do not directly heat the nanoparticle. Instead, the mechanical rotation of the nanoparticle heats its surrounding medium by Brownian motion.

Both, Néel and Brownian relaxation, contribute simultaneously for the particle temperature increase. The effective relaxation time, $\tau_{eff}$, is expressed by equation 2.8.

$$\frac{1}{\tau_{eff}} = \frac{1}{\tau_N} + \frac{1}{\tau_B} \tag{2.8}$$

R.E. Rosensweig, [28], developed an numerical model for the power generation consequent of magnetization change in an AC magnetic field. The result is displayed in equation 2.9.

$$P = \mu_0 \pi \chi_0 H_0^2 f \frac{2\pi f \tau_{eff}}{1+\left(2\pi f \tau_{eff}\right)^2} \tag{2.9}$$

Where $\mu_0$ is the vacuum permeability (4π x $10^{-7}$ H/m), $\chi_0$ is the magnetic susceptibility, $H_0$ is the external magnetic field amplitude and $f$ the external magnetic field frequency. From the previous equation it is obtained that the power generation is proportional to the frequency and to the square of field strength, [29].

In the next chapter it is described the magnetism of the Mn-Zn ferrite.

# Chapter 3: MN-ZN FERRITE

Ferrites are Iron-Oxide based ceramics with a ferrimagnetic behavior. The ferrite family arouse much interest for being a magnetic material with high electrical resistivity, thus, reducing the power loss from Eddy currents and presenting suitable magnetic properties for high-frequency applications. The main cause for their ferrimagnetic behavior is the Iron (III) cations present in its crystal structure along with divalent cations. Ferrites chemical formula is $MFe_2O_4$, where M represents a divalent ion, or multiple divalent ions. The choice of the divalent ion has a great impact on the ferrite properties, as it can change its magnetic and crystallographic structure, as well as other intrinsic properties. Ferrites are usually divided in two categories, hard ferrites, with high values of coercivity and soft ferrites, with low values of coercivity.

Although the discovery of ferrites is attributed to Y. Kato and T. Takey, [30], ferrites have been used by Mankind since the development of the first compasses. Lodestone magnetic properties are caused by the presence of magnetite. Magnetite ($Fe_3O_4$) is a ferrite, with two distinct Iron cations, $Fe^{2+}$ and $Fe^{3+}$, and consequently its chemical formula might also be written as $\mathrm{Fe^{2+}Fe_2^{3+}O_4}$. Nowadays, ferrites are spread through a vast range of applications due to their tunable magnetic properties and low-cost production. Despite these materials were originally envisaged for electronic applications, today they are used for biomedicine, catalysts, pollution control, magnetic shielding, among many other applications, [31] [32].

Mn-Zn ferrite, $(Mn_{1-x}Zn_x)Fe_2O_4$, is a soft ferrite which attracted researchers attention due to its versatility in multiple areas. This ferrite has been highly studied for biomedical applications, such as cancer treatment, via hyperthermia, magnetic resonance imaging and drug-delivery agents, [33]. The technological advances in the medicine field derives from the remarkable properties of this ferrite, for example, biocompatibility, low toxicity, magnetic properties and ease of synthesis in the nanoparticle form, [2].

For this master thesis, the interest in Mn-Zn ferrite is the dependence of the structural and magnetic properties with its composition. During the present chapter, synthesis methods, crystallography and magnetic structure will be explored.

## 3.1 SYNTHESIS OF MN-ZN FERRITE

The widely used synthesis method for bulk ferrites is through solid-state reaction route. This method requires metal oxides ($Fe_2O_3$, $MnO_2$, ZnO) and temperatures above 1000ºC for the reaction to start, [34]. Stable oxides require high temperatures for their reaction, which causes this method to be very expensive. The synthesis of Mn-Zn ferrite in nanoparticle form is usually done via chemical methods, which can be performed close to room temperature, reducing the production cost.

There are multiple synthesis methods for synthesizing Mn-Zn ferrite as nanoparticles. These synthesis methods include: hydrothermal, thermal decomposition, sol-gel auto-combustion, co-precipitation, reverse-micelle, high energy ball milling, among others [35], [36]. During this work, the elected synthesis methods were sol-gel auto-combustion and hydrothermal methods. A brief description of these methods is presented next and the experimental details of both methods are detailed in “Chapter 4: Experimental Procedure”.

The sol-gel auto-combustion method is a chemical synthesis method that relies on the high temperatures generated by a self-sustained combustion to synthesize ceramic oxides. This method is frequently mentioned for Mn-Zn ferrite synthesis, as it produces quality samples in high quantities.

The hydrothermal method is a chemical synthesis method. It relies on the precipitation of Iron Hydroxides precursors through an acid-base reaction. These precursors are then subjected to conditions of high pressure and moderate temperature inside an autoclave, then the ferrite crystals are obtained. This synthesis method is frequently cited for the production of high crystallinity Mn-Zn ferrite. Moreover, the use of chemical additives, control over temperature and the choice of the initial reagents allows the control of synthesis with varied morphologies, particles’ size distribution and dispersion levels [37], [38].

## 3.2 Crystal Structure

The Mn-Zn ferrite belongs to cubic spinel crystal structure (Fd-3m space group, $O_h^7$ or 227), [32], figure 3.1. The spinel crystal structure is composed of a face-centered cubic arrangement of oxygen ions, 32 oxygen atoms per unit cell. In the interstitial space, 8 tetrahedral and 16 octahedral sites are occupied by transition metal cations. The cations are required in order to maintain this structure electrically neutral. The tetrahedral site is frequently called A site, and the octahedral site, B site. Thus, the general spinel crystal structure molecular formula is represented by $AB_2O_4$. respectively,

Literature frequently reports two types of spinel crystal structure, the normal spinel and the inverse spinel. In the normal spinel, the tetrahedral site is occupied by divalent cations (2+) and the octahedral site are occupied by trivalent cations (3+). In the inverse spinel, the tetrahedral site is occupied by trivalent cation and the octahedral site by divalent cations. However, the most frequently configuration for ferrites is the mixed spinel crystal structure, this structure is characterized by a mixture of both normal and inverse spinel crystal structures. Thus, the molecular formula for spinel ferrites must be written as $(M^{2+}_{1-\delta}Fe^{3+}_{\delta})^{tetra}[\,M^{2+}_{\delta}Fe^{3+}_{2-\delta}]^{octa}O^{2-}_{4}$, where $\delta$ represents the inversion parameter, [39]. For instance, in the normal spinel ($\delta$=0) all the divalent cations are in the tetrahedral site and all the trivalent cations are in the octahedral site. In the inverse spinel ($\delta$=1), all divalent cations are in the octahedral site with half of the trivalent cations and the other half of the trivalent cations are in the tetrahedral site. For a matter of simplicity, and since the crystal structure characterization technique (XRD) does not allow such definition, the Mn-Zn ferrite molecular formula will be represented as a normal spinel crystal structure, $(Mn_{1-x}Zn_x)Fe_2O_4$, with the divalent cation ($Mn^{2+}$ and $Zn^{2+}$) in the tetrahedral site and all Iron cations ($Fe^{3+}$) in the octahedral site. Although it must be highlighted, that not only the synthesized samples probably have a mixed spinel crystal structure, but also the Manganese ions can be present in the divalent and trivalent cation states, $Mn^{2+}$ and $Mn^{3+}$. All these parameters are highly dependent of the synthesis method and synthesis conditions, [40].

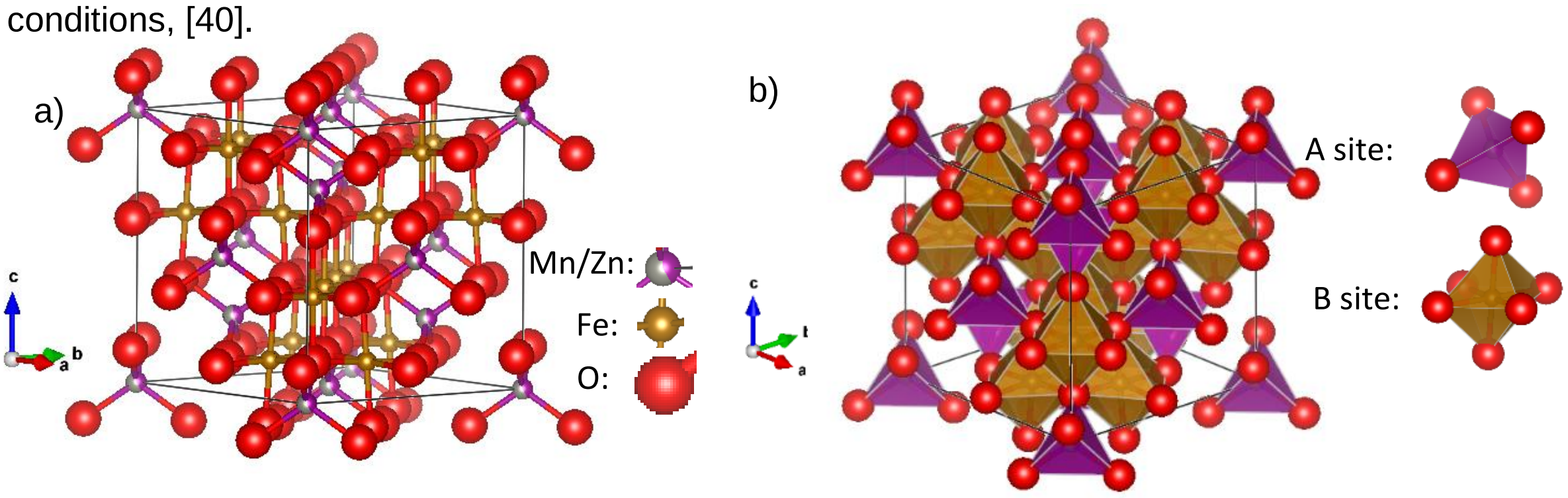

**Figure 3.1 – a)** Normal spinel crystal structure of $Mn_{1-x}Zn_xFe_2O_4$. b) tetrahedral (B) and octahedral (A) sites in Mn-Zn ferrite. Atoms and site captions are in the right side of each image. Vesta software was used for unit cell 3D rendering.

## 3.1 Magnetic Structure

The magnetic structure of Mn-Zn ferrite is the reason for it being highly studied, conferring these compounds excellent properties for new technological applications. Some of the magnetic properties that arouse researchers' interest are the high saturation magnetization, high magnetic susceptibility at high-frequency magnetic fields, low coercivity, low losses through Eddy currents and hysteresis loops, Curie temperature close to room temperature and the main reason, the tuning of all these properties by varying the Zn/Mn ratio.

In order to understand the magnetic structure of this ferrite, is important to understand each atom role in the spinel crystal structure. Four different atoms are present in the Mn-Zn ferrite: Oxygen, Iron, Manganese and Zinc.

The Oxygen anion, $O^{2-}$, has the fundamental role of bounding all the elements together. Its electrons configuration is $[He]2s^22p^6$. Although Oxygen ions do not possess unpaired electrons, and consequently no corresponding magnetic moment, they are responsible for the superexchange interaction between metal cations. Accordingly with P. J. Zaag, [41], Oxygen bounds through a single p-orbital, which makes 180º between metal cations, thus, the shared electrons have opposite spins and are responsible for the antiferromagnetic order.

The Iron cation, $Fe^{3+}$, has an electronic configuration of [Ar]$3d^5$. Due to the high-spin d-orbital, it contributes with magnetic moment for the material, with a value of 5 $\mu_B$. The Manganese cation, $Mn^{2+}$, has an electronic configuration of [Ar] $3d^5$, same as Iron, contributing with approximately the same magnetic moment, 5 $\mu_B$. The Zn cation, $Zn^{2+}$, has an electronic configuration of [Ar]$3d^{10}$. With the valence orbitals fully paired, this cation does not contribute with magnetic moment.

As it might be predicted, the magnetic structure of Mn and Zn ferrites strongly differ, consequence of the different magnetic moment of Mn and Zn cations. An individual analysis is present in the next paragraphs.

The Mn ferrite, composed of magnetic Fe and Mn cations, which are bounded antiferromagnetically through the oxygen anion, thus presenting a ferrimagnetic behavior, figure 3.2.a). The Fe-Fe magnetic moments are aligned, the Mn-Mn magnetic moments are also aligned, however, the Fe-Mn magnetic moments are antiparallelly aligned. This ferrite is expected to present a ferrimagnetic behavior since the Fe cations are twice the amount of Mn cations.

The Zn ferrite, which is composed of magnetic Fe and non-magnetic Zn, bounded antiparallelly through Oxygen bonds. An enhanced overall magnetic moment is expected, since only parallelly aligned Iron moments contribute for the materials' magnetization. However, this ferrite is barely magnetic, as discussed by H. L. Anderson et al, [42], the weak magnetic ordering is responsibility of the inversion parameter as $Fe^{3+}$ cations occupy tetrahedral sites. The Fe magnetic moments in the octahedral site couple antiparallel. It is also reported that a low value of saturation magnetization was obtained, which was caused by Fe located in tetrahedral site, due to the spinel inversion. The antiparallel coupling of Iron cations confers to this ferrite an antiferromagnetic behavior.

The intermediate compositions have a ferrimagnetic behavior due to the presence of Mn cations. An increase of magnetization is expected while going from the antiferromagnetic Zn ferrite to the ferrimagnetic Mn ferrite. Besides, it is also reported by H. L. Anderson et al, that Mn ferrite have the easy axis along the <111> direction.

.

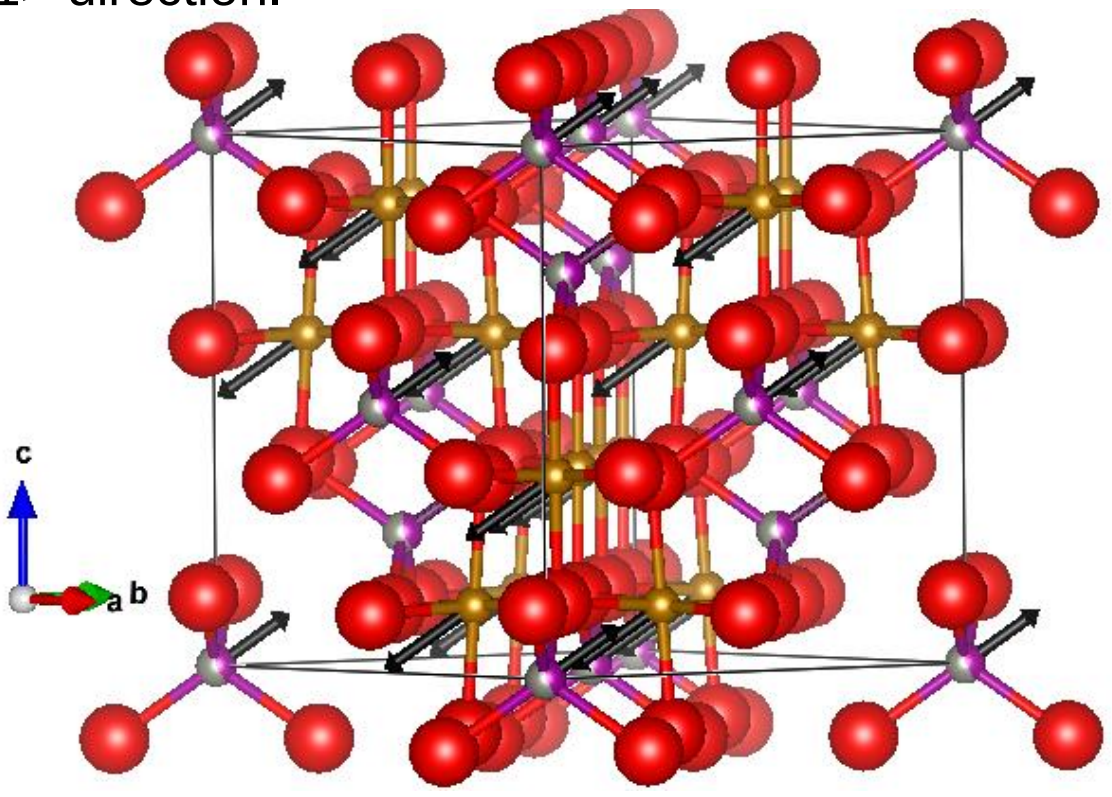

**Figure 3.2** –Magnetic structure of normal spinel $Mn_{1-x}Zn_xFe_2O_4$. Tetrahedral and octahedral magnetic moments are represented by vector aligned antiparallelly in <111>. The unit cell schematic displays a vector length equal for tetrahedral and octahedral sites, which is only true for $MnFe_2O_4$.

The Zn ferrite presents an antiferromagnetic behavior, with a respective Néel temperature, and the Mn ferrite a ferrimagnetic behavior, with a respective Curie temperature. From this point on, when mentioning the full range of compositions of $Mn_{1-x}Zn_xFe_2O_4$, the temperature that these compounds return/cease to have a magnetic order will be denominated "magnetic ordering temperature".

The different magnetic structures of Mn ferrite and Zn ferrite causes its magnetic properties to be highly composition dependent. We aim to take advantage of this dependency to observe the magnetic ordering temperature varying with the ferrite composition, as observed by P. H. Nam, [43]

Next chapter, Experimental Procedure, explores the synthesis methods, characterization techniques and the data treatment methodology.

# Chapter 4: EXPERIMENTAL PROCEDURE

The "Experimental procedure" chapter includes details concerning the sample synthesis and the characterization techniques used, which are XRD, SEM, TEM, SQUID and magnetic induction heating.

## 4.1 SAMPLE SYNTHESIS

### *4.1.1 SOL-GEL AUTO-COMBUSTION METHOD*

The sol-gel auto-combustion method is a synthesis method, capable of producing different nano powders by an exothermic and auto-sustained reaction between metal salts and an organic component. The simplified chemical reaction is written below:

$$2\ Fe(NO_3)_3\,(aq) + (1\text{-}x)\ Mn(NO_3)_2\,(aq) + x\ Zn(NO_3)_2\,(aq) + 2{,}2(2)\ C_6H_8O_7\,(aq) + NH_4OH\,(aq)$$

$$\rightarrow \text{drying} \rightarrow \text{auto-combustion trigger} \rightarrow Mn_{1-x}Zn_xFe_2O_4\,(s) + H_2O\,(g) + CO_2\,(g) + NO_x\,(g) + \text{energy}$$

All nitrates were bought in Chem-Lab, Zn and Fe nitrate have a purity of 98+% and Mn nitrate 97+%, citric acid was bought from Sigma-Aldrich and has a purity of 98+% and ammonium hydroxide at 25% was bought in Chem-Lab. The procedure starts by weighing hydrated nitrates salts of Fe, Mn and Zn, followed by their dissolution in de-ionized water. As presented by the chemical equation, the number of nitrates mols will influence the ferrite composition and its amount. Then, citric acid amount is weighed, dissolved and added in the solution. Citric acid has a fundamental role in this synthesis method, as it works as a reductant for the salts (oxidants). However, there is not a consensus in the literature about the amount of citric acid that should be added. Dong Limin et al, [44], reported ferrite synthesis with 3 mols of citric acid (1 mol of $C_6H_8O_7$ per mol of metal ion), while A. Sutka et al, [45], defended that the oxidation number must be equal to the reductant number, in accordance with the valence electrons of the compounds. Both strategies were tried, and the best results were analyzed.

The aqueous solution of metal salts and organic component is called solution, or sol. The solution is stirred at 400 RPM at a temperature between 60 and 70 $^{o}$C. Then, ammonium hydroxide ($NH_4OH$) is added to the solution dropwise until the solution ph is adjusted to 7. the pH influences the ignition temperature and the fullness of the reaction, [46]. When $NH_4OH$ interacts with the solution, the hydroxide reacts with the acidic solution, releasing heat and leading to the evaporation of water (in vapor form), then ammonium nitrate ($NH_4NO_3$) precipitates together with other amorphous precursors. During the pH adjustment, the sol was kept at a temperature below 70 $^{o}$C (controlled by a thermocouple). The sol was let dry in the hotplate at 120$^{o}$C. Almost fully dehydrated, the solution is now a gel, with the appearance of a black mud. The gel is kept in the oven until fully dehydration takes place.

The auto-combustion is conducted in a hot-plate (inside a fume hood) at a temperature of 250$^{o}$C. When ignition starts, ammonium nitrate starts to decompose, generating heat and ions in the form of a flame. The metallic ions are under the temperature conditions where ferrite synthesis is possible. The combustion temperatures can vary from 600 to 1350$^{o}$C, [45]. Mn-Zn ferrite is created as a combustion ash. The ash is then milled using a mortar until a fine powder is achieved.

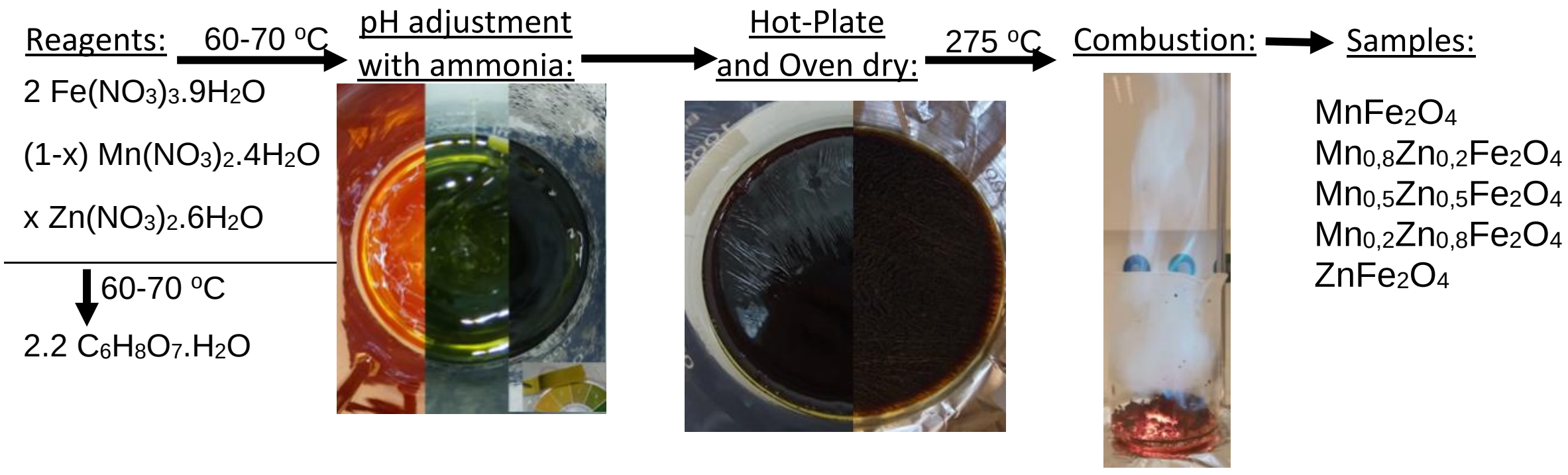

**Figure 4.1 –** Sol-gel auto-combustion method - experimental scheme.

#### 4.1.1.1 SAMPLES

Samples of Mn-Zn ferrite, $Mn_{1-x}Zn_xFe_2O_4$, of composition x = 0; 0.2; 0.5; 0.8 and 1 were synthesized. It was noticed the presence of secondary phase of zinc oxide (ZnO) for $Mn_{0.2}Zn_{0.8}Fe_2O_4$. A secondary phase is not desired because it could compromise the main ferrite properties. Efforts for its reduction were made. Table 4.1 shows the synthesis parameters and how these parameters were manipulated.

| Parameter | Control | Parameter | Control | Parameter | Control |
|---|---|---|---|---|---|
| Ferrite composition (x) | 0 - 1 | Citric acid amount | 1; 2.2; 3 mol | pH adjustment | Fixed (7) |
| Combustion temperature | Free | Drying temperature | 0 - 120 | Drying time | 0 – 4 days |
| Combustion time | Free | Combustion atmosphere | free | Temperature set | $R_T$, $C_6H_8O_7$, $NH_4$ |

**Table 4.1** – Available synthesis parameters and respective control. Controlled parameters are identified by a range of used values. Free means that there was no control over the parameter.

Multiple testing for $Mn_{0.2}Zn_{0,8}Fe_2O_4$ revealed that the samples that did not undergo the dehydration process presented higher single-phase nature (>88%). The results presented in the next chapter for sol-gel auto-combustion samples did not undergo the oven dehydration process. The amount of citric acid used is 2.2 mol. The temperature was increased before citric acid was added to the solution. After, samples were milled using a mortar to obtain a finer powder.

### *4.1.2 HYDROTHERMAL METHOD*

Hydrothermal method is a chemical method for fine powders synthesis. Its advantages over other methods are the simplicity of the technique, control over the reaction and acquisition of high-quality nanoparticles, as elected from I. Sharifi et al, [20] (table 5). The simplified chemical equation is the following:

$$2\ Fe(NO_3)_3\,(aq) + (1-x)\ Mn(NO_3)_2\,(aq) + x\ Zn(NO_3)_2\,(aq) + y\ NaOH\,(aq) + \text{pressure} +$$

$$+ \text{temperature} \rightarrow (1-z-w)\ Mn_{1-x}Zn_xFe_2O_4\,(s) + y\ NaNO_3\,(aq) + z\ Fe2O3\,(s) + w\ FeOOH\,(s)$$

The experimental procedure follows the recipe presented in the article by Xin Li et al, [47]. It starts by calculating the stoichiometric masses of the nitrates salt to obtain 4mmol of Mn-Zn ferrite. The salts are dissolved in 80 mL of de-ionized water. The amount of NaOH, 1.3877mg per 15mL, these values were obtained in Xin Li et al article. NaOH initially defines the pH of the solution, settled to 12, and its exothermic contact with the water creates a series of compounds, some soluble in water, others insoluble. See more about hydrothermal precursors in appendix A: Autoclave time.

The autoclave is kept in the oven for 6h at 180ºC, the autoclave is filled with a maximum of 120mL and have a capacity of 150mL. This causes the solvent (water) to enter a supercritical fluid state. In this state, liquid or gas are indistinguishable. The pressure inside the PTFE container should surpass 3MPa, the maximum advice temperature is below 220ºC, [48]. This extreme conditions of controlled pressure and temperature allows the solution ions to crystalize in Mn-Zn ferrite and with the proper time, temperature and pH, minimize secondary phases. Figure 4.2 is a schematic of the hydrothermal method.

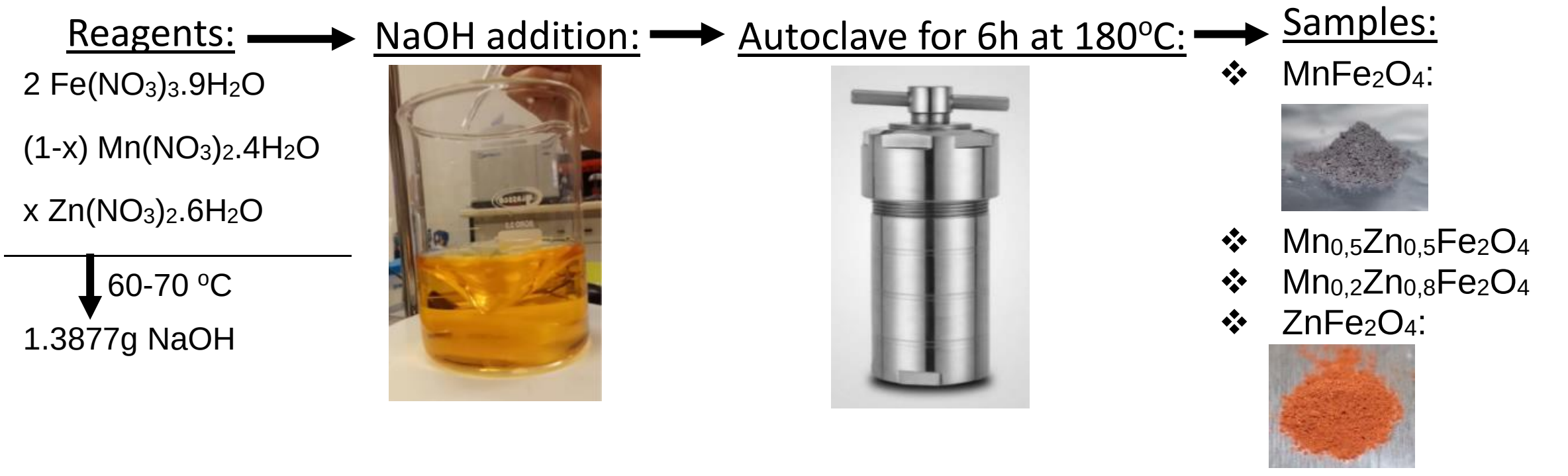

**Figure 4.2** – Hydrothermal method - experimental scheme.

#### 4.1.2.1 SAMPLES

Hydrothermal samples with $Mn_{1-x}Zn_xFe_2O_4$ (x = 0; 0.5; 0.8; 1) compositions were prepared at constant temperature (180ºC), constant pH (12) and variable time in autoclave (0, 6 and 21 hours). Samples synthesized for 6h presented the highest percentage of pure phase, for this reason, only the results of these samples will be included in this thesis. Samples prepared by the autoclave procedure for 21h hours are not included in the core of this thesis but can be found in Appendix A: Autoclave Time

## 4.2 EXPERIMENTAL TECHNIQUES

### *4.2.1 X-RAY DIFFRACTION (XRD)*

To measure sample's crystallinity and inspect the crystallographic phases present the X-ray diffractometer Pan'Analytical EMPYREAN was used. Figure 4.3.a) is a picture of the diffractometer and 5.2.b), [49], a schematic of the Bragg-Brentano geometry.

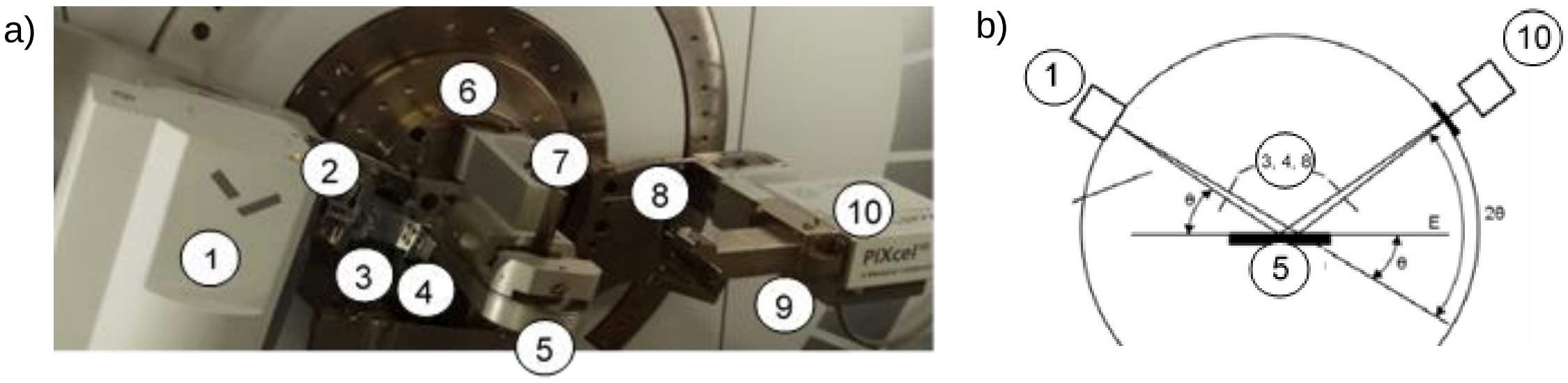

**Figure 4.3 –** a) X-Ray diffractometer and components. b) Bragg-Brentano geometry: 1 – Cathodic rays' tube; 2 – collimator; 3 and 4 – Slits; 5 - Sample holder; 6 – Goniometer; 7 – Beam knife; 8 – Slit; 9 – Collimator and Nickel filter; 10 – X-Ray photodetector.

The obtained XRD diffraction patterns, were accomplished by directing an X-ray beam to the sample. Samples parallel atomic planes (with associated Miller indices) reflect the beam. The X-rays coming from parallel atomic planes interfere, due to X-ray wavelength being of the same scale as the distance between atomic planes (Angstroms). The diffraction pattern is the result of constructive and destructive interference of x-ray waves coming from the sample. By scanning the spherical diffraction pattern that surrounds the sample, maximum intensity peaks are displayed revealing interatomic distances, d. The interatomic distance can be calculated using Bragg law, equation 4.1. Figure 4.4 presents a schematic of the Bragg law with two X-ray beams diffracted in parallel crystallographic planes.

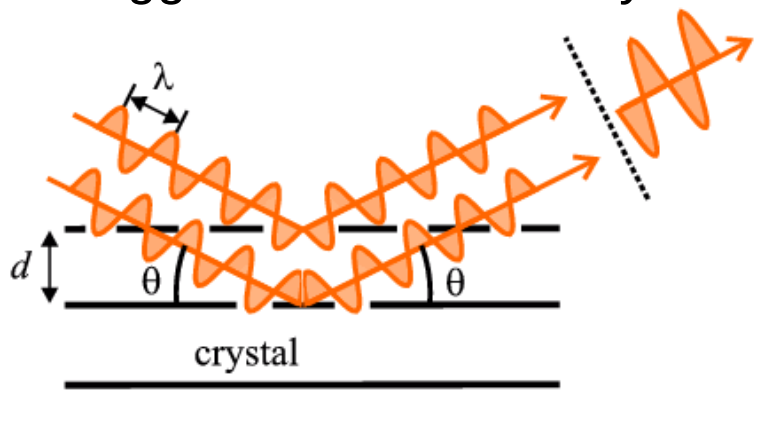

$$2\,d\,sen(\Theta) = n\,\lambda \qquad (4.1)$$

**Figure 4.4 –** Bragg law: X-ray diffraction schematic.

The X-ray beam is generated in a Copper anode when a high energetic beam of electrons is directed against it, exciting the Cu atoms, which emit X-rays in k-α and k-β wavebands. Further in the diffractometer, a nickel filter attenuates the k-β peak, whereas kα emission remains. k-α emission is composed of k-α1 wavelength of 1.540598 Å and a less intense emission of k-α2 at 1.544418 Å (typically k-α2/k-α1 ~0.5). The sample rotates ϴ and the photodetector also moves, ϴ, radially, with the sample. The diffractogram x-axis units are in 2ϴ, which is the angle between emitted beam and detected beam. This configuration is known as Bragg-Brentano, figure 4.3.b). Slits and beam knife purpose is to direct the beam to the sample, controlling the area of interaction and the beam area that reaches the detector. The sample-holder rotates around itself to ensure the maximum detection from all orientations present in the analyzed powder.

The X-ray diffraction pattern gives different information about the crystal structure. Using Highscore Plus, Fullprof Suite and Origin softwares, information about the crystal phases, lattice constants and crystallite sizes were obtained. The diffractogram analysis is explained below.

#### 4.2.1.1 SAMPLE PREPARATION

The powder was placed in a Silicon wafer (Si single crystal), which was cut in a specific crystal direction to prevent the appearance of Silicon diffracted peaks at 2ϴ angles inferior to 100º.. The samples are gently pressed against the wafer to ensure a smooth surface and then are placed below the beam knife. The powders should be fine, in order to ensure that the diffracted beam is a result of the randomly oriented atomic planes. The sample surface should be smooth for the diffraction to occur at the same height throughout the whole sample.

#### 4.2.1.1 PHASE IDENTIFICATION

Different crystallographic structures were identified with resource to High Score Plus software. The diffractogram is composed of several peaks, each peak corresponding to an interplanar distance, identified by a Miller index. Every crystal structure presents a group of Miller indices. To identify the materials, present in the sample, High Score Plus compares the sample diffractogram with a database of diffractograms and rates their proximity by analyzing the similarity between Miller indices of different atomic structures. The user is then allowed to manually select which atomic structural phases are present in the diffractogram.

The percentage of phases was estimated by Rietveld refinement with Fullprof Suite software package, as it offers control over fitting and consequently better fitting results. Rietveld refinement requires initial information of the present phases in order to enable the fit, such as atomic positions, elements, space group, lattice constants and others. This information is available in free data bases, such as Crystallographic Open Database, used in this work. The structural information, or .cif files, were obtained and imported into Fullprof Suite in order to create a .pcr file for each phase. The .pcr files of all phases are then merged in a single .pcr file. The final .pcr file is then loaded to Fullprof and fitted with the diffractogram. The output is .sum file with crystallographic information. The percentage of phases is automatically calculated based on the scale of each diffractogram. Figure 4.5.a shows the Rietveld refinement of single-phase Mn Ferrite from sol-gel auto-combustion method. Figure 4.5.b) shows the Rietveld refinement of $Mn_{0.2}Zn_{0.8}Fe_2O_4$ of hydrothermal method, the refinement of both spinel crystal structure and impurity (hematite).

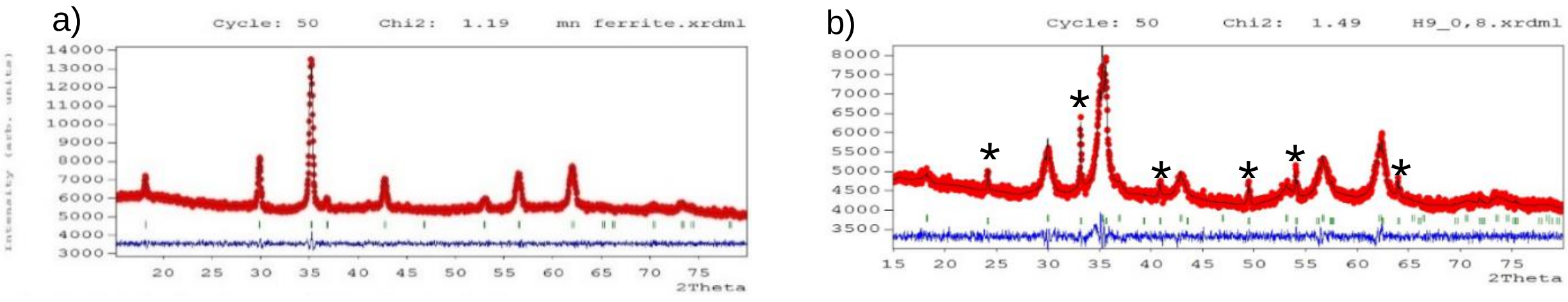

**Figure 4.5 –** Rietveld refinement of: a) $MnFe_2O_4$ (sol-gel) and b) $Mn_{0.2}Zn_{0.8}Fe_2O_4$ (hydrothermal). Asterisks identifies $Fe_3O_4$ phase. Red dots are for experimental data, black line is the Rietveld refinement, blue line is the difference between data and fitting and green lines identify the peaks position. Chi2 is the goodness of fitting.

#### 4.2.1.1 LATTICE CONSTANT

The lattice constant was also fitted using Fullprof suite software. The lattice constant is calculated with resource to the interplanar distances of a unit cell, the interplanar distance is related to the peaks' positions in the 2θ-axis, as previously mentioned. Changes of interplanar distances are visible as shifts in the peak position (2ϴ). FullProf Suite software uses the atomic positions and space group information to calculate the unit cell lattice constant.

The lattice constant is sensitive to different phenomena, such as synthesis conditions, pressure, temperature, interatomic interactions and nanoparticles size. Also, nanoparticles differ from the bulk material as the ratio surface/volume is larger in nanoparticles, being more likely to suffer strains due to the pressure of the surrounding medium. Lattice constant in bulk samples also change when pressure is applied, however, a bulk material lattice constant does not change as much as the same material in nanoparticle form, [50].

#### 4.2.1.1 Other Results from Rietveld Refinement

The main results obtained from Rietveld refinement are the phases quantification, their lattice constants and atomic positions. However, the Rietveld refinement from Fullprof Suite software outputs other fitted parameters. Some of these parameters are identified in table 4.2. This information is available in the .sum file exported after the Rietveld refinement.

| Atomic Parameters: | Cell Parameters: | Fitting parameters: | Others: |
|---|---|---|---|
| Atomic Position | a, b, c | scale factor | Volumic mass |
| Occupancy | alpha, beta, gama | Peaks position and FWHM | Temperature effects |
| | | unit cell volume | Sharpest peaks: position and FWHM |

**Table 4.2 –** Crystal phases information exported from the Rietveld refinement.

The Rieveld refinement quality is identified for each phase and for the complete diffractogram by $R_{WP}$, $R_{exp}$ and $\chi^2$ (Chi squared), [51]. Several iterations with different fitting parameters were performed in order to decrease these values, increasing the fit quality.

#### 4.2.1.2 Crystallite Size and Strain

Crystallite size is defined as the mean size of the single crystals that compose the sample. A single crystal is the coherent length throughout which the unit cell repeats periodically with the same orientation. When an X-ray beam interacts with parallel atomic planes it creates an interference pattern. If an atomic plane repeats itself throughout a long distance, the more X-rays reflected from that plane will interfere constructively, and an increase in the intensity and narrowing of the peak is expected.

It was G.K. Williamson and W. H. Hall who merged the Scherrer equation (for crystallite size) and Stokes and Wilson equation (for strain), creating the Williamson-Hall equation, [52]. The Williamson-Hall equation (H-W) relates the experimental band broadening of the peaks with the crystallite size and strain as displayed in equation 4.1, [53].

$$\beta_{TOTAL} = \beta_{SIZE} + \beta_{STRAIN} = \frac{K.\lambda}{\mathrm{d}_{XRD}.\cos(\theta)} + 4.\eta.\tan(\theta) \quad (4.1)$$

Where $\beta_{TOTAL}$ is the experimental FWHM, $\beta_{SIZE}$ and $\beta_{STRAIN}$ are the FWHM contribution of crystallite size and strain, respectively. K is the Debye-Scherr constant (~0.94 for spherical nanoparticles), $\lambda$ is x-rays wavelength, $\mathrm{d}_{XRD}$ is the crystallite size, $\eta$ is the crystal strain and $\theta$ is the angle (in radians) of the peak. For this analysis Origin software was used for peak fitting and calculations. Peaks are fitted using a pseudo-Voigt function. The fit automatically outputs the FWHM, or $\beta_{TOTAL}$, and the peak position, $2\theta$. Figure 4.6.a is the fitting of a peak with a pseudo-voigt function. This procedure is repeated for a characteristic set of peaks (typically the most intense). Hence, the obtained fitting results are plotted according with the Williamson-Hall equation by making $y = \frac{\beta_{TOTAL}.\cos(\theta)}{K.\lambda}$, and $x = \frac{4.\mathrm{sen}(\theta)}{K.\lambda}$. In this plot, slope is m = $\eta$ and the interception is b $= \frac{1}{\mathrm{d}_{XRD}}$. Figure 4.6.b presents the linearization using W-H equation for Mn Ferrite.

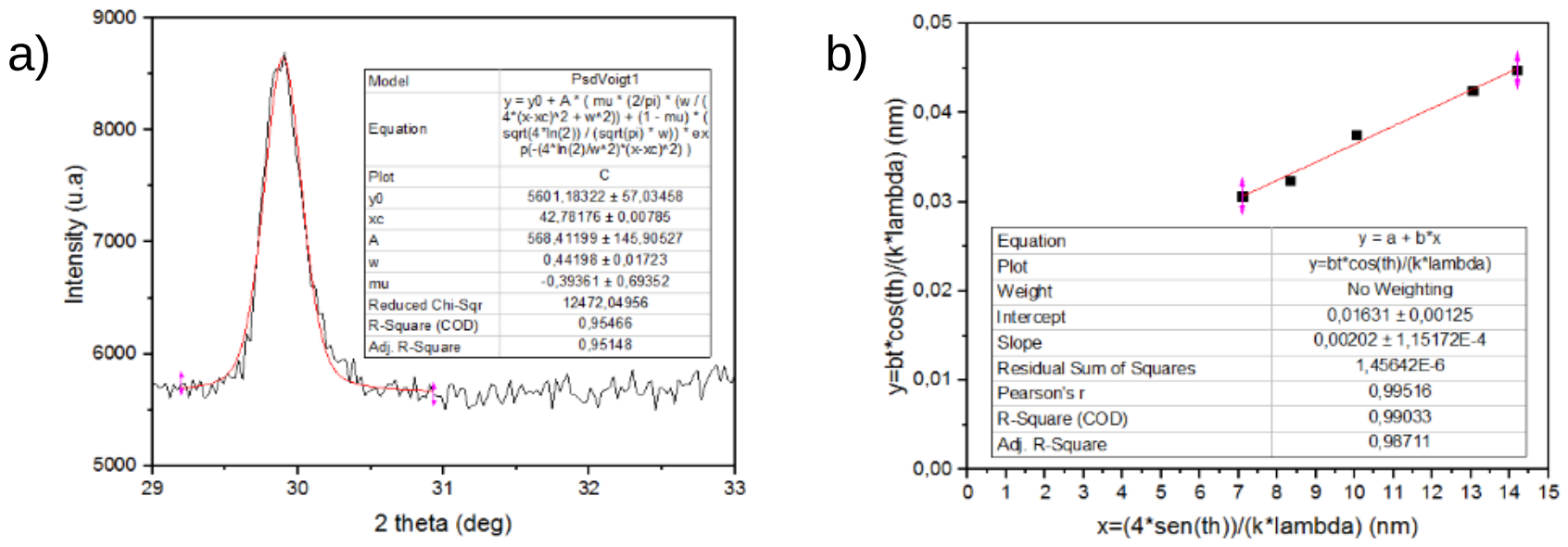

**Figure 4.6 –** a) Peak fitting using Origin. b) Williamson-Hall linear fit.

#### 4.2.1.3 SYSTEMATIC ERRORS

For Williamson-Hall analysis, fitting was performed with Lorentz, Gaussian and pseudo-Voigt functions. The one which presented a better value of $R^2$ was the pseudo-Voigt function. All the fitted peaks presented a $R^2$ value greater than 90%.

The Williamson-Hall analysis is based on the linearization described in section 4.2.1.2. As in every linearization the value $R^2$ is  and has an erratic contribution for the calculated properties, this means that the crystallite size has an error associated with the linearization. When linearized in Origin, it presents the errors associated with the interception value, the error in crystallite size was calculated as presented in equation 4.2.

$$\Delta D_{XRD} = \frac{\partial}{\partial b} D_{XRD} \, . \Delta b = \frac{1}{b^2} \, . \Delta b \tag{4.2}$$

Crystallite size values, $d_{XRD} \pm \Delta d_{XRD}$, are presented in the crystallite size plots.

An amorphous/nanocrystalline material has very small crystallites which results in very broadened peaks, sometimes just very spread humps. Amorphous phases are not identified by XRD, usually contributing as background for the diffraction pattern.

##### *4.2.1.3.1 Experimental Line Broadening*

The experimental line broadening is an equipment limitation that affects the Williamson-Hall analysis. The experimental line broadening can be experienced while making XRD in a monocrystal. If the crystal size is orders of magnitude larger than the wavelength, $\beta_{TOTAL}$ should approach zero (a Dirac-delta function), however this is not observable as the diffraction pattern peaks will always have a FHWM greater than zero. This phenomenon occurs for multiple reasons, for example due to the detector limited aperture or atoms not being in their equilibrium position (thermal kinetics, for example).

The experimental line broadening was calculated for a monocrystal of Lanthanum Hexaboride ($LaB_6$). Being a monocrystal, the FWHM of the peaks should give a Dirac-delta function. The analysis of this crystal is important to understand de error of $\beta_{TOTAL}$ and the maximum coherence length detectable by this diffractometer. The peak fitting for this sample is present in figure 4.7.

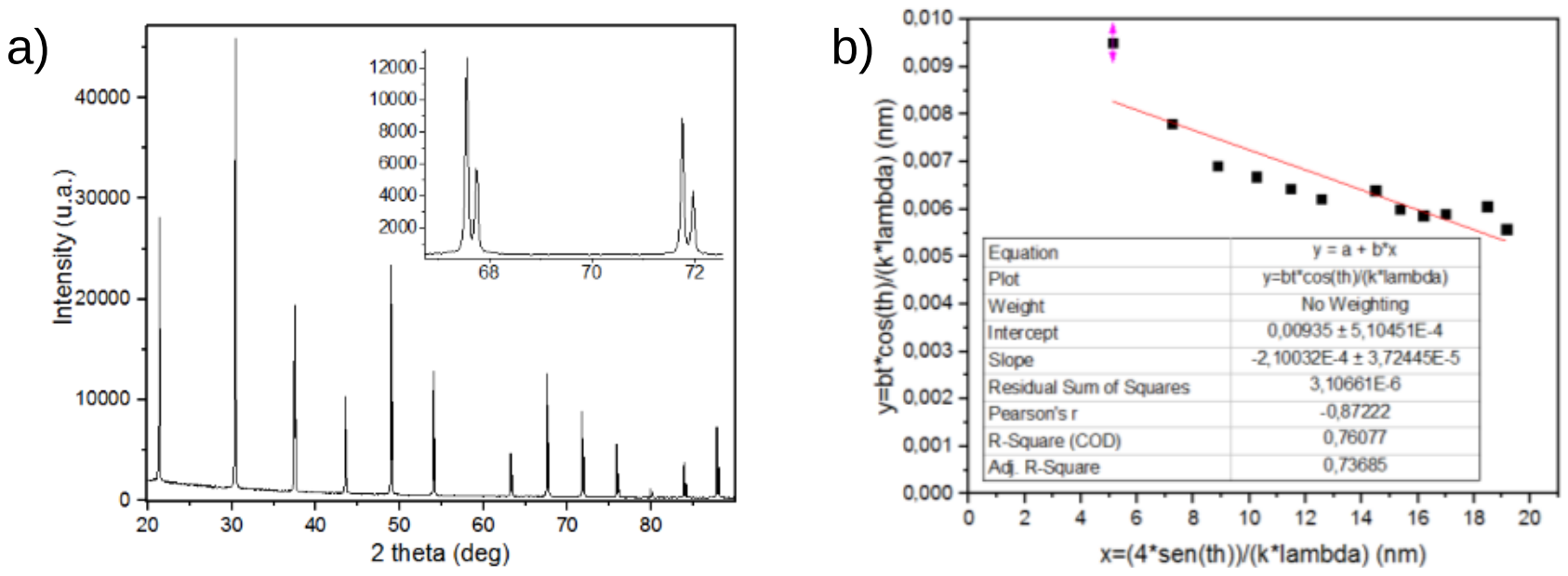

**Figure 4.7 –** (a) Diffractogram of $LiB_6$ with inset for K-α1 and K- α2 distinguished. (b) Linear fit of Williamson-Hall equation.

From the Williamson-Hall analysis performed on this diffractogram, it was obtained the maximum size of crystallite detectable by this diffractometer: 120.62 ± 4.55 nm. The peaks are broadened as the angle increases: from 15 to 80 degrees the FWHM value varies from 0.00952º to 0.0059º (in 2ϴ).

### *4.2.2 SCANNING ELECTRON MICROSCOPY (SEM) AND ENERGY DISPERSIVE SPECTROSCOPY (EDS)*

Scanning Electron Microscopy (SEM) and Energy Dispersive Spectroscopy Scanning electron microscope (SEM) is a technique that relies on electromagnetically accelerated electrons and their inelastic collision with the sample for imaging processing. Energy dispersive spectroscopy (EDS) is based on radiation emission from the sample, when excited with high energy electrons. During this work, the SEM equipment available, and capable of analyzing magnetic nanoparticles, was Hitachi S-4100 system, coupled with an EDS sensor. The manual of a close model, S-4800, can be found in [54].

#### 4.2.2.1 SCANNING ELECTRON MICROSCOPY (SEM)

The scanning electron microscope (SEM) is a complex tool for nano and micro-scale imaging. It consists on the detection of electrons emitted from a sample when interacting with an electron beam. The SEM consists of a vacuum chamber containing an electron gun, a series of electromagnetic lens, apertures and detectors. The high energy electrons are produced in the electron gun and accelerated by a potential difference. The high-speed electrons are then selected in an aperture and go through a series of condensing lens and apertures for beam focusing. The beam is finally deflected by the deflecting lenses, in order to scan the sample, while the beam intensity is controlled by the last aperture. Sample rotation and tilting motors allows sample observation from various angles. It is schematically represented in figure 4.8.a.

When the electrons reach the sample, figure 4.8.b), different interactions take place within the depth of penetration. Electrons will inelastically collide with the sample lattice, which will originate different emissions: backscattered electrons, secondary electrons, Auger electrons, characteristic x-ray, and others. Image formation occurs when the sample is scanned. At each sample point, a quantity of backscattered and secondary electrons is detected, [54]. The final image pixel intensity is defined by the number of detected electrons.

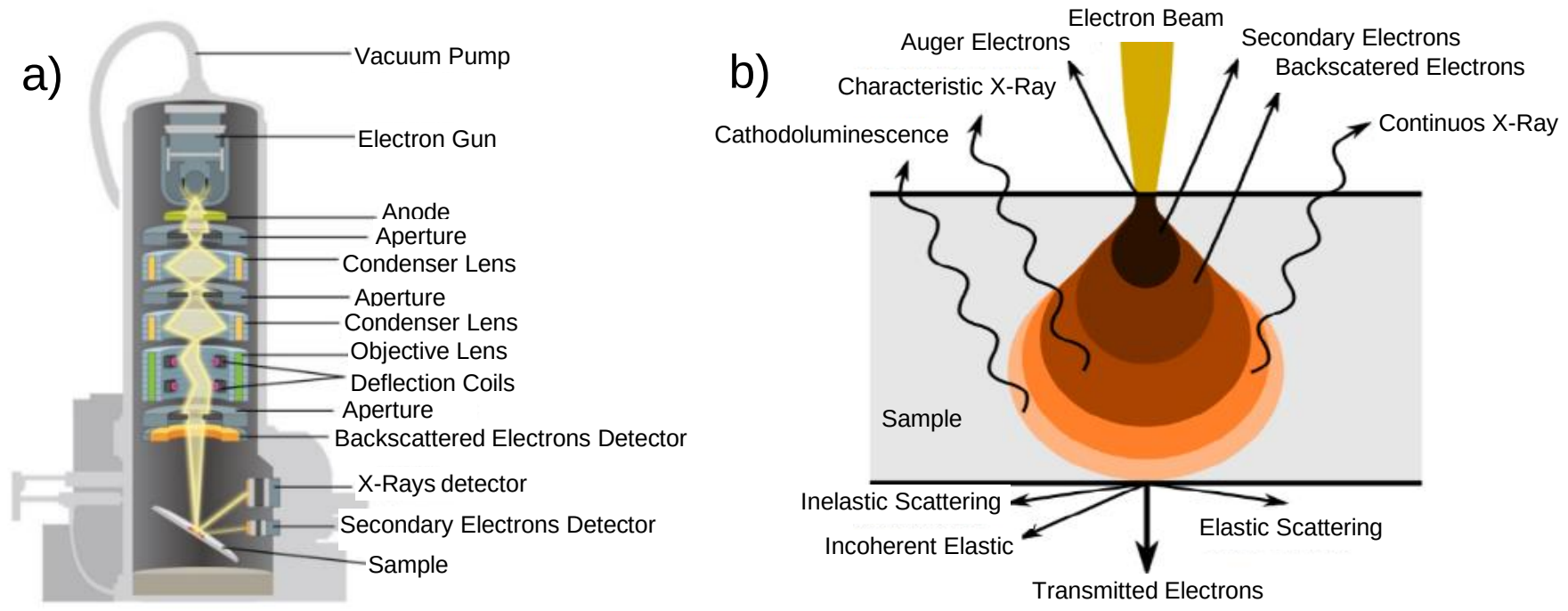

**Figure 4.8 -** a) Electrons flow within TEM, its components and detectors. b) Different possible interactions between the sample and the electron beam.

#### 4.2.2.2 SAMPLE PREPARATION

The magnetic powders were held by carbon tape, on top of a silicon wafer, and then coated with carbon. Carbon coating is performed by putting the samples in a vacuum chamber and sublimating a carbon wire by Joule effect. The carbon coating is necessary due to the insulator nature of the powders. If samples were not coated with carbon, they would accumulate charge when interacting with the electron beam, deflecting the electron beam and saturating the image brightness to a limit where no particles would be distinguishable.

The samples were then attached to SEM pre-chamber to be subjected to vacuum and introduced in SEM main chamber for visualization.

#### 4.2.2.3 Size Distribution Measurements

The nanoparticles were imaged using SEM, the ImageJ software was employed to measure the particle area. Figure 4.9.a depicts the image before counting and figure 4.9.b after counting, where only particles with a clear limit were measured. The area values and number of particles were then treated using Origin software. The particle area data was converted into radius (assuming spherical shape) and then diameter.

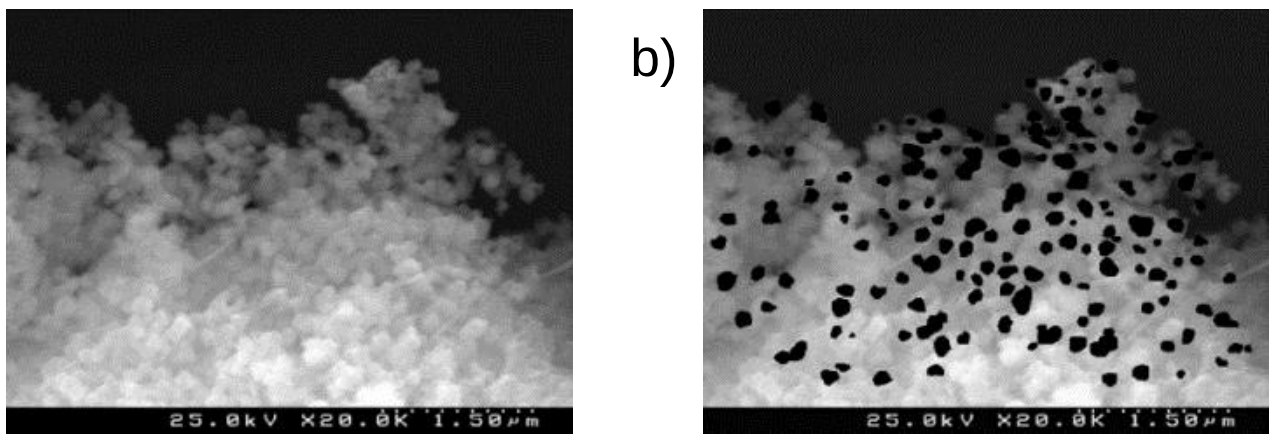

**Figure 4.9 –** Particle counting using ImageJ. Black areas are the particles counted for size statistics.

### *4.2.3 Energy Dispersive Spectroscopy (EDS)*

Energy dispersive spectrometer (EDS) is a sensor attached to SEM capable of distinguish different X-ray energies. As previous mentioned, once the electron beam interacts with the sample, different types of emission can occur. The EDS sensor receives x-ray atomic characteristic emissions from the sample and group them accordingly with their energy.

X-ray characteristic emission is created by each atom when a high energy electron interacts with it. The electron beam excites electrons from the inner electronic levels. Consequently, electrons from outer orbitals occupy their place, releasing energy in the form of x-rays, with an amount of energy dependent of the energy between energy levels. The released radiation is element-specific and hence enables the indexing of the obtained spectrum peaks with respective element.

EDS results in a spectrum, yield photons versus energy (keV), as represented in figure 4.10.a). Databases of x-ray characteristic emission for every element allows the microscope software to perform a fitting of the experimental spectrum whose outcome is the atomic percentage fractions present in the sample, as exemplified in figure 4.10.b).

a)

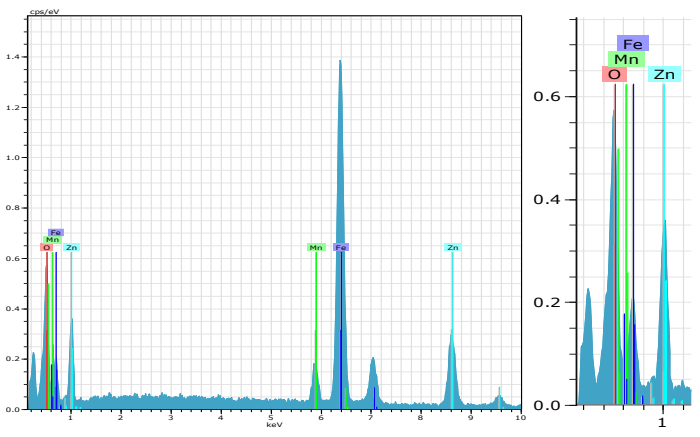

b)

| El | AN | Series | unn. C [wt.%] | norm. C [wt.%] | Atom. C [at.%] | Error (1 Sigma) [wt.%] |
|---|---|---|---|---|---|---|
| O | 8 | K-series | 19,65 | 26,17 | 56,22 | 3,49 |
| Mn | 25 | K-series | 3,01 | 4,01 | 2,51 | 0,14 |
| Fe | 26 | K-series | 38,21 | 50,91 | 31,33 | 1,05 |
| Zn | 30 | K-series | 14,20 | 18,91 | 9,94 | 0,46 |
| | | Total: | 75,06 | 100,00 | 100,00 | |

**Figure 4.10 –** EDS results for $Mn_{0.2}Zn_{0.8}Fe_2O_4$. a) EDS spectrum – legend is: red for O, green for Mn, light blue for Zn and dark blue for Fe. b) EDS results table.

#### 4.2.3.1 Systematic Errors

For lighter elements like Oxygen, the EDS presents a considerable error. For this reason, when analyzing the samples composition, it was assumed that the Oxygen was fully integrated in ferrite.

### *4.2.4 EDS Analysis*

For stoichiometry estimation, it was assumed that the Oxygen has 4 elements per molecule and the sum of all metal ions results in a total of 3 elements, accordingly with ferrite metal/Oxygen ratio: $Mn_{1-x}Zn_xFe_2O_4$. The composition of the sample was then calculated by normalizing the metal ions and multiplying them by 3. Equation 4.3 shows the stoichiometry calculus, per element, for every sample.

$$\#Mn = 3\frac{\%Mn}{\%Fe+\%Mn+\%Zn}\ ;\ \#Fe = 3\frac{\%Fe}{\%Fe+\%Mn+\%Zn}\ ;\ \#Fe = 3\frac{\%Fe}{\%Fe+\%Mn+\%Zn} \quad (4.3)$$

Missing elements, or excess, are represented by the difference between the measured fraction of elements and the expected fraction of elements.

### *4.2.5 TRANSMISSION ELECTRON MICROSCOPY (TEM)*

Transmission electron microscope (TEM) is a device that relies on electromagnetic accelerated electrons being transmitted through a sample for imaging it. TEM is a powerful device for nanoparticles imaging due to its high resolution, in some cases (high-resolution TEM), capable of imaging the atomic lattice of the sample, [55]

The working principle of TEM is the following: electrons are created in a hot-filament (electron gun) and accelerated by anodes in sample direction, then a group of condensing lenses focus and direct parallelly the beam towards the sample. The high-energy electrons, that reach the sample, have a short wavelength, which allows them to diffract in the atomic lattice, thus creating a reciprocal-space image or a real-space image. Changing the focus of the condensing lenses the user can change between a reciprocal-space or a real-space image, [56]. During this work, the TEM equipment available, and capable of magnetic nanoparticles visualization was H9000 Hitachi, figure 4.11.

a) b) c)

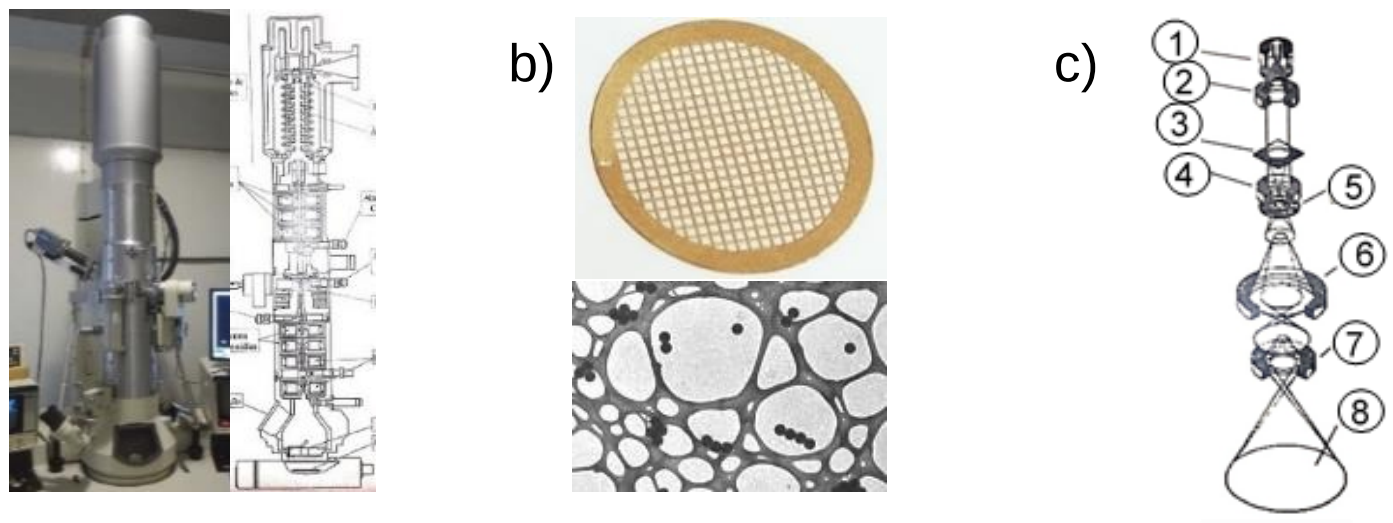

**Figure 4.11 –** a) TEM and its components. b) Sample-holder (Carbon grids) macro and micro scale c) Electrons' path in TEM, components used: 1 – Electrons' source; 2 – Condensing leans; 3 - Sample; 4 – Objective lens; 5 – Objective lens aperture; 6 – Intermediate lens; 7 – Projector lens; 8 – Fluorescent display.

#### 4.2.5.1 SAMPLE PREPARATION

The preparation of nanoparticles for TEM starts with the dilution of a small fraction of powder in ethanol, then the prepared solution goes to an ultrasonic bath to favor particle dispersion in ethanol and prevent particle agglomeration. When the powders are sufficiently diluted a TEM support grid is dive inside the solution and is let to dry. The used TEM uses Copper, or Carbon, grids as support for the nanoparticles. The grids must be conductive in order to prevent electrical charging of the sample. The grids were mounted in TEM support and attached to the microscope.

#### 4.2.5.2 SIZE DISTRIBUTION

Nanoparticles' area was acquired using ImageJ software. This software outputs the counts and the area of a manually surrounded nanoparticles. These results allowed the estimation of the nanoparticles' diameter which is then displayed in a histogram of counts as a function of particle size, figure 4.12. The mean grain size, $\langle d_{TEM} \rangle$ is obtained by adjusting the size distribution to a Lorentz function. The standard deviation, σ, is obtained using the equation 4.4, where μ is the mean size and $x_i$ is the size of each nanoparticle.

a) b)

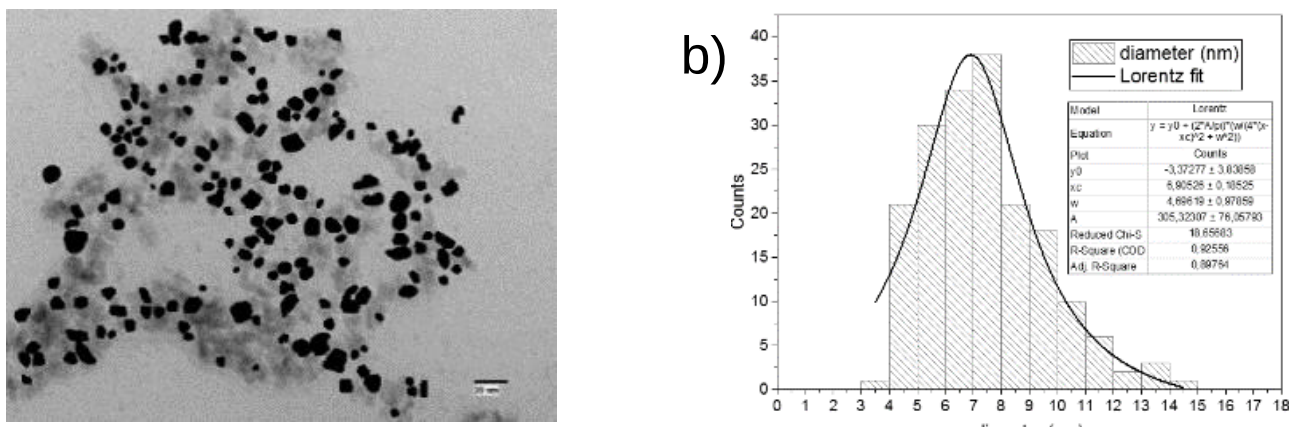

$$\sigma = \sqrt{\frac{1}{N}\sum_0^n (x_n - \mu)^2} \qquad (4.4)$$

**Figure 4.12 –** a) Nanoparticle count; b) Lorentz fit.

### *4.2.6 SUPERCONDUCTING QUANTUM INTERFERENCE DEVICE (SQUID)*

The Superconducting Quantum Interference Device (SQUID) is a highly sensitive magnetometer with a Josephson junction which allows the measurement of magnetic fluxes down to the $10^{-17}$ Wbm. Its incomparable sensitivity makes it the election magnetometer for a vast range of applications, from measurements of materials magnetic properties to 3D mapping of the brain.

The model of the used SQUID is a MPMS-3 (Quantum Design, Inc.), figure 4.13.a), with a magnetization sensibility of 5x10$^{-8}$ emu and capable of producing magnetic fields up to 7T, [57]. Figure 4.13.b) is a schematic of the main chamber, with the 4 superconducting coils presented, connected to a Joseph-junction, [58].

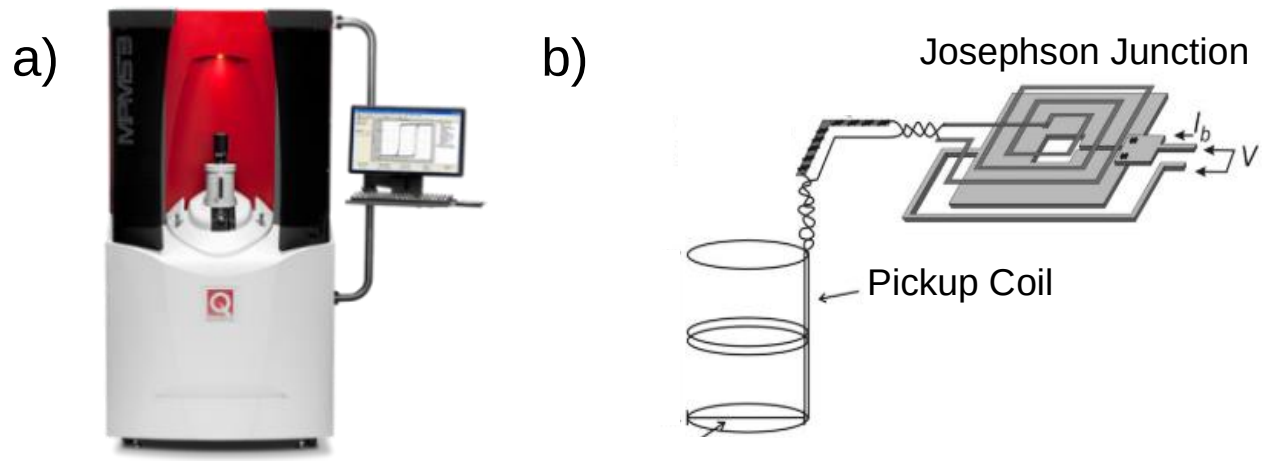

Figure 4.13 – (a) SQUID MPMS-3 (Quantum Design, Inc.). (b) Schematic of the SQUID superconducting coils connected to a Josephson-junction.

A SQUID is a complex and versatile magnetometer capable of working in a vast range of temperatures, fields and materials. It is composed of many different components, sensors and four main coils. Superconducting wires are connected to the induction coils and to the Josephson-junctions. A DC current creates a magnetic field, magnetizing the sample. The sample is measured by the Josephson-junctions as it oscillates in the vertical axis of the magnetometer. While moving, the sample inducts a magnetic field flux which is detected by the Josephson-junctions. A series of measurements, and sequences, can be pre-programmed using the magnetometer software. Inside the program sequence it is possible to make measurements while changing variables. For the purpose of this work, temperature, T, and applied field, H, are the variables and sample magnetization, M, is outputted by the magnetometer as M(T) and M(H).

The Josephson-junction working principle is based on the Josephson effect. A Josephson-junction is made of 2 superconducting layers separated by a thin insulator layer. Josephson predicted mathematically how the supercurrent phase changes when tunneling through the insulator layer, [59]. The Josephson-junction presents an impedance and capacitance, thus it resonates at some frequencies. The flux change generated by the vibrating samples is acquired by the Josephson-junction as a phase difference between the junction sides, resulting in a destructive or constructive interference of spin-waves. This extremely sensitive device is the reason why SQUID magnetometer is capable of distinguishing magnetic fluxes of the magnetic flux quanta order (h/2e), [60].

The function of the nitrogen and helium reservoirs is to cool the Josephson-junction and the induction coils (keeping them in a superconducting state) and to control the sample temperature, the magnetometer is also provided with resistors for heating. The vacuum pump is connected to the main chamber and to the antechamber. The main chamber is always kept in vacuum, the antechamber must be in vacuum before passing the sample to the main chamber. The vacuum pump is an important component of the magnetometer for various reasons: the magnetometer is composed of small tubes for cooling liquids delivery, the tubes are so thin that they can easily clog if a solid compound freezes in the tube. Also, at liquid helium temperature (boiling point at ~4,4 K) some gaseous compounds can crystalize, for example oxygen, that in the solid state presents a magnetic transition from antiferromagnet to ferromagnet at 43 K, [61]- oxygen impurity can contribute with a parasite signal.

A series of measurements, and sequences, can be pre-programmed using the magnetometer software. The used sequences consisted in measurement of the M(H) and M(T) curves of Mn-Zn ferrite.

#### 4.2.6.1 Sample Preparation

Powder samples are encapsulated and weighed. Due to their ferromagnetic behavior they are held into a sample holder of brass with Kapton tape - both have a paramagnetic behavior with low susceptibility in order not to interfere with the sample signal. The sample holder is kept in the antechamber until vacuum levels of main chamber and antechamber are equal. The sample is then moved to the main chamber and centered between coils with resource to a small applied field and a M(z) curve. Once centered, oscillation amplitude and frequency must be adjusted to the magnetic behavior of the sample in order not to saturate the detector. Then the measuring sequence is selected and the measurement can start. Samples measurement sequence starts with M(T) curves and then M(H) curves.

#### 4.2.6.1 SQUID Data Analysis

The SQUID was used for obtaining M(H) curves at 5, 300 and 380 K and M(T) curves between 5 and 380K, with an applied field of 100 Oe. Both data sets give information about the magnetic properties of the samples at different temperatures. The extracted information from M(H) curves was saturation and remnant magnetization, coercive field, hysteresis area and magnetic susceptibility. The magnetic ordering temperature and blocking temperatures were obtained from the M(T) curves (when visible). All magnetization (in emu) were normalized to the sample mass in order to obtain the magnetization per mass (emu/g).

The M(H) curves give information on how the materials magnetization change with the applied field. These curves are taken by starting the magnetic field at 70000 Oe, measurements are made while decreasing the field until -70000 Oe and inversing it to 70000 Oe. The data acquisition is logarithmic close to 0 T in order to have more points in this region and hence decrease the error on the coercive field and remnant magnetization estimations close to zero-field. The linear magnetic susceptibility, $\chi$, is calculated by linearizing of M(H) slope for high fields. this constant is related with paramagnetic and antiferromagnetic states, as other misalignment of spins.

The measured M(T) curves reveal how the materials magnetization change with the evolution of temperature. Two M(T) curves were taken, the zero-field cool (ZFC) and field cool (FC). For the ZFC measurement the sample is cooled until 5 K without a magnetic field applied. In this procedure the sample is cooled down without any preferred direction for the magnetic particles. Then, a field of 100 Oe is applied to the sample and measurements starts, the time between measurements is approximately 11 s. FC measurements start at room temperature, when a field of 100 Oe is applied to the sample, followed by the sample cooling down. Measurements start when the sample achieves 5/6 K. The measurements are taken while increasing the temperature, maintaining the 100 Oe magnetic field). FC curve must be acquired after the ZFC, as if it was the opposite the samples might keep a preferred magnetic orientation which is not desired for the zero-field cool measurements.

The magnetic ordering temperature was estimated by the linearization of Curie-Weiss law at the paramagnetic state. Figure 4.14 shows the used linearization and the H/M(T) plot. From the linearization it is possible to obtain a close value of the Curie temperature, the calculated value is $\Theta_p$, as assigned in the graph.

a) $$\chi = \frac{C}{T - \Theta_P} \Leftrightarrow \chi(T - \Theta_P) = C \Leftrightarrow$$

$$\Leftrightarrow T = \frac{C}{\chi} + \Theta_P$$

$$y = T \; ; \; x = \frac{1}{\chi} = \frac{H}{M}$$

$$m = C \; ; \; b = \Theta_P$$

b) 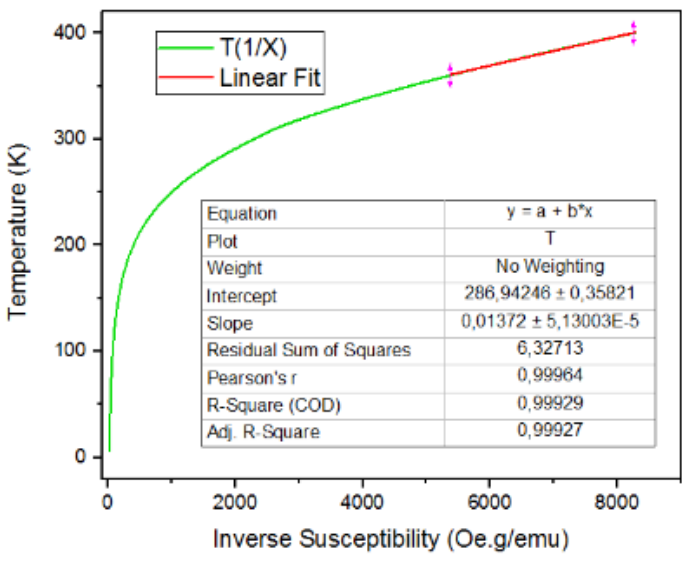

**Figure 4.14 –** a) Curie-Weiss law linearization to obtain $\Theta_p$. b) Determining $\Theta_p$ using the temperature dependence of the inverse susceptibility for $ZnFe_2O_4$.

The mean blocking temperature was calculated with resource to the maximum of the temperature derivative of the difference between FC and ZFC magnetization curves, [53], $\frac{\partial(FC-ZFC)}{\partial T}$. The different magnetization of the FC-ZFC curves is due to the blocked states. The derivative have a maximum in the mean blocking temperature, $\langle T_b \rangle$, and the maximum temperature of the distribution agrees with the irreversibility temperature, [10].

The susceptibility, $\chi_P$, was obtained from the M(H) curves by linearizing the hysteresis loop at high field intensity, in the regime where the magnetization is almost saturated. By subtracting the susceptibility from M(H) curves, it is possible to decompose the ferrimagnetic hysteresis loops in a ferromagnetic loop and a paramagnetic line.

#### 4.2.6.2 SYSTEMATIC ERRORS

The SQUID magnetometer is a very sensitive device, that fits the data points while acquiring, for this reason the measurements are reliable. However, the machine is calibrated for specific shape and size, as the sample shape influences the materials magnetization and the flux lines passing through the Josephson-junction. This said, a systematic error was detected when samples magnetization was measured with resource to 2 different measuring modes, DC and VSM. The sample magnetization must remain the same independently of the measuring mode, using this principle a correction was proposed by [62]. All data were corrected accordingly with personal communication of the correction.

### *4.2.7 MAGNETIC INDUCTION HEATING*

Magnetic induction heating is a technique that measures the temperature change of a sample while submitting it to an AC magnetic field. Heating a sample with an AC magnetic field is a phenomenon based on the sample hysteresis loops loss, i.e. the inversion of the sample magnetic moment releases energy in the form of heat. Consecutive inversions of the sample magnetization will generate energy in heat form which is read by a temperature sensor. Magnetic hysteretic loss is the main energy generation mechanism, whereas other magnetic loss mechanisms were previous mentioned in section 2.5.

The magnetic induction setup was developed by Dr. Nuno João Silva from "CICECO – Aveiro Institute of Materials", it is homemade and has an extensive apparatus. Figure 4.15.a) displays a schematic of the system and figure 4.15.b) presents the inductor coil.

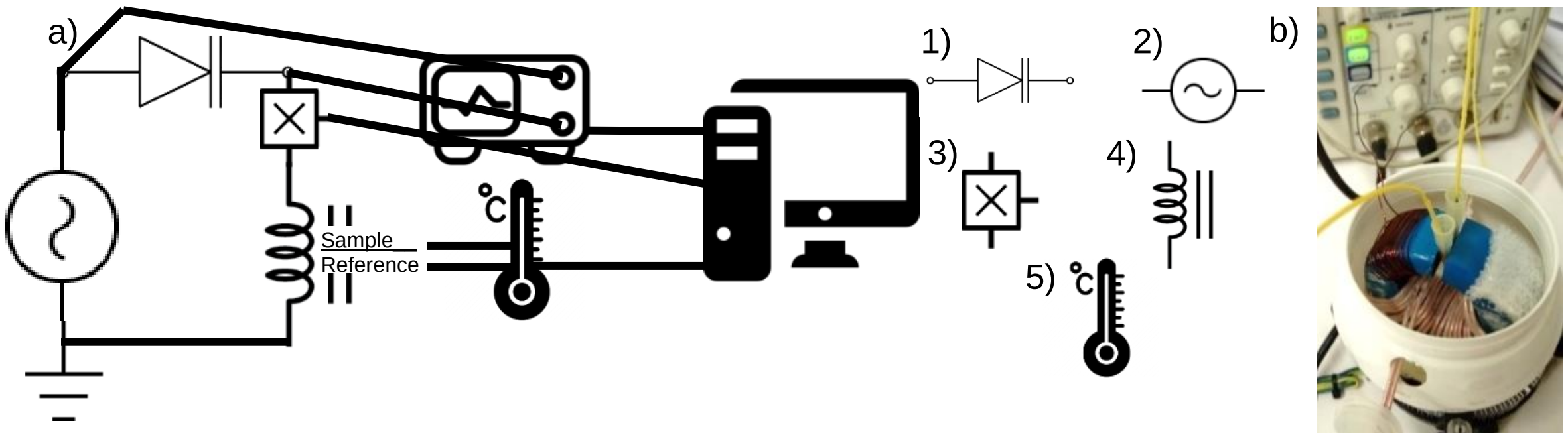

**Figure 4.15 –** a) Magnetic induction heating experimental setup. 1) amplifier with a resonant circuit. 2) Signal generator; 3) Hall sensor; 4) Induction core; 5) Fiber optics thermometer. b) inductor core

The system is composed of a magnetic inductor core in a ring shape with an aperture where the sample and reference are placed. The aperture, which is seen in figure 4.15.b), has an important role, as the magnetic field lines between this space are parallel, this ensures that, both sample and reference, are subjected to the same intensity of magnetic field.

The current flow starts in an AC generator, goes through an operational amplifier, then into a capacitors system in order to achieve resonance. The described system has the function of increasing the current intensity passing by the coil, which is directed to the sample by the inductor core. The magnetic field generated by the magnetic core is controlled by the electric current that flows in the coil strapped around the core. For a better control of the magnetic flux, a Hall-sensor is attached to the wire that turns around the coil, measuring the potential caused by an inducted flux.

For temperature measurements, the magnetic induction heating system has two temperature sensors, both are composed of an optical fiber with a silica end. Both temperatures (sample and reference) as a function of time are recorded. It is important that the reference (empty) sample holder has a temperature sensor of its own, because, the AC magnetic field will not only heat the sample but also the induction core. The inductor core temperature, T(t), is subtracted from the sample signal.

#### 4.2.7.1 Sample Preparation

Mn-Zn ferrite powders are placed into an organic, non-metallic and non-magnetic sample holder. The reference sample holder remains empty. Due to the polymeric nature of the sample holder the maximum temperature reached was 80ºC. The frequency of the generator is then set into resonance, at 364 kHz, and the magnetic field can be controlled using the system software (a matlab GUI for controlling the AC generator was used).

Both sample holders are placed in the induction core aperture, with a temperature sensor inside each of them. The measurements can now start.

#### 4.2.7.2 Data Analysis

The raw data of a magnetic induction heating experiment is shown in figure 4.16. The data file a matlab figure file. To extract the data, a matlab script was developed.

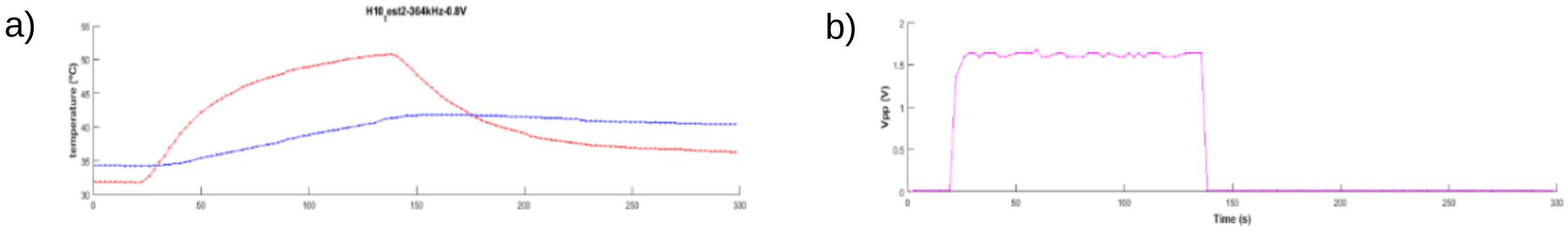

**Figure 4.16 –** Magnetic induction heating results as a function of time. a) Red line for sample. Blue line for sample-holder. b) Hall sensor Voltage p-p (V).

The difference in the initial temperature is associated with differences in both sensors and it does not compromise the experiment, as the analysis is based on the sample initial slope. The blue line shows the reference sample holder increasing the temperature - it happens due to the increasing induction core temperature.

For plotting the sample heating rate, dT/dt, the reference curve was subtracted from the sample curve, as this is representative of how much sample heated more than the reference. Then, the initial temperature was added to the difference to show the real temperature values achieved by the sample under induction heating mechanism. The heating rate is then obtained by linearizing the initial slope of the magnetic induction heating results, as shown in figure 4.17

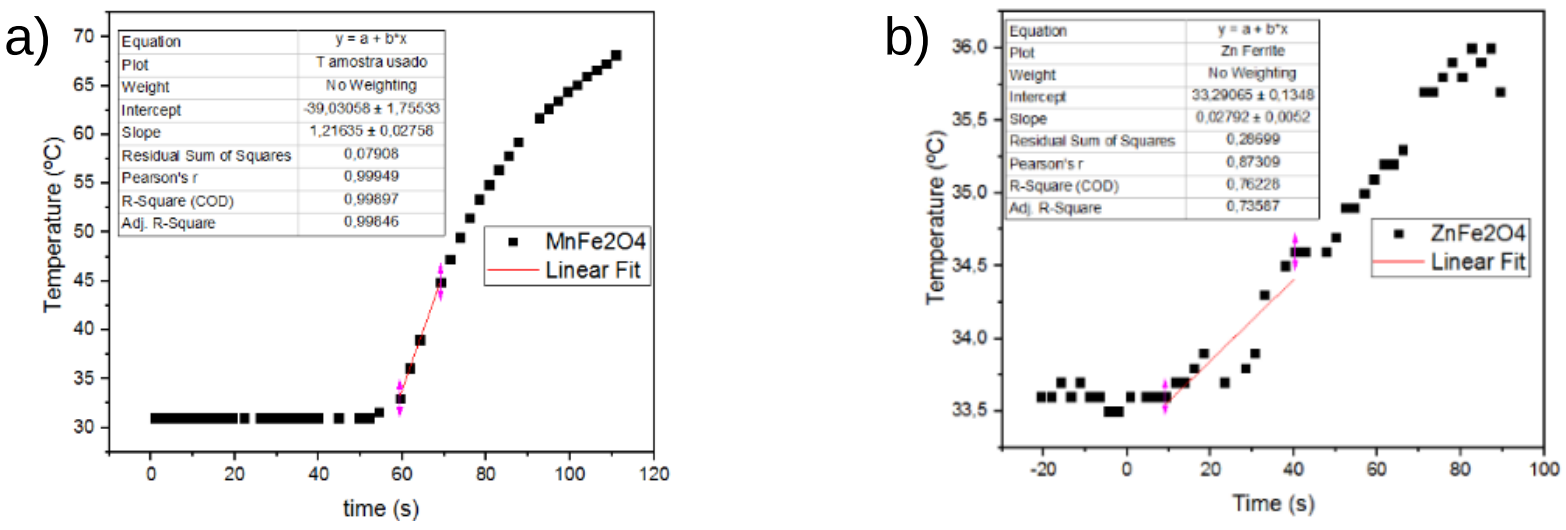

**Figure 4.17 –** Linear fit of the initial slope for $MnFe_2O_4$ and $ZnFe_2O_4$.

SAR (Specific Absorption Rate) values are calculated with resource to equation 4.5, [2]:

$$SAR = C.\frac{\Delta T}{\Delta t}.\frac{1}{m_{sample}} = c.\frac{\Delta T}{\Delta t} \quad (4.5)$$

Where C is the heat capacity of the medium, $\frac{\Delta T}{\Delta t}$ is the rate change of the temperature (given by the initial slope of the magnetic induction heating graph) and $m_{sample}$ is the sample mass. Heat capacity is usually associated with the medium where the particles are suspended, however, the experiment was made only with powder, for this reason, the heat capacity, C, must be calculated with resource to the specific heat, c, the relationship is $c = \frac{C}{m}$. The value of specific heat for Mn-Zn ferrite is 750 J $Kg^{-1}$ $K^{-1}$, accordingly to [63] and [64].

# Chapter 5: RESULTS AND DISCUSSION

The results of sol-gel auto-combustion method and hydrothermal method were independently analyzed by the characterization techniques detailed in the previous chapter. After results of sol-gel auto-combustion method a brief discussion is displayed. Follows the results of hydrothermally synthesized samples, which are presented in higher detail due to the superior quality of this samples. The discussion of hydrothermally prepared samples is presented after its results. The discussion on the hydrothermal samples intends to correlate results and finishes with a general review of the Mn-Zn ferrite nanoparticles.

## 5.1 RESULTS OF SOL-GEL AUTO-COMBUSTION METHOD

Sol-gel auto-combustion samples were characterized via XRD, SEM, SQUID and TEM. Analyzed samples have the composition of $Mn_{1-x}Zn_xFe_2O_4$ (x= 0; 0.2; 0.5; 0.8; 1) and were not dehydrated. Most results are compared with the results from "Magnetic and Optical Properties of $Mn_{1-x}Zn_xFe_2O_4$ Nanoparticles" of A. Demir et al, [65]. This article is referenced due to the similarities in experimental procedures, full composition range (x = 0 to 1), similar characterization techniques and coherent results.

### *5.1.1 STRUCTURAL CHARACTERIZATION*

X-ray diffraction was the technique used to explore the crystallographic order and distinguish crystallographic phases. Figure 6.1 shows the XRD patterns for all the compositions.

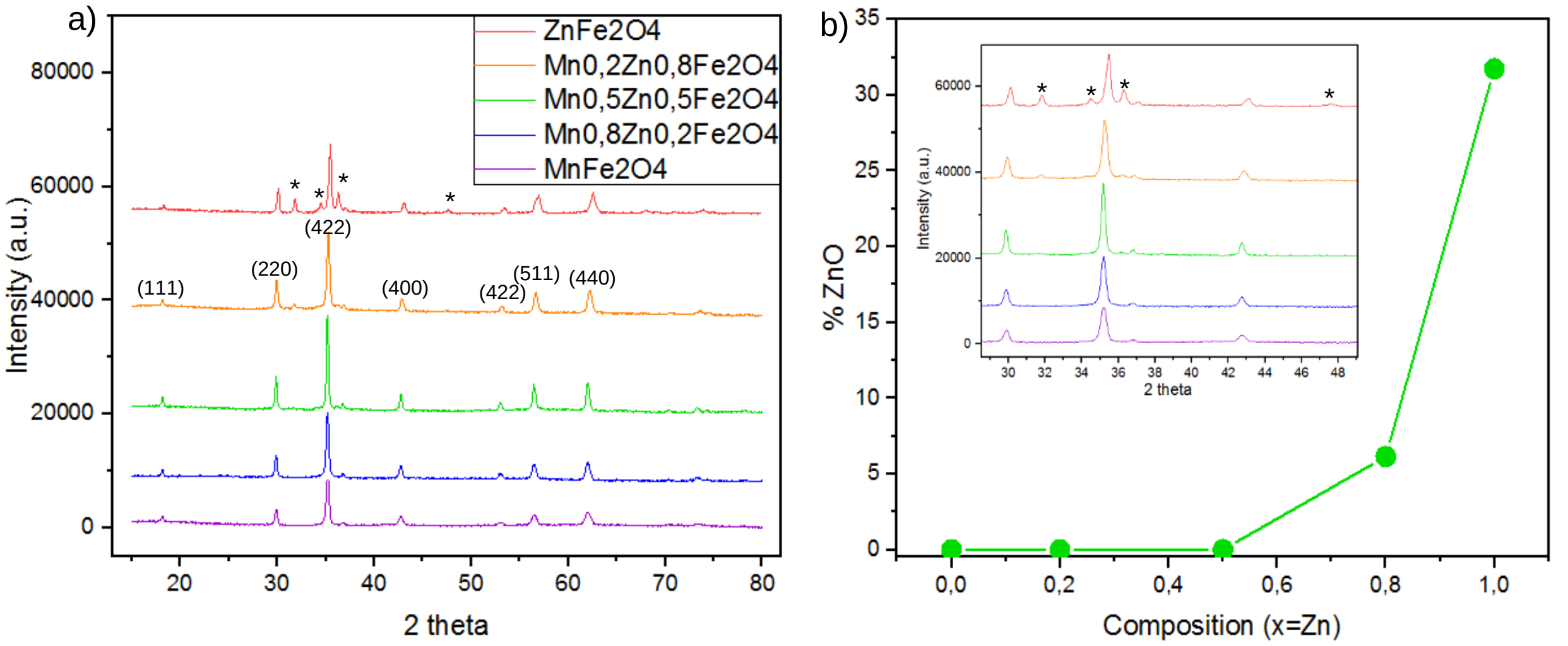

**Figure 5.1 –** a) XRD patterns of $Mn_{1-x}Zn_xFe_2O_4$. Miller indices of Mn-Zn ferrite spinel crystal structure are displayed. ZnO impurity peaks are represented by *. b) percentage of ZnO impurity as a function of Zn content, inset graph shows the evolution of ZnO peaks in the diffractogram.

The diffractograms displayed in figure 5.1.a present the peaks of the spinel crystal structure, space group fd-3m belonging to Mn-Zn ferrite, and the respective Miller indices are indexed. The secondary phase, Zincite (ZnO), peaks are marked with *. The percentage of ZnO is composition dependent, once the samples with higher Zn/Mn ratio present higher percentage of impurity phase, figure5.1.b). $ZnFe_2O_4$ is the composition which presents the highest percentage of impurity phase (32%). Magnetite ($Fe_3O_4$) can be present in the samples, but it is not distinguishable from Mn-Zn ferrite diffractogram, since it belongs to the same space-group and has the same peaks positions.

A shift of the main peak of Zn ferrite for higher angles in comparison with Mn ferrite was noted, which predicts a smaller lattice constant for Zn ferrite, figure 5.2.a). The lattice constant was obtained via Rietveld refinement. Figure 5.2 is a comparison between the obtained lattice parameters and crystallite sizes with those obtained by A. Demir et al, [65].

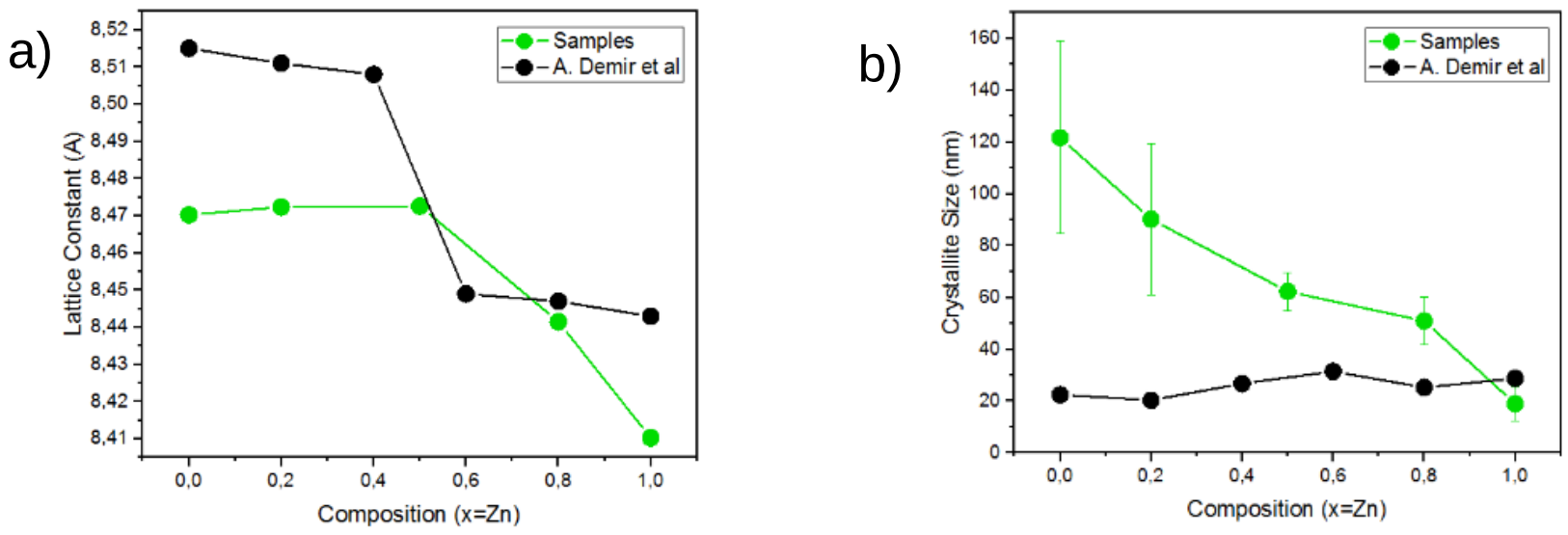

**Figure 5.2 –** a) Lattice constant as a function of the Zn content; b) Crystallite size as a function of Zn content. Green line is for synthesized samples. Black line is for bibliographic samples from A. Demir et al.

Lattice parameter is in rough agreement with A. Demir et al, varying from 8.47 to 8.41 Å with the increase of Zinc content. The crystallite sizes for samples with higher Mn content are more displaced from A. Demir et al values than ferrites with higher Zn content.

Zincite (ZnO) lattice constant was obtained from the Rietveld refinement for both samples with more impurity phase (x = 0.8 and 1). ZnO has an hexagonal crystal structure, space-group: $P6_3mc$, lattice constant values for a(=b) of 3.250 Å and c of 5.207 Å, in [66]. The Rietveld refinement values for ZnO lattice constant were a(b) = 3.25 and 3.25 Å, and c = 5.21 and 5.27 Å, respectively for x=0.8 and 1 samples. Lattice constants for both compositions agree with bibliographic values, [66], [67]

### 5.1.2 *Morphological and Chemical Characterization*

A selection of the best SEM images obtained for sol-gel auto-combustion samples are shown in figure 5.3.

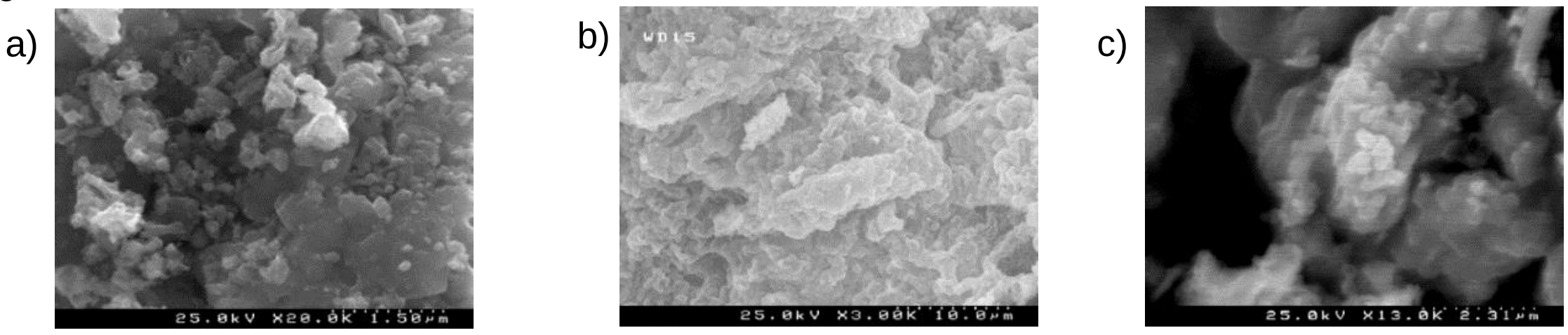

**Figure 5.3 –** SEM images of samples**:** a) $MnFe_2O_4$; b) $Mn_{0.2}Zn_{0.8}Fe_2O_4$; c) $ZnFe_2O_4$.

SEM images cannot distinguish individual nanoparticles even with the maximum amplification achievable (x20k, scale of 1.5 µm). All images present agglomerated particles. The morphology of the powders is similar between compositions, it is also similar with the samples reported by A. Demir and R. Gimenes et al.

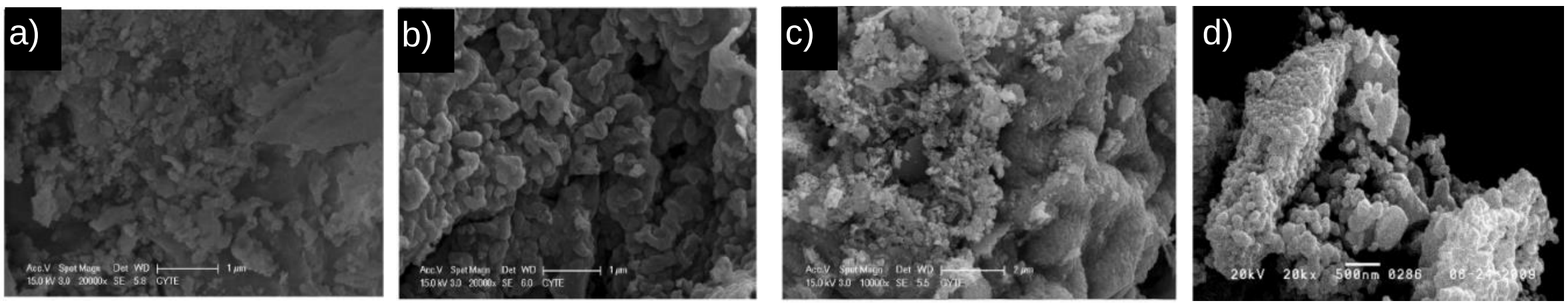

**Figure 5.4 –** SEM images from literature: (a), (b) and (c), from A. Demir et al, showing the $MnFe_2O_4$, $Mn_{0.2}Zn_{0.8}Fe_2O_4$ and $ZnFe_2O_4$, respectively. Image (d), from R. Gimenes et al, with $Mn_{0.8}Zn_{0.2}Fe_2O_4$.

The EDS results allowed to estimate the metal ions ratio in the samples, figure 5.5.

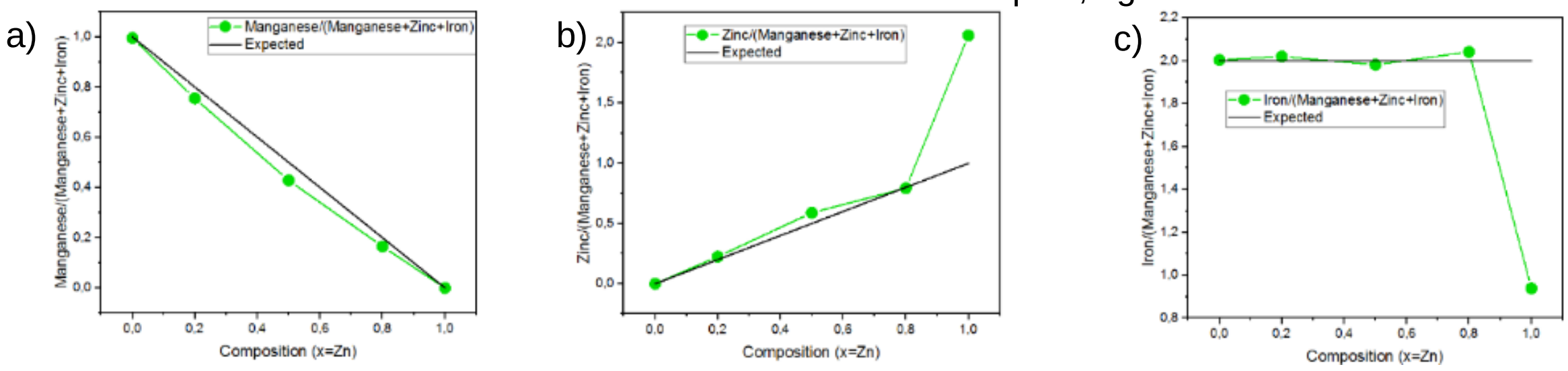

**Figure 5.5 –** Metal ions ratio. a) Mn ratio; b) Zn ratio; c) Fe ratio. Green line is the sample ratio and black line is the expected ratio of the predicted stoichiometry.

By the analysis of figure 5.5 it's visible that the largest deviation from the expected values occurs for $ZnFe_2O_4$. Zinc in excess (twice the expected) and Iron lacking (50%) might be a consequence of excess of Zinc and lack of Iron during the synthesis, contributing for the formation of ZnO impurity and lack of ferrite phase. Mn is easily incorporated by the samples, with almost no deviation from the expected values. The calculated ratios by EDS strongly correlate with the percentage of zincite impurity calculated via XRD.

### 5.1.3 *MAGNETIC CHARACTERIZATION*

The powders were analyzed in SQUID magnetometer. The powders magnetization was measured as a function of applied field (M(H) curves) at different temperatures, figure 5.6. The magnetization as a function of temperature at 100 Oe (M(T) curves) is presented in figure 5.8.

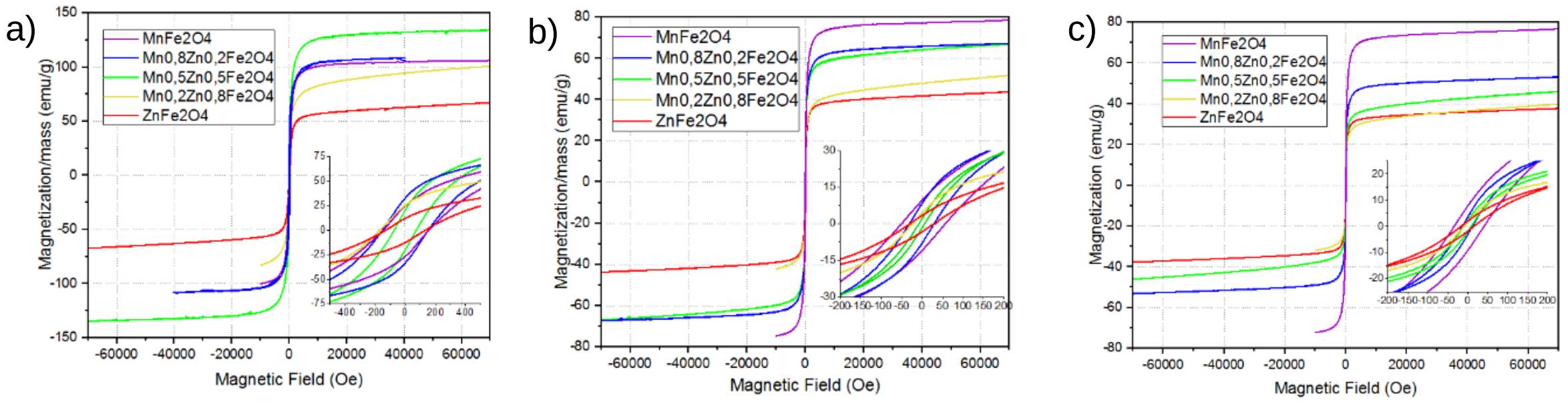

**Figure 5.6 –** M(H) curves at: 5K a), 300K b) and c) 380K for Mn1-xZnxFe2O4 (x = 0; 0.2; 0.5; 0.8; 1). Inset figure shows the hysteretic loops of the samples.

The M(H) loops are expected to present an increase of saturation magnetization from Zinc to Manganese compositions, accordingly with A. Demir et al, [65]. This is expected due to the magnetic moment of Mn ion being superior than Zn ion. Mn ferrite has the highest $M_S$ at 300 and 380 Kelvin but, at 5K, $Mn_{0.5}Zn_{0.5}Fe_2O_4$ saturation magnetization surpasses $MnFe_2O_4$. $M_S$ values vary from 77.8 to 43.8 emu/g with the increase of Zn/Mn ratio, at room temperature, figure 5.7.a).

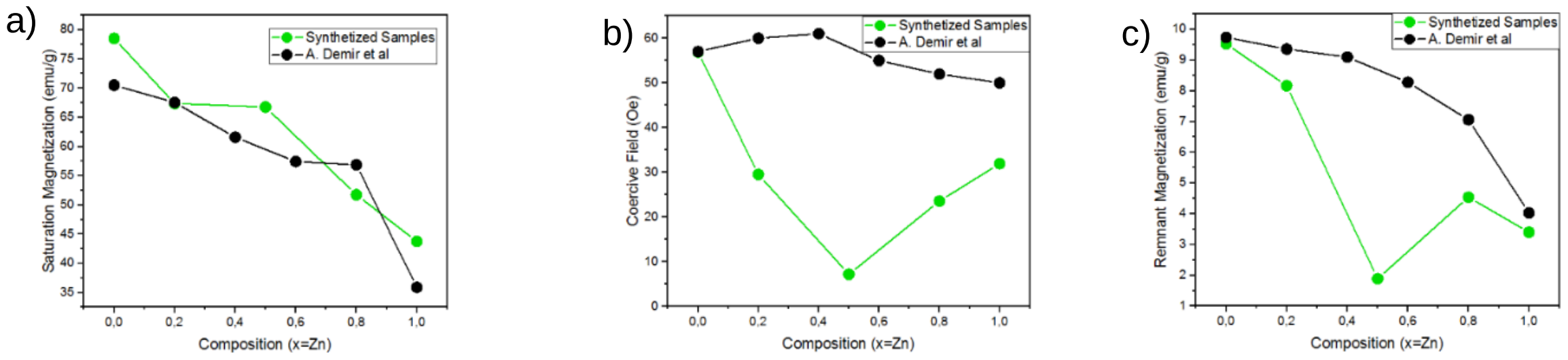

**Figure 5.7 –** Magnetic data obtained from M(H) curves at 300K. (a) Saturation Magnetization; (b) Coercive Field; (c) Remnant Magnetization. Green lines are for synthesized samples and black lines for the comparison with samples of A. Demir et al.

Values of coercive field, $H_C$, and remnant magnetization, $M_R$, do not present the expected tendency as can be seen in figure 5.7.b) and c), against Demir et al obtained values. Remnant magnetization generally seems to follow the tendency observed in literature, [65] [68], however, the composition x=0.5 disrupts the decreasing behavior.

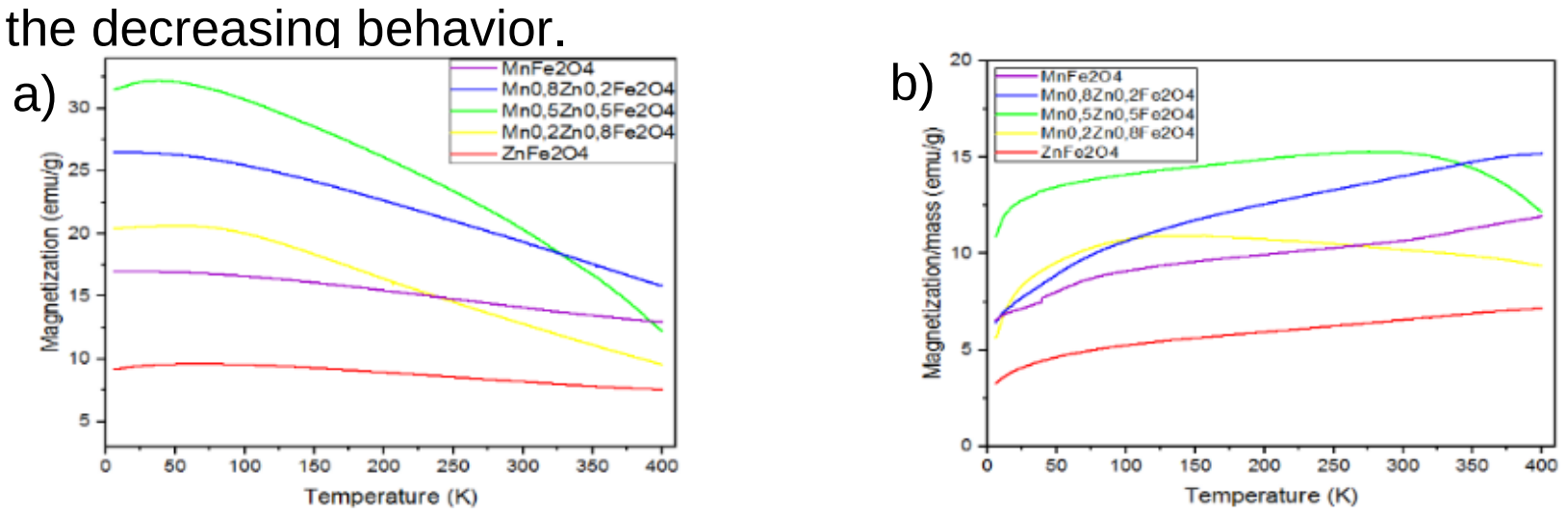

**Figure 5.8 –** M(T) curves measured at 100 Oe. a) Field cooled. b) Zero-field cool.

The FC curves show a decrease of magnetization with the increase of temperature, which is expected for ferrimagnetic materials. In accordance with the literature on sol-gel auto-combustion samples, the magnetic ordering temperature occurs only for higher temperatures (above the measurement range, 400K) unless thermal treatments, encapsulation or combustion in alternative mediums is performed, [67], [69], [70], [71], [72]. This result is expected but not desired for the purpose of this thesis. Then, sol-gel auto-combustion samples were analyzed via TEM.

### *5.1.4 Particle Size Characterization*

Two samples were selected to be analyzed by TEM microscope: $Zn_{0.2}Mn_{0.8}Fe_2O_4$ due to low quantity of secondary phase and $ZnFe_2O_4$ due to high quantity of secondary phase. In Figure 5.9, two obtained images for both samples are displayed.

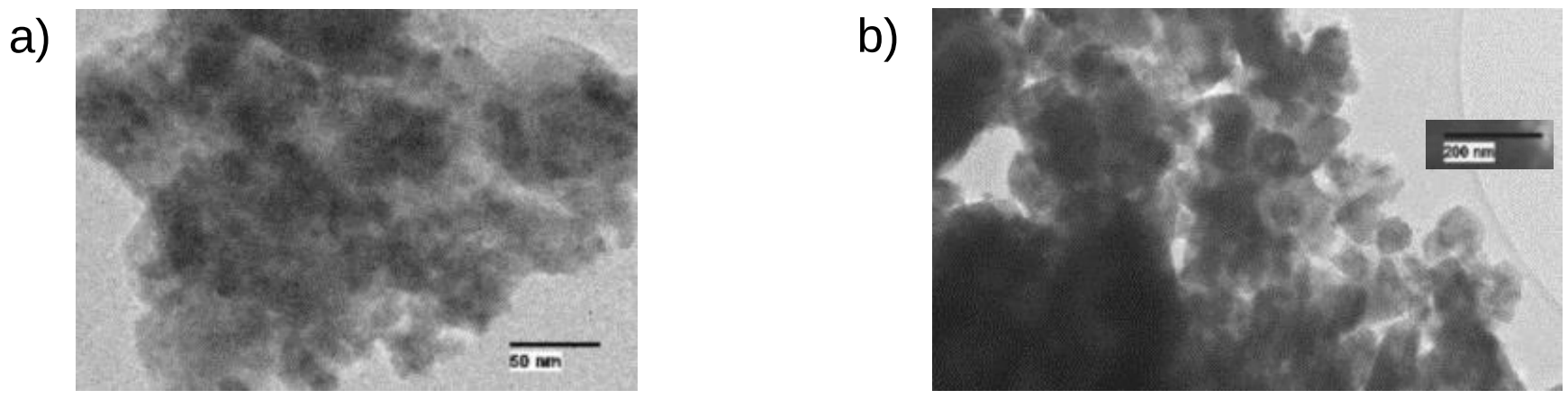

**Figure 5.9 –** TEM images for sol-gel auto-combustion samples. a) $Zn_{0.2}Mn_{0.8}Fe_2O_4$. b) $ZnFe_2O_4$.

Both images show agglomerated nanoparticles. For the first image, $Zn_{0.2}Mn_{0.8}Fe_2O_4$, it is noticeable that the image is blurry, probably due incomplete evaporation of the solvent (ethanol). Whereas in the second image, $ZnFe_2O_4$, is was not possible to distinguish single nanoparticles for higher amplification, and a higher amplification had to be used. The wide size particle distribution is clear, also the larger particle size, which can be consequence of counting agglomerated nanoparticles. Particle counting and diameter analysis confirms the previous sentence, figure 5.10.

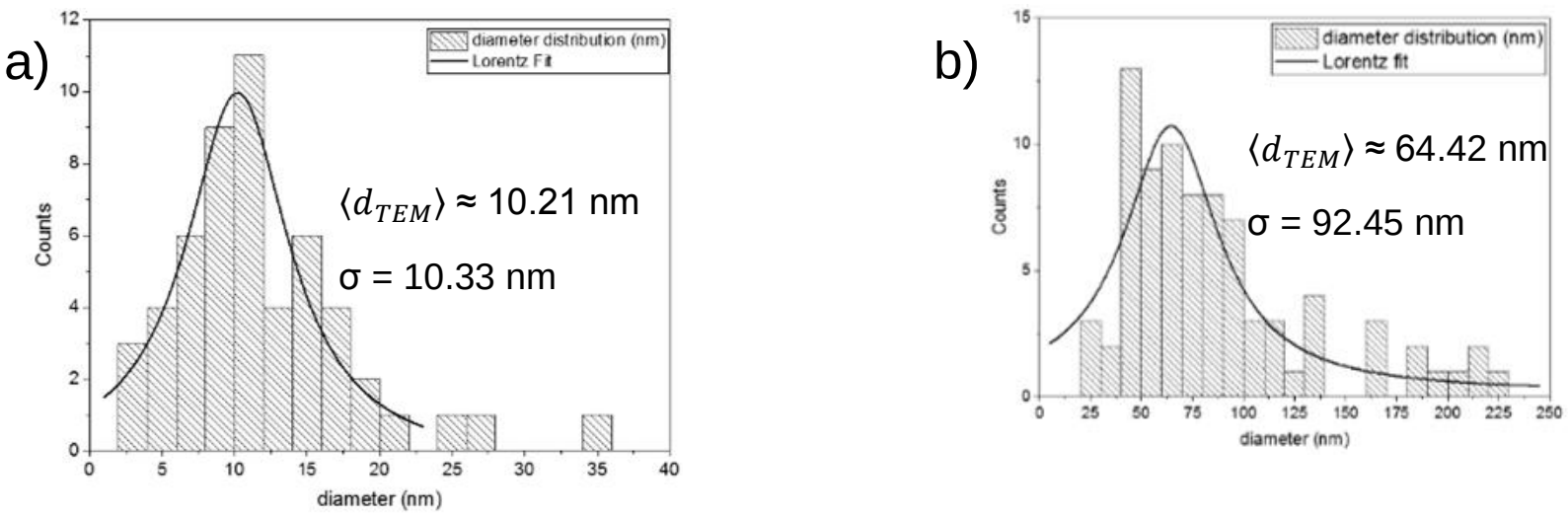

**Figure 5.10 –** Size distribution of a) $Zn_{0.2}Mn_{0.8}Fe_2O_4$ and b) $ZnFe_2O_4$.The distribution was adjusted with a Lorentz function.

From the particle counting, it is visible the mean size is ~10nm for $Mn_{0.2}Zn_{0.8}Fe_2O_4$ and ~64nm for $ZnFe_2O_4$. The particle size distribution for $ZnFe_2O_4$ is wider (σ ~ 92 nm) that its mean size. $Mn_{0.2}Zn_{0.8}Fe_2O_4$ size distribution it's of the same scale of the mean size (σ ~ 10 nm). Such a wide size distribution leads to a consequential wide distribution of the magnetic properties (Tc, Msat, etc..) and ultimately impacts in their magnetic hysteretic losses because of their strong dependence on the particle size. Consequently, the correct assessment of these important magnetic parameters becomes impractical or nearly impossible. Difficulties controlling the particle size will difficult the control of all magnetic and structural properties and therefore all the further work.

## 5.2 Discussion of Sol-Gel Auto-Combustion Method

The sol-gel auto-combustion method is a reliable synthesis method to produce oxide samples in large amounts. This method requires high control over the combustion in order to obtain reproducible samples. The combustion is dependent of multiple synthesis parameters which difficult the synthesis.

The XRD results show the expected spinel crystal structure of the Mn-Zn ferrite. Impurity phase of Zincite increases in percentage as the content of Zn increases in the ferrite, the Zn ferrite has 32% of ZnO. Lattice constant is in rough agreement with bibliographic values. The crystallite size diverged from the expected values towards larger crystallite sizes.

The EDS results showed a good incorporation of the metal ions in the samples, except for $ZnFe_2O_4$. The SEM images present a morphology similar with literature.

M(H) curves displays saturation magnetization in agreement with the bibliographic behavior, however, coercivity and remnant magnetization are below the expected values probably due to equipment limitations. M(T) curves did not present the magnetic ordering temperature in the measurement range, 5-400K. TEM analysis was employed and the particles have a wide size distribution.

TEM revealed that the synthesized nanoparticles have a wide size distribution. This result demystifies why the magnetic characterization do not present the expected results. In particular the most important result for this work purpose, the magnetic ordering temperature dependence with Zinc content, is not clearly definable in the measured FC curves. Also, a wide size distribution would be a problem in a magnetic induction heating experiment, nanoparticles of different size in the sample would heat up by different mechanisms and achieve different temperatures. Furthermore, the wide size distribution also constitutes a disadvantage for biomedical applications. For instance, in hyperthermia applications, a narrow size distribution is required to ensure that the functionalization of the nanoparticles occurs similarly for every particle.

The cause of the undesired results might be the uncontrolled combustion temperature, atmosphere and the multi-step procedure which enables room for mistakes. If the objective was simply the synthesis of Mn-Zn ferrite, it would have been accomplished and the samples have good enough properties for a ferrofluid or a bulk application. For the main objective of this work, self-regulated heating via Curie temperature, the synthesized samples via sol-gel auto-combustion method are not suitable.

The undesired results obtained via sol-gel auto-combustion method lead to the change of the synthesis method. I. Sharifi et al, [20], published a paper with a review on various synthesis methods for Mn-Zn ferrite, where co-precipitation, thermal decomposition, microemulsion and hydrothermal methods are compared (table 5 of the paper resume them), from the description of these different synthesis methods hydrothermal was the elected as the alternative to sol-gel auto-combustion method, as it a simple and inexpensive technique which enables a narrower particle size distribution and a finer control over particle shape.

| Synthetic method | Synthesis | Reaction temp. [°C] | Reaction period | Solvent | Surface-capping agents | Size distribution | Shape control | Yield |
|---|---|---|---|---|---|---|---|---|
| Co-precipitation | Very simple, ambient conditions | 20–90 | Minutes | Water | Needed, added during or after reaction | Relatively narrow | Not good | High/ scalable |
| Thermal decomposition | Complicated, inert atmosphere | 100–320 | hours–days | Organic compound | needed, added during reaction | very narrow | very good | high/ scalable |
| Microemulsion | Complicated, ambient conditions | 20–50 | Hours | Organic compound | Needed, added during reaction | Relatively narrow | Good | Low |
| Hydrothermal synthesis | Simple, high pressure | 220 | Hours-days | Water-ethanol | Needed, added during reaction | Very narrow | Very good | Medium |

**Table 5.1 –** Synthesis methods summarized by H. Shokrollahi et al.

## 5.3 RESULTS OF HYDROTHERMAL METHOD

Hydrothermal method is adopted due to the controlled environment where nanoparticles are grown. The analyzed samples were synthesized following the autoclave procedure for 6 hours at a temperature of 180ºC. The chosen compositions of the produced samples were x= 0; 0.5; 0.8; 1; $Mn_{1-x}Zn_xFe_2O_4$, similarly to sol-gel method. Hydrothermal method results are in great accordance with the samples of P. H. Nam et al, [43], in “Effect of zinc on structure, optical and magnetic properties and magnetic heating efficiency of $Mn_{1-x}Zn_xFe_2O_4$ nanoparticles”.

### *5.3.1 STRUCTURAL CHARACTERIZATION*

The crystal structure of hydrothermally synthesized samples was investigated using the X-ray diffraction pattern of the powders. XRD patterns are presented in figure 5.11 for the compositions of $Mn_{1-x}Zn_xFe_2O_4$ x = 0; 0.5; 0.8; 1.

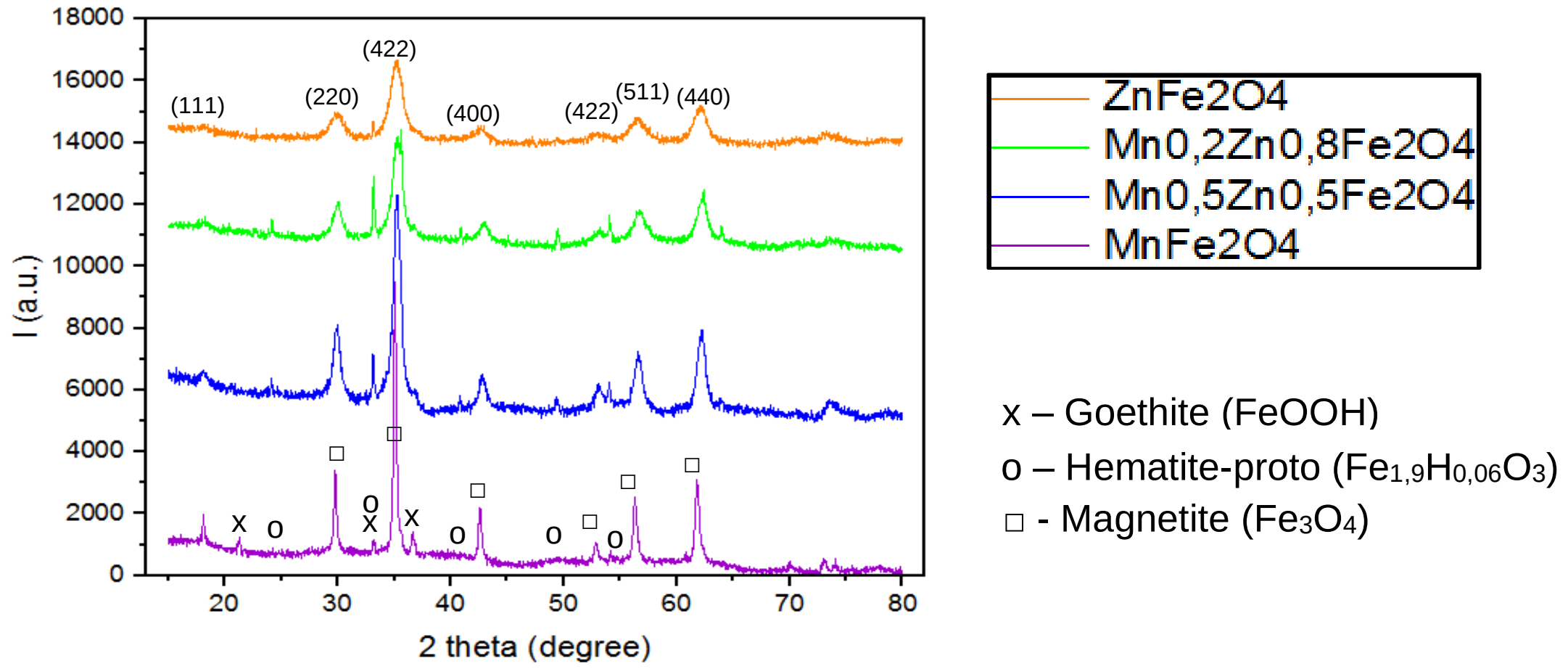

**Figure 5.11** – XRD pattern of $Mn_{1-x}Zn_xFe_2O_4$ with x = 0; 0.5; 0.8; 1.Spinel phase peaks are signed by the respective Miller indices, impurities peaks of Magnetite, Hematite-proto and Goethite are marked with open squares, circles and “x”s, respectively.

Rietveld refinement of the presented samples outcome that the single phase spinel structure fraction is above 88% for all samples, as plotted in figure 5.11.a. Impurities of Magnetite ($Fe_3O_4$), Hematite-proto ($Fe_{1.9}H_{0.06}O_3$), and Goethite (FeOOH) are present in every sample at low quantities (below 12%); $MnFe_2O_4$ sample presents mostly goethite, whereas all the other samples seem to have both Hematite-proto and Goethite. The previously mentioned Magnetite phase is indistinguishable of spinel crystal structure and for this reason it is not easily distinguishable using XRD.

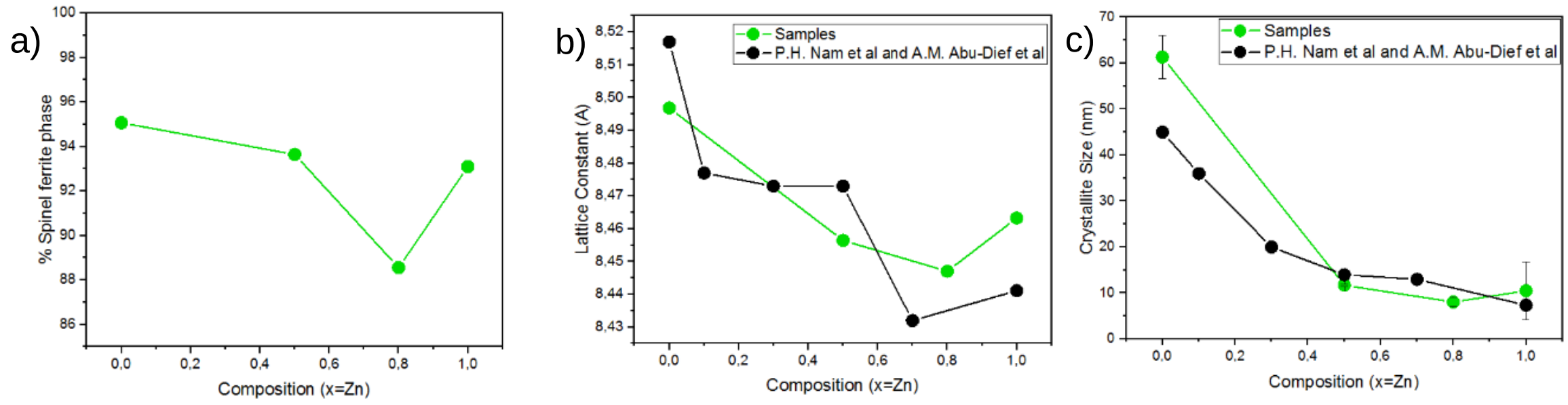

**Figure 5.12** – Crystallographic parameters a) Percentage of spinel ferrite phase. b) lattice constant. c) crystallite size.

The lattice constant of these ferrites, refined by Rietveld refinement method in Fullprof Suite software, reveals a shrinkage of the lattice constant from 8.5 to 8.46 Å with the increase of Zinc content in ferrite, as shown in figure 6.11.b. The cause of the lattice constant shrinkage is the smaller ionic radius of $Zn^{2+}$ (0.74 Å) in comparison with $Mn^{2+}$ (0.82 Å), [73], [74] .

The crystallite size varies from 61 to 11 nm with the increase of Zinc content in ferrite. Zinc decrease results are obtained by, [43], [68]. The explanation of this phenomenon was given by [75] and [68], which attributes the larger crystallite size for Mn ferrite as responsibility of Mn ions forming 2+ and 3+ cations, which means that they have a chance to be incorporated in octahedral and tetrahedral vacancies, while the Zinc ions only form $Zn^{2+}$ cations and the probability of being incorporated is smaller, thus, Zn ferrite has a smaller probability of creating particles larger than Mn ferrite, ,.

Based on the lattice parameter, the unit cell volume was calculated, $V = a^3$, since it’s a cubic cell. The ferrite density was also calculated based on the mass and site occupancy of the elements composing the ferrite: Zinc ($10.9\times10^{-25}$ kg), Manganese ($9.12\times10^{-26}$ kg), Iron ($9.27\times10^{-26}$ kg) and Oxygen ($1.33\times10^{-26}$ kg). The site occupancy of unit cell is 8 for Zn or Mn, 16 for Fe and 32 for O. The calculations showed that the ferrite density increases from 4300 $kg/m^3$ to 4580 $kg/m^3$ with the increase of Zn/Mn ratio, which is in agreement with the trend of the unit cell volume decrease with Zinc content, and the fact Zinc is heavier than Manganese. A similar trend was observed in the literature [76].

### 5.3.2 MORPHOLOGICAL AND CHEMICAL CHARACTERIZATION

Morphology and size distribution of $Mn_{1-x}Zn_xFe_2O_4$ nanoparticles were studied with resource to SEM. The characterized samples have x = 0; 0.5; 0.8 and 1. Image 6.12 present the obtained images for different samples.

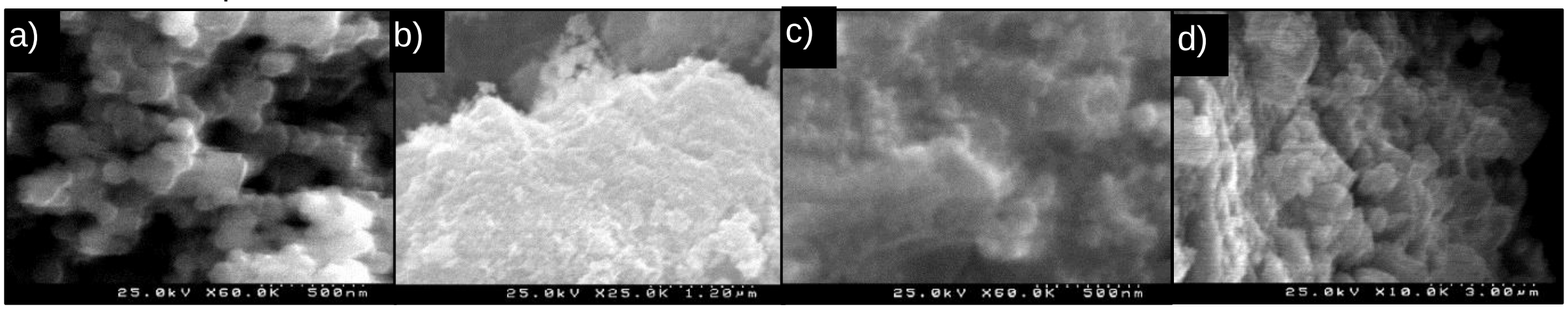

**Figure 5.13 –** SEM images of all compositions: (a) $MnFe_2O_4$; (b) $Mn_{0.5}Zn_{0.5}Fe_2O_4$; (c) $Mn_{0.2}Zn_{0.8}Fe_2O_4$; (d) $ZnFe_2O_4$. Poor definition in SEM images was obtained for intermediate compositions (x = 0.5 and 0.8). For Mn and Zn ferrites, the image shows enough definition for counting agglomerated nanoparticles.

SEM images of hydrothermally prepared samples present agglomerated nanoparticles, as the microscope amplification is not high enough to distinguish single nanoparticles. For the Mn to count and estimate the particles diameter (with resource to ImageJ software). The Mn ferrite has a mean agglomerate size between 100 and 125 nm, comparing with the crystallite size, around 60 nm, suggest a mean particle size composed of 2 crystallites. For other compositions, the definition of the images is not high enough to distinguish agglomerates.

EDS analysis was carried out as explained in the experimental procedure chapter. The ratio of each metal ion is compared with the expected stoichiometry of the ferrite. Figure 5.14 presents the results of EDS analysis for all metal ions with respect to the expected values.

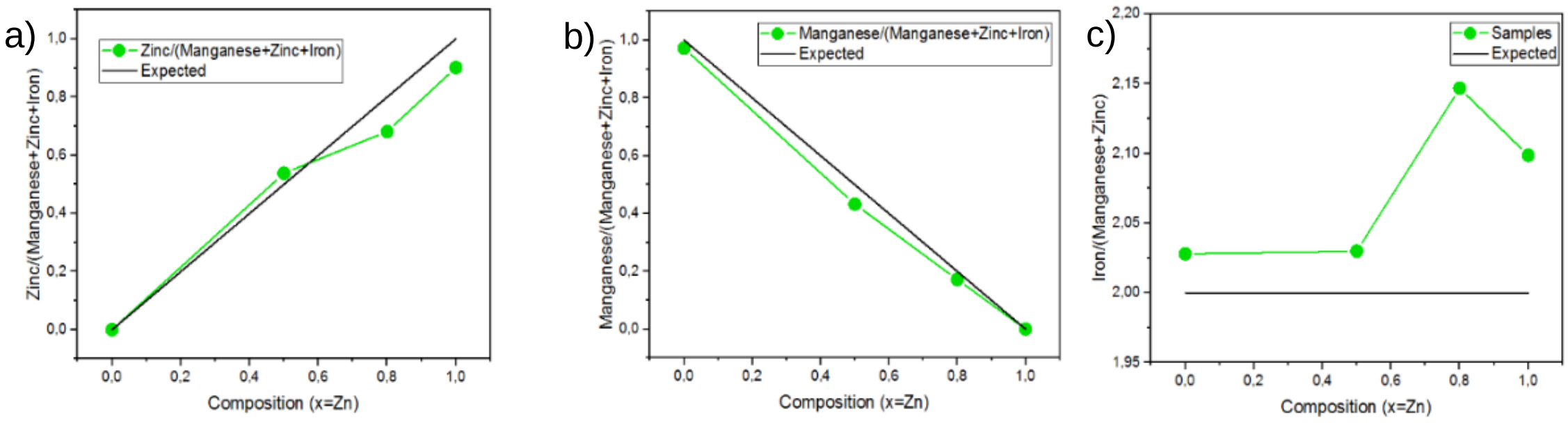

**Figure 5.14 –** Metal ions relative quantity times 3 (number of metal ions) and comparison with the expected values (Fe = 2; Mn = 1-x; Zn = x).

The EDS results show the deviations on the fraction of elements in the samples. In the Zinc element plot (6.14.a)), larger deviations are found in the samples with higher content of Zinc (up to 15% of deviation). Manganese plot is in good agreement with the expected quantity of Manganese. Iron metal ratio reveals an excess of Iron for all samples, in particular samples with higher Zn/Mn ratio present higher excess of Iron.

The samples with larger deviations from the expected are those with higher Zn content. Still, all the deviations are inferior to 15% of change in elements relative quantity.

### 5.3.3 MAGNETIC CHARACTERIZATION

The hydrothermally prepared samples were analyzed via SQUID magnetometer with M(H) loops measured at 5, 300 and 380 K, figure 5.15. The dependence of their magnetization with the increase of temperature in FC-ZFC curves is present in figure 5.17. Samples of $Mn_{1-x}Zn_xFe_2O_4$ (x = 0; 0.5; 0.8; 1) were analyzed.

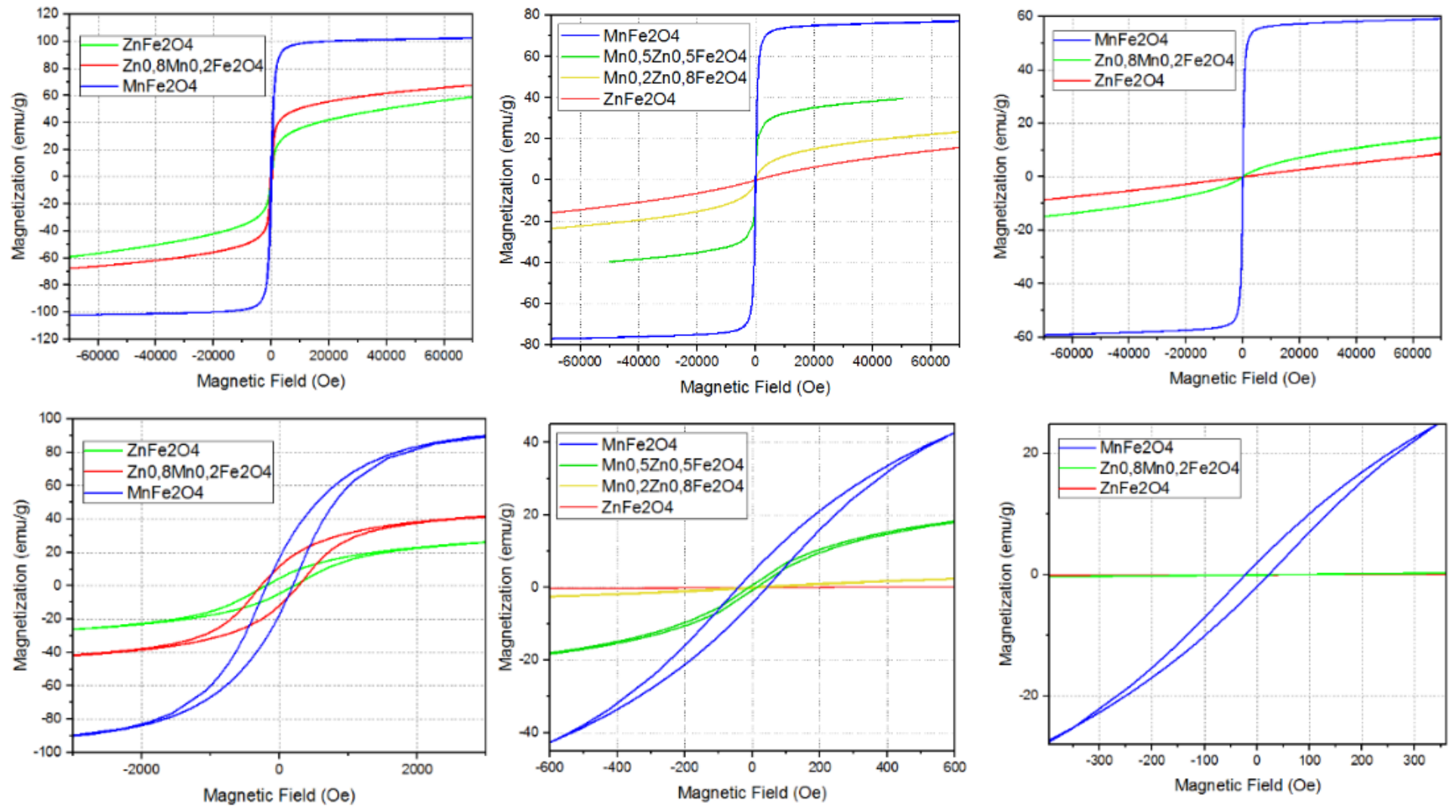

**Figure 5.15 –** M(H) curves. From left to right column 5, 300 and 380 K respectivelly. In the upper row a full scan is shown (from -70000 to 70000 Oe), in the bottom row an amplification of the M(H) curves is made for hysteresis loops analysis.

Different types of magnetic information can be extracted from the M(H) curves, as explained in “Chapter 4: Experimental Procedure”. Figure 6.16 show the evolution of the magnetic parameters $M_S$, $M_R$ and $H_C$, with the increase of Zn content ratio, at room temperature.

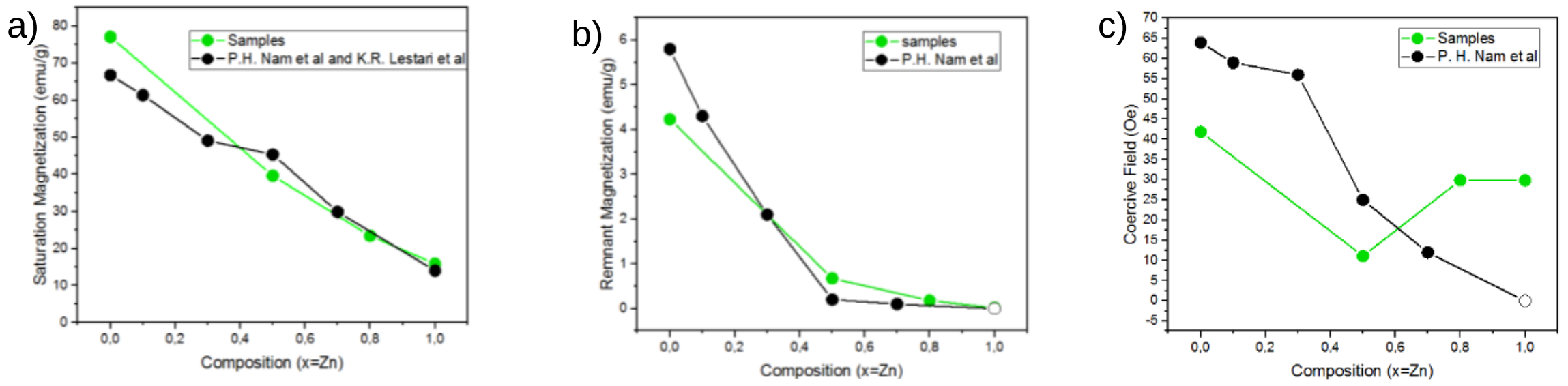

**Figure 5.16 –** Saturation magnetization (a), remnant magnetization (b) and coercive field (c) at 300K. Green lines are for samples, black lines are samples from literature. Black open circles are for values which were not explicit in literature.

Figure 5.16 shows a decrease in saturation magnetization, from 79 to 19 emu/g, and remnant magnetization, from 5 to 0 emu/g, with the increase of Zn/Mn ratio. The obtained values are in good agreement with P.H. Nam et al, [43], and were complemented by [77]. Coercive field does not present the bibliographic reported behavior, in which $H_C$ is almost zero for $ZnFe_2O_4$ sample, [78]. The obtained values for the coercive field might be due to the remnant currents in the SQUID superconducting coil, since the samples with x = 0.5 and 0.8 appear to be superparamagnetic (theoretically having no coercivity) and x= 1 is in the paramagnetic regime, having no hysteresis and no coercive field.

For a better understanding of the dependence of magnetization with temperature (M(T) curves) figure 5.17.a) shows the FC (field cooled) and ZFC (zero-field cool) measurements.

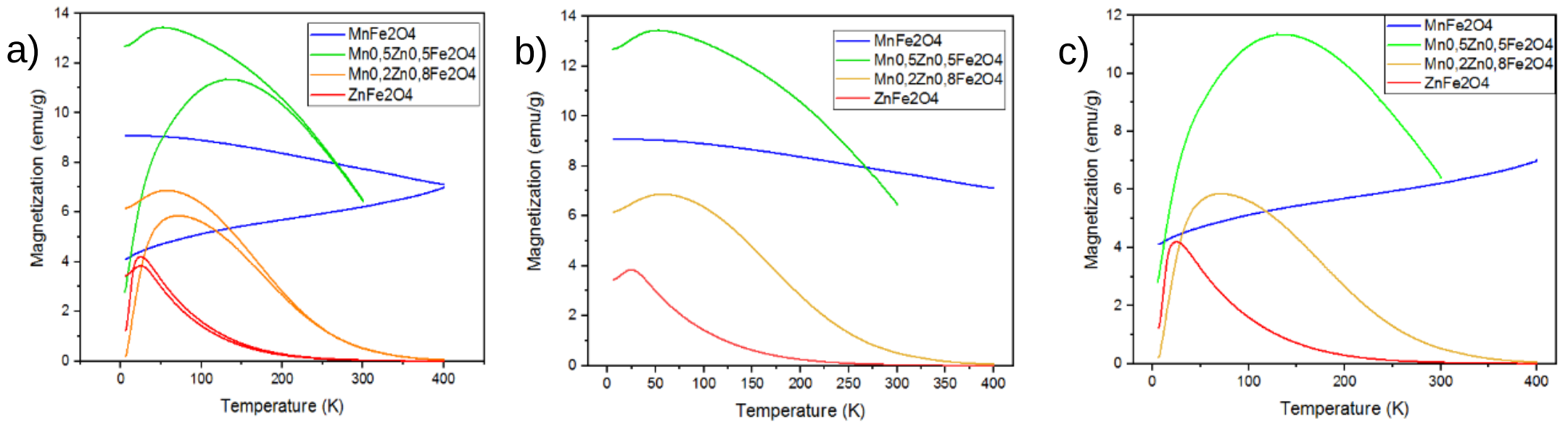

**Figure 5.17 –** M(T) measurements – FC and ZFC (a), FC (b) and ZFC (c). Range of temperature from 5 to 400K, except for $Mn_{0.5}Zn_{0.5}Fe_2O_4$ which goes from 5 to 300 K.

The Mn-Zn ferrite presents a strong dependence of the magnetization with temperature. In a general analysis, all FC magnetization decrease with the increase of temperature, this is related to the transition from ferromagnetic to paramagnetic phase. At higher temperatures more nanoparticles are in the paramagnetic state, thus reducing the sample magnetization. The magnetic ordering temperature dependence with ferrite composition is the basic property to develop a self-regulated heater with tuning capacity. The magnetic ordering temperature decreasing with the increase of Zn/Mn is observable in the FC measurements, the magnetization drops at lower temperatures when Zn content is higher in ferrite.

Analyzing the FC curves, it is noticeable that the intermediate composition (x=0,5) has a higher magnetization than the Mn ferrite, for temperatures below 275 K. Above this temperature, Mn ferrite has the highest magnetization. Also, by the FC curves it is possible to estimate the magnetic ordering temperature, relying on $\Theta_p$, as explained in “Chapter 4: Experimental Results”. A decrease of the magnetic ordering temperature with the increase of Zinc content, from ~610 to 250 K is noticeable in these curves. The $MnFe_2O_4$ Curie temperature value is from P. H. Nam et al. Going through the FC curve from higher to lower temperature, an increase in magnetization is clear as the temperature drops. This occurs as the paramagnetic particles (above the Curie temperature) undergo a paramagnetic-ferrimagnetic transition. The magnetization increases until it reaches a maximum and then drops, the drop of magnetization might be related to inter-particle interaction, [10]. The blocking of less anisotropic nanoparticles is dependent of the orientation of previously blocked nanoparticles.

The ZFC measurement starts by cooling the samples without an applied magnetic field. At high temperatures, the superparamagnetic nanoparticles change their magnetization rapidly as they possess more thermal energy than anisotropic energy, with relaxation time, $\tau$. While cooling without field, the nanoparticles are free to rearrange their magnetization in order to decrease the total magnetization of the system. When cooled below the blocking temperature, the nanoparticle magnetization is said blocked, which means that the thermal energy takes longer to change the nanoparticles’ magnetization and the system stabilizes with a low magnetization value. When the measurement starts, a magnetic field of 100 Oe is applied and the temperature is increased. The increasing of temperature will allow the smaller nanoparticles (with low block temperature) to unblock and thermally change their magnetization in the applied magnetic field direction. This is the cause for the magnetization increasing until a maximum.

An approximate value for magnetic ordering temperature is obtained by analyzing the paramagnetic phase of the FC curve with the Curie-Weiss law, as explained in “Chapter 4: Experimental Procedures”. Figure 5.18.a) is the normalization of the FC curves, for comparison improvement. $Mn_{0.5}Zn_{0.5}Fe_2O_4$ Curie temperature was estimated by extrapolation of the normalized FC curve for this sample.

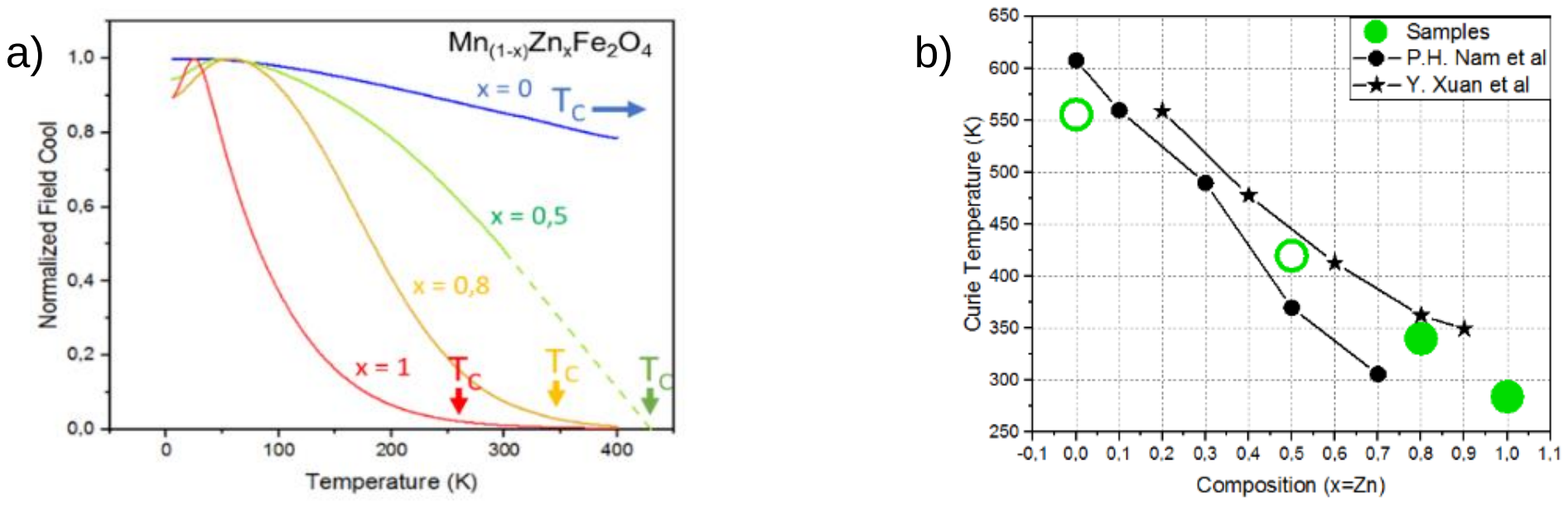

**Figure 5.18 –** a) Normalized FC curve, magnetic ordering temperatures represented with arrows. b) magnetic ordering temperatures obtained via Curie-Weiss law are represented in closed green circles. Open circles are the values indirectly obtained.

Figure 5.18.b) shows the decrease of the magnetic ordering temperature as the Zn/Mn ratio increases and compared with P.H Nam et al and Y. Xuan et al, [43] and [79]. These authors introduce a linear dependency of the Curie temperature with the composition, which was used to extrapolate the Curie temperature of $MnFe_2O_4$, 556 K. This value was not considered, for further analysis the Curie temperature of $MnFe_2O_4$ is considered as above 400 K.

Analysis of the blocking temperature, ferromagnetic M(H) curves at 300K and susceptibility are present in figure 5.19.

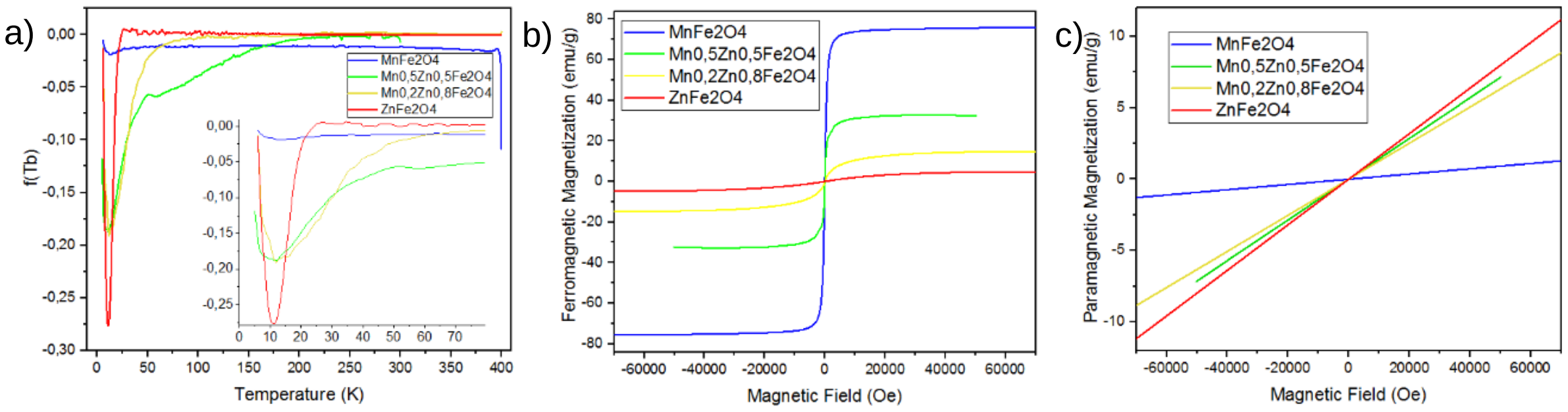

**Figure 5.19 –** a) Blocking temperature distributions. b) Ferromagnetic contribution of M(H) curves. c) Paramagnetic contribution of M(H) curves.

The blocking temperature distribution is present in figure 5.19.a). This figure has an inset for a clearer observation of the $T_b$ distribution near 0 Kelvin. By the analysis of this figure it is possible to see that the narrowest $T_b$ distribution happens for the Zn ferrite, with a mean blocking temperature of 11K. For Mn ferrite, the blocking temperature it is not in the 5-400K range. $Mn_{0.2}Zn_{0.8}Fe_2O_4$ and $Mn_{0.5}Zn_{0.5}Fe_2O_4$ have a wide $T_b$ distribution and the $\langle T_b \rangle$ is hard to precise, for this reason it was not assumed.

Figure 5.19.b) shows the hysteresis loops subtracted by the linear paramagnetic component. This figure represents the ferromagnetic component of the synthesized samples and will be used to determine the superparamagnetic behavior of the nanoparticles.

Figure 5.19.c) shows the susceptibility of the M(H) curves. It is visible that the Zn ferrite has the highest contribution, at 300 K, which is expected, since the Zn ferrite has the lowest magnetic ordering temperature. Mn ferrite has the lowest paramagnetic contribution, which is also expected due to the ferrimagnetic contribution of the Mn ions. For the intermediary compositions, $Mn_{0.5}Zn_{0.5}Fe_2O_4$ has a higher susceptibility than the $Mn_{0.2}Zn_{0.8}Fe_2O_4$.

To clarify if any sample is in the superparamagnetic regime, the M/Ms(H/T) was plotted for 300 and 380 K. Accordingly with [10], the M/Ms(H/T) plot at different temperatures and above the irreversibility temperature should superimpose into a universal Langevin curve. This review article also refers that in order to obtain a better Langevin fit, the susceptibility should be subtracted from the M(H) curves, which is equivalent to M(H) curves of figure 5.19.b). The cause for the susceptibility above the irreversibility temperature might be caused by the buildup of paramagnetic moments on the particles' surface. The obtained results are present in figure 5.20 along with the Langevin fit performed by Origin.

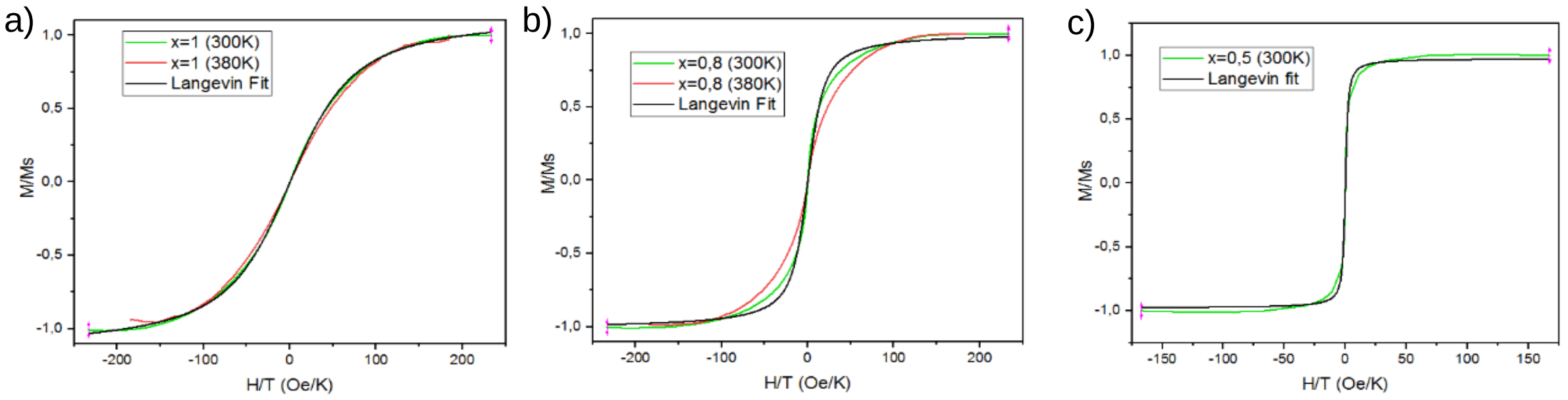

**Figure 5.20 –** $M/M_S(H/T)$, at 300 K and 380 K, and Langevin fit for a) $ZnFe_2O_4$ b) $Mn_{0.2}Zn_{0.8}Fe_2O_4$ and c) $Mn_{0.5}Zn_{0.5}Fe_2O_4$.

It was verified that the $ZnFe_2O_4$ sample, figure 5.20.a), superimpose into a universal Langevin curve at 300 and 380 K, thus revealed to be in the superparamagnetic state. The low initial susceptibility of these curves is because of the measurement temperature, almost all nanoparticles are above the magnetic ordering temperature. Both samples, $Mn_{0.2}Zn_{0.8}Fe_2O_4$ and $Mn_{0.5}Zn_{0.5}Fe_2O_4$, do not superimpose into a universal Langevin curve, thus, they are not in the superparamagnetic state.

### 5.3.4 *Particle Size Characterization*

Hydrothermally prepared samples of $Mn_{1-x}Zn_xFe_2O_4$ of composition x = 0; 0.8 and 1 were analyzed via TEM. Multiple amplifications were used for inter-composition analysis and higher amplifications were used to construct the nanoparticles size distribution. Figures 5.21 are a selection of the obtained images for the mentioned compositions.

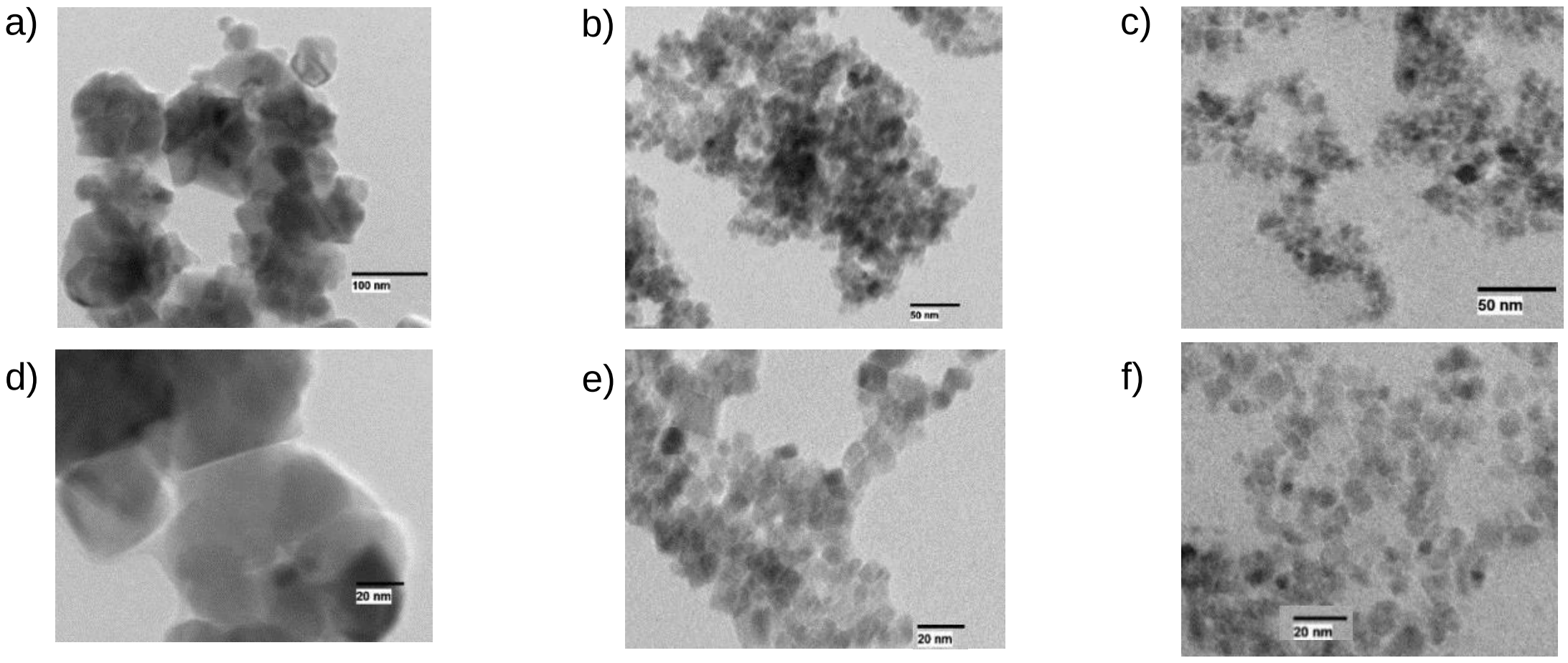

**Figure 5.21 –** TEM images of hydrothermally synthesized $Mn_{1-x}Zn_xFe_2O_4$. Upper row images have a 50 nm scale (except $MnFe_2O_4$) and bottom row a 20nm scale. a) and d) is $MnFe_2O_4$; b) and e) is $Mn_{0,2}Zn_{0,8}Fe_2O_4$; c) and f) is $ZnFe_2O_4$

A brief analysis to the TEM images of the synthesized samples reveal the tendency previously observed in crystallite size by XRD: a decrease in grain size with the increase of Zn/Mn ratio. All images present agglomerated nanoparticles, which is expected due to their magnetic behavior. ZnFe2O4 nanoparticles are more disperse than the other samples which might be a consequence of the higher paramagnetic-like behavior associated with this sample

TEM images of $MnFe_2O_4$ revealed that the nanoparticles form agglomerates in a flower arrangement. In the higher amplification image, it is distinguishable a cubic shape nanoparticle. The higher amplification image also reveals that the flower shape agglomerates are formed of spherical nanoparticles. Irregular particle shapes were also found. The images of higher amplification were used to determine the size distribution of the samples, figure 5.22.

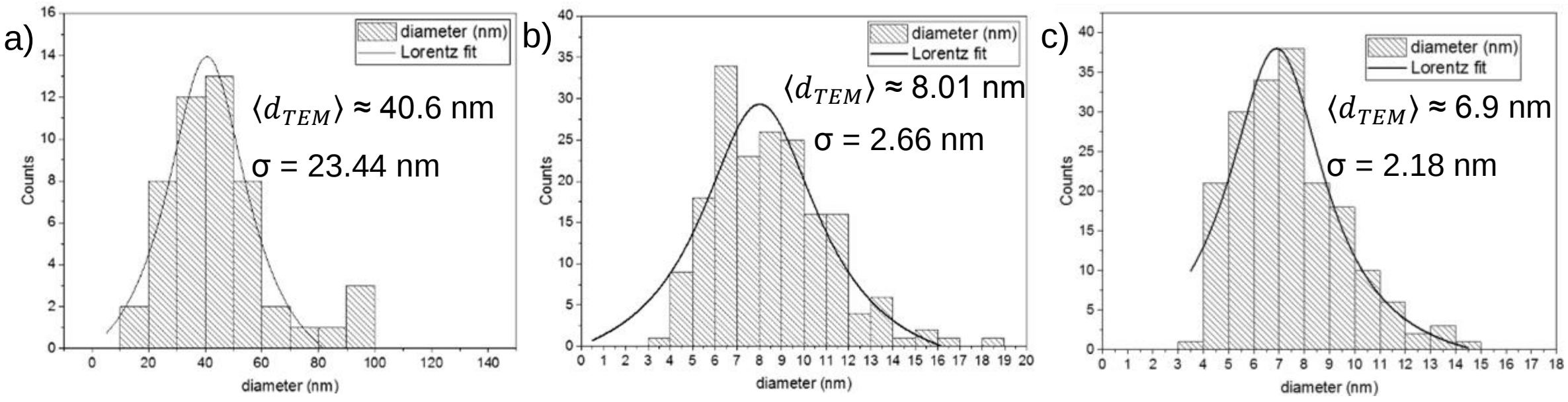

**Figure 5.22 –** Size distribution of $Mn_{1-x}Zn_xFe_2O_4$ x = 0; 0.8; 1 in TEM images. Mean grain size and standard deviations for each sample. a) $MnFe_2O_4$; b) $Mn_{0.2}Zn_{0.8}Fe_2O_4$; c) $ZnFe_2O_4$.

Particles counting revealed a decrease in mean grain size with increase of Zn/Mn ratio, from 40.6 nm to 6.9 nm, $MnFe_2O_4$ to $ZnFe_2O_4$. The standard deviation is also reduced with the shrinking of the distribution.

### 5.3.5 *Heat Generation*

The hydrothermal samples analyzed in the magnetic induction heating setup were measured without the use of a solvent, i.e. the dry ferrite powders were measured. For this reason, the following measurements are introductory to further studies, as they are incomparable with literature, where typically, the induction heating performance is measured for NPs immersed in a liquid solvent.

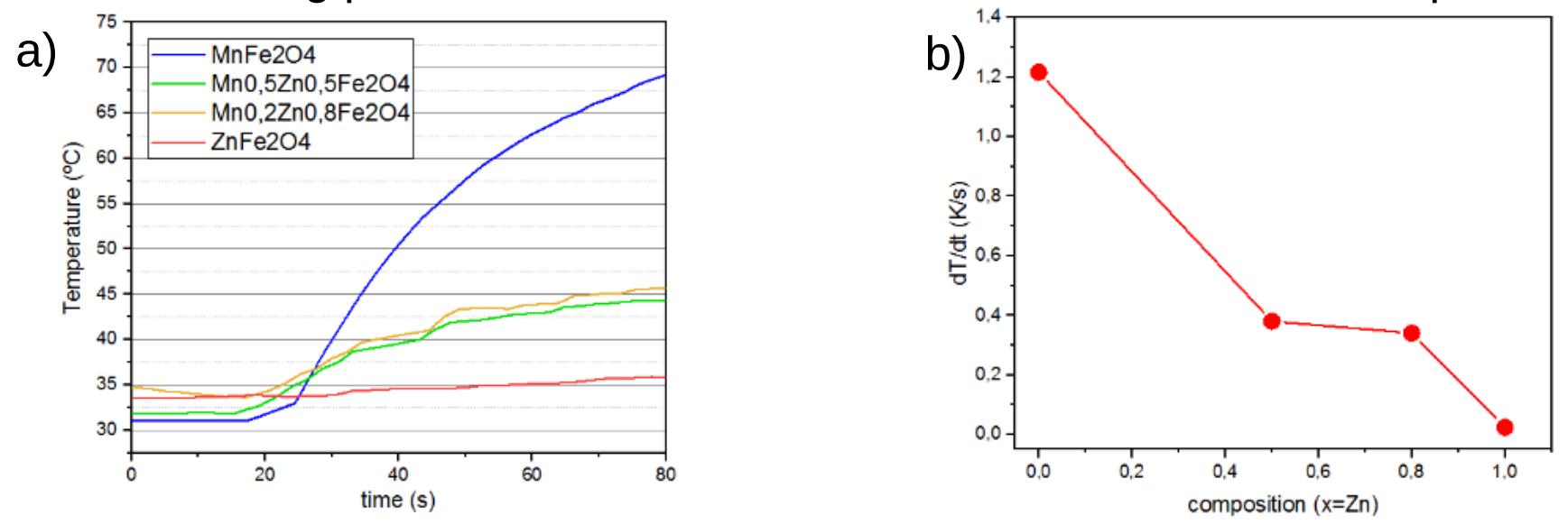

**Figure 5.23 –** Magnetic induction heating results with an AC magnetic field of 25mT at a frequency of 364KHz. a) Temperature as a function of time for all compositions; b) Time derivative of temperature composition dependence.

Figure 5.23.a shows the magnetic induction heating results and figure 5.23.b the dT/dt extracted from the initial slope of the magnetic induction heating curves. From these images it is visible that the heating rate decreases with the increase of Zn content in ferrite.

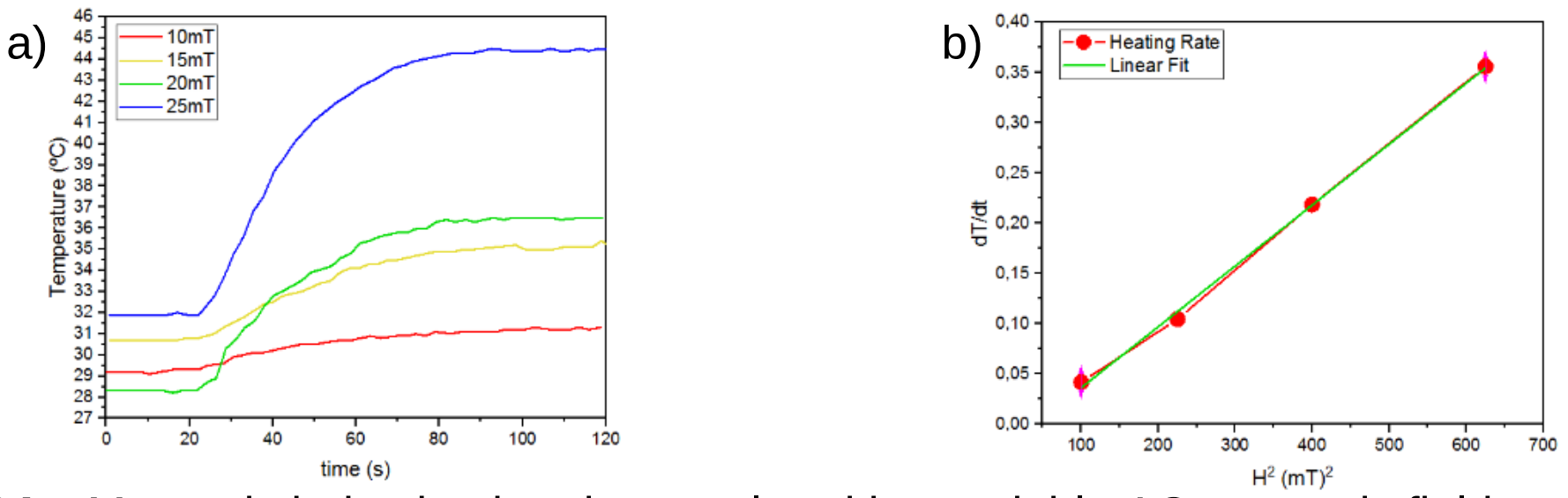

**Figure 5.24 –** Magnetic induction heating results with a variable AC magnetic field amplitude at a frequency of 364KHz. a) Temperature as a function of time for $Mn_{0.5}Zn_{0.5}Fe_2O_4$; b) Heating rate quadratic dependence of the magnetic field amplitude.

Figure 5.24.a) shows the temperature increasing of $Mn_{0.5}Zn_{0.5}Fe2O4$ with different applied magnetic field. Figure 5.24.b) proves the linear dependency of the heating rate with the squared applied field, as predicted by R.E. Rosensweig, [28], equation 2.9.The linearization has a slope of $6.06x10^{-4}$, an interception of -0.02 and $R^2$ of 0.998.

## 5.4 Discussion of Hydrothermal Samples

Let us start by looking at the samples from a structural point of view. All synthesized samples are nano-powders of $Mn_{1-x}Z_xFe_2O_4$, spinel ferrite almost single phase, with impurities percentage below 12%. The most common impurity found by XRD is hematite-proto, which is present in every composition except $MnFe_2O_4$ which has a secondary phase of goethite. The presence of impurities is also suggested by EDS: the excess of iron found by EDS is closely related with the XRD impurities (mainly iron oxides) - the comparison can be found in figure 5.25.

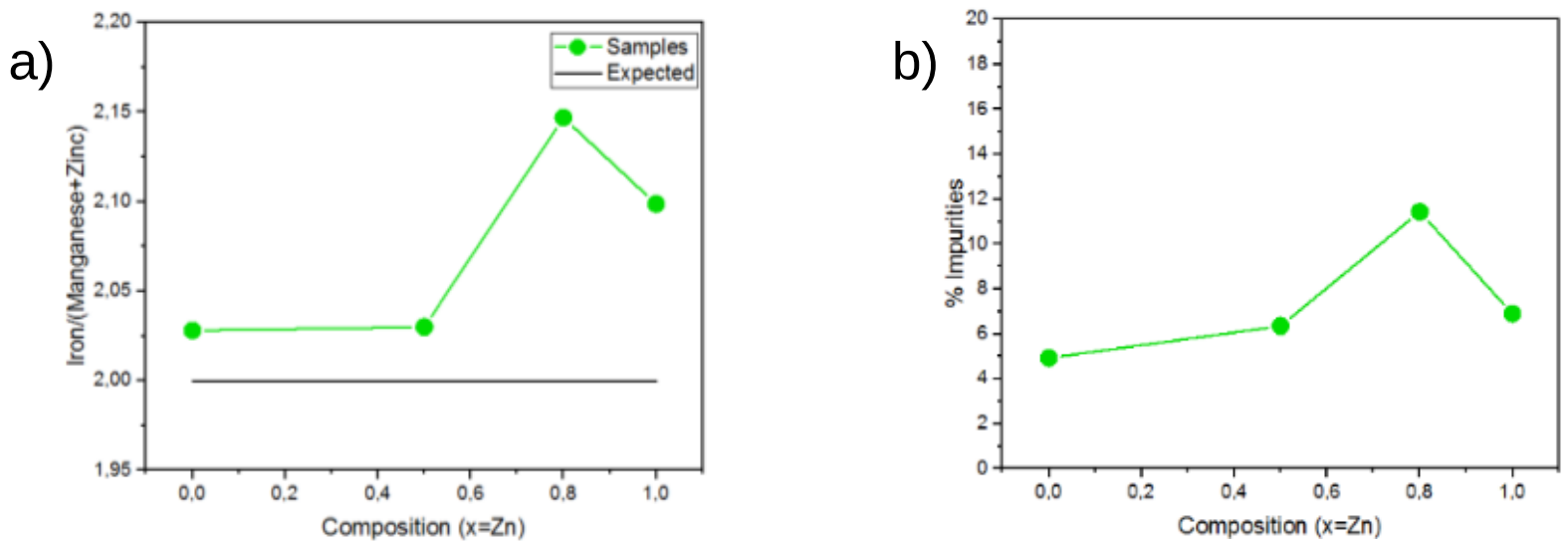

**Figure 5.25 –** (a) Iron metal ions ratio (EDS) and (b) % of impurity phase (XRD).

Impurity phases of are found in low percentage (below 12%) in the samples. Due to the residual phases of goethite and hematite and the fact that both impurities have an antiferromagnetic nature, their interference in the magnetic analysis is not considerable. Additionally, Hematite has the Néel transition below 955 K and remains antiferromagnetic at 6 K, [80]. Goethite is also antiferromagnetic with a Néel temperature around 400 K and remains antiferromagnetic at temperatures lower than 5 K, [81]. Magnetite was not identified in magnetic analysis, since its saturation magnetization increases as crystallite size increases, [82]. Similarly with Mn ferrite in [83]. Thus, it is indistinguishable from Mn ferrite and it is not present in Zn ferrite, or it would increase its $M_S$ value.

The magnetic properties of the nanoparticle are mainly dependent of the composition, however, the nanoparticles' size also as a significant role. The particle size is roughly calculated via Williamson-Hall equation, when assuming every particle as a single crystal. Though this is not always true, it is a good approximation due to the nanoparticles scale. The most reliable size measurement of the nanoparticles was performed via TEM, which revealed a constant size distribution decreasing with the increase of Zn/Mn ratio and reasonable standard deviations. Also, it revealed a particle size smaller than crystallite size, which is contradictory at first, but if the errors associated with the measurement techniques are considered, the values are not disagreeing. A crystallite size of the same order of the particle size (or superior) suggests that most nanoparticles are single crystal.

For structural and magnetic discussion, the crystallite size was used, due to the same compositions being analyzed via SQUID and XRD. The crystallite/grain size has a great influence in the magnetic properties of these ferrites, as discussed in section 2.3.1. Crystallite size, saturation magnetization, remnant magnetization and coercive field are displayed in figure 5.2.6

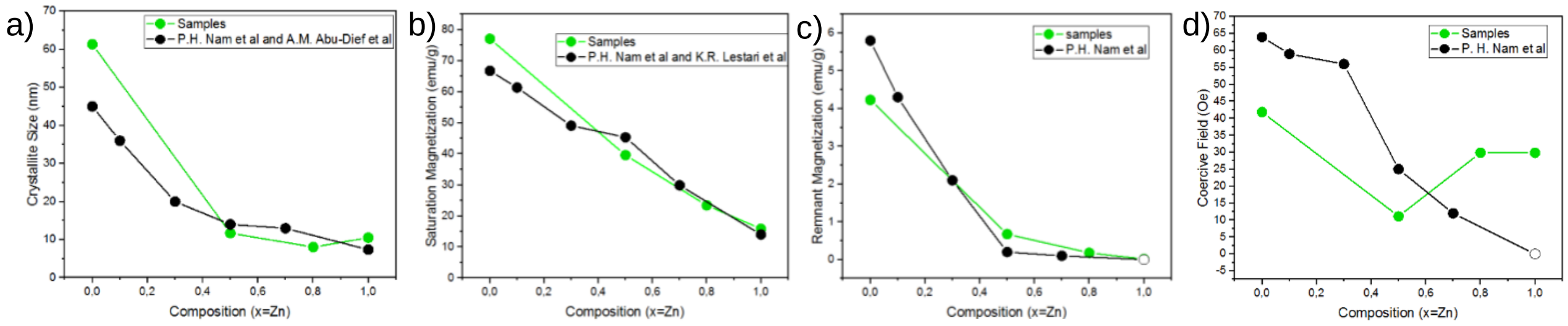

**Figure 5.26 –** Crystallite size influence on the saturation magnetization, remnant magnetization and coercive field.

As previously mentioned, the decrease of the crystallite size with the increase of Zn/Mn ratio is due to the lower probability of incorporation for $Zn^{2+}$ than to $Mn^{2+/3+}$. Zinc ions do not contribute for the ferrite magnetization as they do not have unpaired electrons, furthermore, they contribute for an antiparallel alignment of Iron magnetic moment, thus, causing all magnetic parameters to decrease with the Zn increase.

The main cause for saturation magnetization to decrease is the incorporation of Zinc ions. However, crystallite size also has a relevant role: larger crystallite sizes, means smaller surface/volume ratio and consequently less spin canting contribution, which leads to a higher value of $M_S$. Z. X. Tang et al showed that the increase of crystallite size favors an increase of saturation magnetization for $MnFe_2O_4$, in [83]. This correlates with larger crystallite size found in synthesized $MnFe_2O_4$ sample with the value of saturation magnetization being superior to the ones found by P. H. Nam et al.

The remnant magnetization decreasing is mainly dependent from the $Zn^{2+}/Mn^{2+}$ ratio increasing. The $M_R$ decreasing is also affected by the crystallite size decreasing, as shown by Y. Su et al, [84], and M. H. Mahmoud, [85]. The obtained values for the synthesized samples closely agree with P. H. Nam et al samples, however, the small values of remnant magnetization can easily be affected by the remnant field present in the superconducting coils.

The coercive field is composition and grain size dependent. The composition dependence of coercive field can be observed in P. H. Nam et al samples. The relationship between grain size and coercivity was studied by M. Vopsaroiu, [86], for FeCo films, the larger the grain size, the higher the coercivity. In Mn-Zn ferrite, the increase of Zinc decreases the crystallite size, thus, a general decrease in coercive field is expected. In order to understand the reason why this is not observed we should look at the samples from an individual perspective. $MnFe_2O_4$ has the largest crystallite and the higher coercivity, which agrees with the previous citations. $Mn_{0.2}Zn_{0.8}Fe_2O_4$ and $ZnFe_2O_4$ present an ambiguous comparison: the higher Zn composition the smaller the expected coercivity, however the crystallite size is larger for $ZnFe_2O_4$ having a coercive field similar with $Mn_{0.2}Zn_{0.8}Fe_2O_4$. The intermediate composition, $Mn_{0.5}Zn_{0.5}Fe_2O_4$, presents the higher deviations from the other synthesized samples but a smaller deviation from the reference values. The reason for this might be the measurement in a different SQUID equipment, also, the M(H) was measured at a maximum applied field of 5T (in comparison with 7T of all other samples). A possible cause for the coercive field of $Mn_{0.5}Zn_{0.5}Fe_2O_4$ to have a smaller value might be caused by a lower value of remnant field in the SQUID superconducting coils

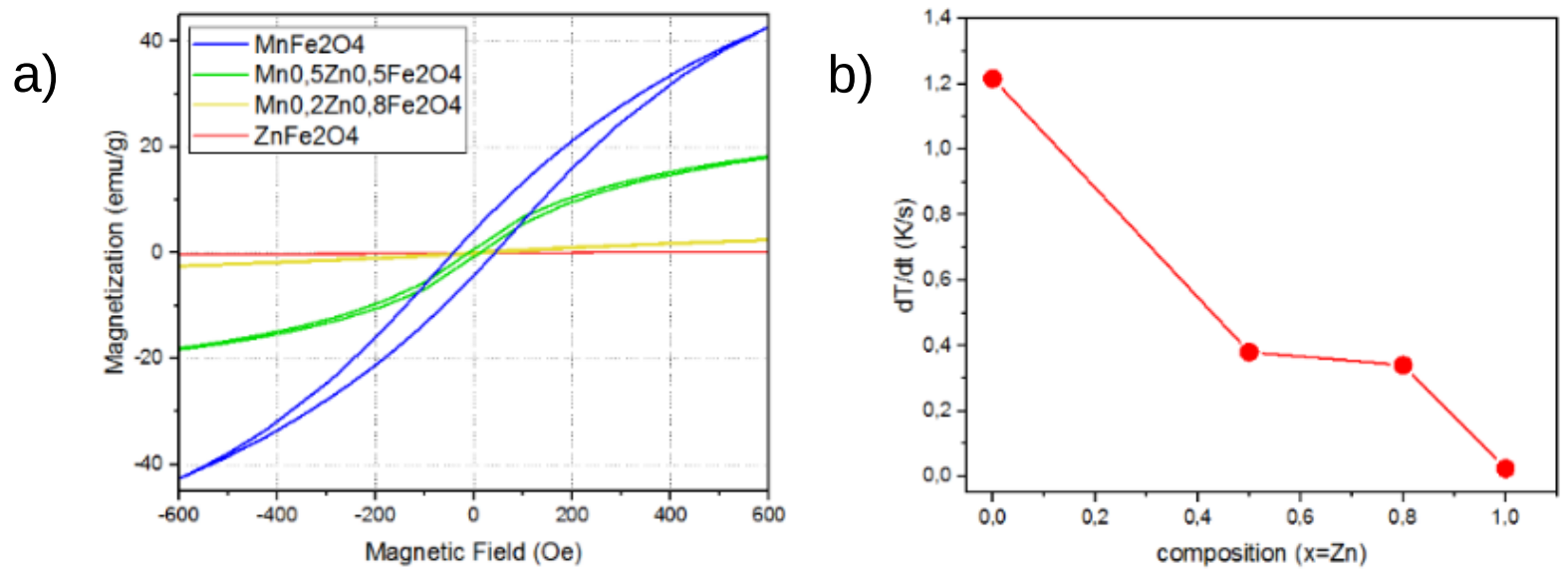

**Figure 5.27 –** Comparison of a) hysteretic loops area with b) heating rate.

The magnetic induction heating experiment shows an initial heating rate decreasing with the increase of Zn/Mn ratio. The heating rate main contributions are the hysteretic loops loss and Néel relaxation. The experiment was made without a liquid medium, which means that the Brownian motion do not contribute for the nanoparticles heating. The higher heating rate belongs to $MnFe_2O_4$, which has the larger hysteresis area. $Mn_{0.5}Zn_{0.5}Fe_2O_4$ and $Mn_{0.2}Zn_{0.8}Fe_2O_4$ have close values of heating rate, the hysteresis of both samples is considerably small, thus the observable heating can occur via hysteric losses and Néel relaxation. The heating rate for Zn ferrite is close to zero as this ferrite is completely paramagnetic at room temperature.

Crystallite size, blocking temperature and anisotropic constant are also related, accordingly with equation 5.1, [10].

$$T_B = \frac{K_{Eff}V}{\ln\left(\frac{\tau_M}{\tau_0}\right)Kb} \qquad (5.1)$$

This time interval of recorded data is 11 s, $\tau_M$, where $\ln\left(\frac{\tau_M}{\tau_0}\right)$=23.12. Where $K_{Eff}$ is anisotropic constant for Mn ferrite as found in [53], 1.24 kJ/m$^3$, and $K_{Eff}$ for Zn ferrite, 46 kJ/m$^3$ in [87], the intermediary composition's $K_{Eff}$ was obtained by linear interpolation between these two values. The anisotropic constant is highly dependent on the size, shape, temperature and number of magnetic moments of a nanoparticle. In the mentioned articles both ferrites have a similar size. From the above equation several correlations were tried: for the volume calculus, the mean blocking temperature, $\langle T_B \rangle$, was used. For the $T_B$ calculus the crystallite volume was used. The $K_{Eff}$ was calculated with resource to both $D_{XRD}$ and $\langle T_B \rangle$. For the conversion of crystallite size ($D_{XRD}$) to volume and from volume to calculated size ($D_{CALC}$) were performed assuming spherical nanoparticles. The results are presented in Table 5.1.

| Composition (x=Zn) | $D_{XRD}$ (nm) | $\langle T_B \rangle_{FC/ZFC}$ (K) | $K_{Eff_BIBILIO}$ (J/m$^3$) | | $D_{CALC}$ (nm) | $\langle T_B \rangle_{CALC}$ (K) | $K_{Eff_CALC}$ (J/m$^3$) |
|---|---|---|---|---|---|---|---|
| 1 | 10.50 | 15.21 | 4600 | | 12.6 | 8.7 | 7997 |
| 0.8 | 8.05 | 11.85 | 3932* | | 12.2 | 3.4 | 13817 |
| 0.5 | 11.70 | 12.21 | 2916* | | 13.7 | 7.7 | 4640 |
| 0 | 61.31 | >400 | 1240 | | >58.1 | 469.0 | >1057 |

**Table 5.2 -** Estimation of single domain size, blocking temperature and anisotropic constant from the acquired magnetic and structural data. Asterisks means that the value was obtain from linearization.

The calculus of the single domain size, $D_{CALC}$, with resource to the mean blocking temperature presents agreeing results, corroborating the XRD estimation and following the same tendency with Zn content. The estimation of the mean blocking temperature with resource to the crystallite size results: the largest deviation is for $Mn_{0.2}Zn_{0.8}Fe_2O_4$ (280%), the lowest deviation is for $Mn_{0.5}Zn_{0.5}Fe_2O_4$ (72%). It was possible to estimate the blocking temperature of $MnFe_2O_4$ (433K) which is above the FC-ZFC measurement range, however, the single domain size was not performed due to the inexistence of a mean blocking temperature in FC-ZFC curves. In resume, the overall results are in rough agreement with the measured values.

$T_{Irr}$ was used for calculating the largest single domain size. Also, using the $D_{XRD}$ value, it was possible to estimate the anisotropic constant of the highest temperature blocked nanoparticles, table 5.2.

| Composition (x=Zn) | $D_{XRD}$ (nm) | $T_{Irr}$ (K) | $K_{Eff_BIBLIO}$ (J/m$^3$) | | $D_{max_CALC}$ (nm) | $K_{Eff_MAX_CALC}$ (J/m$^3$) |
|---|---|---|---|---|---|---|
| 1 | 10.50 | 23.8 | 4600 | | 14.7 | 12513 |
| 0.8 | 8.05 | 76 | 3932* | | 22.8 | 88618 |
| 0.5 | 11.70 | 196 | 2916* | | 34.5 | 74833 |
| 0 | 61.31 | >400 | 1240 | | >58.0 | >1057 |

**Table 5.3** – Estimation of single domain size and anisotropic constant for the blocked nanoparticle at higher temperature.

The observed increase tendency of the calculated domain size with the increase of Mn content in ferrite is on agreement with the crystallite size tendency estimated by Williamson-Hall analysis and TEM analysis.

The anisotropic constant estimated with resource to the crystallite size and the irreversibility temperature does not follow the linear tendency assumed. The calculus of $K_{Eff}$ depends mostly on the nanoparticles' composition, yet, the size and shape and of the nanoparticles, also plays a role, as explained in section 2.3.1. The hydrothermally synthesized nanoparticles have a relatively narrow size distribution and its shape varies between spherical and cubic.

### *5.4.1 Compositional Analysis*

$MnFe_2O_4$ is ferrimagnetic, the $Mn^{2+/3+}$ magnetic moment aligns with the iron lattice inside the ferrite, creating a ferrimagnetic single domain. Mn ions have a canted spin due to the magnetic moment being weaker than iron magnetic moment, this canting categorizes the Mn ferrite as a ferrimagnet. Mn ferrite has the highest saturation and remnant magnetization, coercivity, Curie and Block temperatures and the highest heating rate of the Mn-Zn ferrite family. Structurally, it has the largest crystallite size and lattice constant, with smaller density.

Ferrimagnetism and paramagnetism at room temperature, occur simultaneously for $Mn_{0.5}Zn_{0.5}Fe_2O_4$ and $Mn_{0.2}Zn_{0.8}Fe_2O_4$. The ferrimagnetic behavior of these samples comes from the Iron lattice and the Mn sublattice. The Zinc does not contribute with an individual magnetic moment, however, it influences Fe spins by shrinking the lattice. For both ferrites, figure 5.20 shows that the M/Msat(H/T) curves do not collapse into a universal Langevin curve, which means that these ferrites are not in the superparamagnetic state at room-temperature.

Superparamagnetism and paramagnetism occurs for $ZnFe_2O_4$ at room temperature. The Zn ions in the spinel structure do not contribute with magnetic moment for the ferrite, instead they contribute for the shrinkage of the crystallite. In this ferrite only the antiparallel Iron atoms contribute with magnetic moment, thus, $ZnFe_2O_4$ is antiferromagnetic below the Néel temperature. From the 5K M(H) curve, it was observed that the magnetization does not saturate even at 70000 Oe, this might indicate spin canting at the surface of the nanoparticles. At room temperature, and at higher temperatures, these nanoparticles are above their magnetic ordering temperature, which makes most of them paramagnetic. The nanoparticles that still contribute with magnetic moment are in the superparamagnetic state.

# Chapter 6: CONCLUSIONS

During this work nanoparticles of Mn-Zn ferrite were synthesized by two different methods: sol-gel auto-combustion method and hydrothermal method. Despite both methods are suitable for the nanoparticle synthesis, the hydrothermally prepared samples present better crystallinity and magnetic properties than the sol-gel auto-combustion samples.

The hydrothermally synthesized samples revealed dependence of all structural and magnetic properties with the Zn/Mn ratio. The XRD revealed the expected spinel crystal structure with high single-phase percentage (>88%). Lattice constant, determined by Rietveld refinement, decreases from 8.50 to 8.46 Å from Mn ferrite to Zn ferrite. The Williamson-Hall analysis presents the crystallite size decreasing from 61 to 11 nm, with the increase of Zn in ferrite. SEM images present agglomerated nanoparticles. TEM images showed mean particle size varying, from 41 to 7 nm, with the Zn/Mn ratio increase. Much narrower particle size distributions were obtained in comparison with the samples of sol-gel auto-combustion method. SQUID results showed that the increase of Zn content in ferrite decreases saturation magnetization (79 to 19 emu/g) and remnant magnetization (5 to approximately 0 emu/g). The M(T) curves, FC-ZFC, revealed that the mean blocking temperature and the irreversibility temperature also decrease with the increase of Zn/Mn ratio. From the magnetic induction heating experiment, higher heating rates were obtained for higher Mn content. Table 6.1 resumes the structural and magnetic properties of the hydrothermally synthesized Mn-Zn ferrite measured at 300K.

| | Composition (x=Zn) | 1 | 0.8 | 0.5 | 0 |
|---|---|---|---|---|---|
| XRD | Pure Phase (%) | 93.1 | 88.56 | 93.65 | 95.07 |
| | a (Å) | 8.46 | 8.44 | 8.46 | 8.50 |
| | $D_{XRD}$ (nm) | 10.50 | 8.05 | 11.70 | 61.31 |
| | Density (kg/m$^3$) | *4581 | *4562 | *4478 | *4301 |
| TEM | $\langle d_{TEM} \rangle$ (nm) | 6.9 | 8.01 | - | 40.6 |
| SQUID | $M_S$ (emu/g) | 15.87 | 23.42 | 39.6 | 77.07 |
| | Hc (Oe) | 29.84 | 29.88 | 11.13 | 41.85 |
| | $M_R$ (emu/g) | 0.01 | 0.02 | 0.67 | 4.23 |
| | $\Theta_P$ (K) | 284 | 340 | 420 | *556 |
| | $\langle T_B \rangle_{FC/ZFC}$ (K) | 15.2 | 11.9 | 12.2 | - |
| | $T_{Irr}$ (K) | 23.8 | 76 | 196.9 | >400 |
| Magnetic Induction Heating | $\frac{\partial T}{\partial t}$ (K/s) | 18 | 255 | 285 | 912 |

**Table 6.1 –** Properties of hydrothermally prepared $Mn_{1-x}Zn_xFe_2O_4$ measured at 300 K**.** Asterisks marks value that were not measured directly.

The hydrothermally synthesized nanoparticles of $Mn_{1-x}Zn_xFe_2O_4$, revealed a decreasing of the magnetic ordering temperature, from ~556 K (estimated) to ~284 K, with x value increasing. Moreover, the decrease of the heating rate might also be correlated with the magnetic ordering temperature, the increase of Zn/Mn ratio leads to a lower $\Theta_P$ and a lower heating rate. Although it was not proved in a magnetic induction heating experiment, in principle, the magnetic ordering temperature of the synthesized nanoparticles can be used as a self-regulated mechanism of heating.

## 6.1 Future Work

Future work guidelines are the optimization of the synthesized ferrites by hydrothermal method. For instance, increasing the single-phase percentage by finding the optimum time and temperature for the autoclave procedure. Organic solvents can be added during the synthesis in order to increase the nanoparticles dispersion, [37]. The initial reagents affect the nanoparticles size and shape, [38], thus, trying different compounds might be interesting to study the size and shape effect on the nanoparticles' magnetic properties.

Although the tuning of the magnetic ordering temperature was shown, the self-regulated induction heating mechanism was not proven during the induction heating experiment, probably due to the non-adiabaticity of the system. A system with higher thermal insulation could be used for this purpose. In order to obtain a more precise measurement of the magnetic ordering temperature, an experiment of AC magnetic susceptibility as a function of temperature could also be performed, [88]. Another idea is to use a polymeric matrix to fix the nanoparticles during the magnetic induction heating measurement, preventing them to rotate. This would mitigate the Brownian relaxation and only the Néel relaxation contribution would be measured, [89].

An interesting future application for the self-regulated heating nanoparticles is a self-pumping magnetic cooling device, as the one presented by V. Chaudhary et al, [3]. For this application, the nanoparticles should be in the ferrofluid form. The fluid is in a close circuit between a heat load and a heat sink. The magnetic fluid is attracted, by a magnet, to a heat load, where it will heat above its Curie temperature. Once above the Curie temperature, the fluid cease to be attracted to the magnet and cold fluid takes its place. Thus, creating a self-pumping fluid, or a negative viscosity fluid. The hot ferrofluid is then driven to the heat sink, where it will cool down and regain its magnetic properties. This system not only allow the heat transport between heat load and heat sink but can also be used for generating energy. The self-propelled magnetic fluid can be used to induce an electromotive force in a coil, generating a potential difference, or even, by adding turbines in the fluid path it will force the turbines to rotate. It works as transducer from mechanical to electric energy.

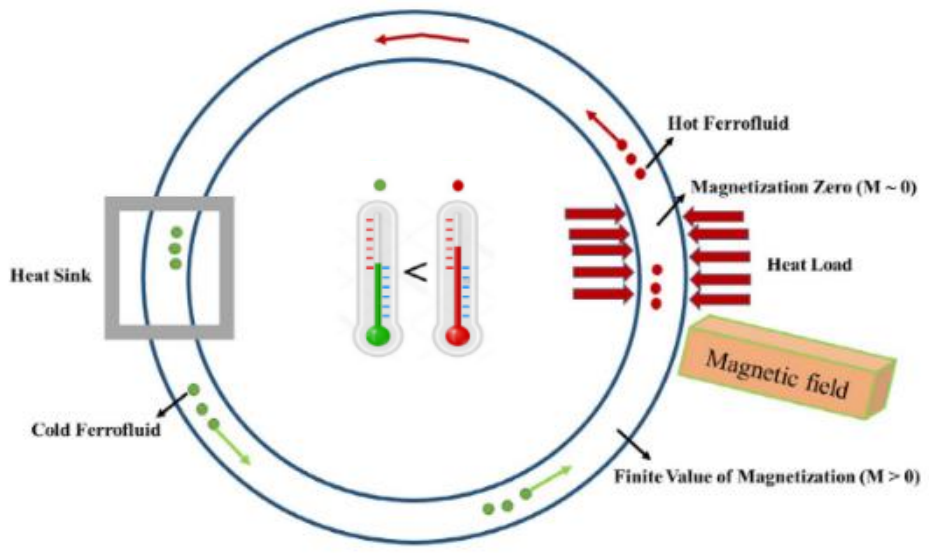

**Figure 6.1 –** Schematic of a self-pumping magnetic cooling device by V. Chaudhary et al.

Another future work guideline is the application of the synthesized nanoparticles in other structural forms, such as, functionalized nanoparticles, ferrofluids, or bulk materials. Playing with the nanoparticles' composition, and consequently with the self-regulated heating mechanism, this nano ceramic can find its use in a vast number of applications.

## REFERENCES

[1] V. Sechovský, *Encycl. Mater. Sci. Technol.*, pp. 5018–5032, 2001.

[2] S. V. Jadhav *et al.*, *J. Alloys Compd.*, vol. 745, pp. 282–291, 2018.

[3] V. Chaudhary, Z. Wang, A. Ray, I. Sridhar, and R. V. Ramanujan, *J. Phys. D. Appl. Phys.*, vol. 50, no. 3, 2017.

[4] T. Bayerl, M. Duhovic, P. Mitschang, and D. Bhattacharyya, *Compos. Part A Appl. Sci. Manuf.*, vol. 57, no. 2014, pp. 27–40, 2014.

[5] T. Waeckerlé, H. Fraissé, B. Boulogne, and S. L. Spire, *J. Magn. Magn. Mater.*, vol. 304, no. 2, pp. 844–846, 2006.

[6] L. B. de Mello, L. C. Varanda, F. A. Sigoli, and I. O. Mazali, *J. Alloys Compd.*, vol. 779, pp. 698–705, 2019.

[7] B. Odom, D. Hanneke, B. D'urso, and G. Gabrielse, *Phys. Rev. Lett.*, vol. 97, no. 3, pp. 6–9, 2006.

[8] Charles Kittel, *Introduction to Solid State Physics*, no. 8. 2005.

[9] H. P. Myers, *Introductory Solid State Physics*, Second Edi., no. 2. Taylo & Francis, 1997.

[10] and J. C. D. M. Knobel, W. C. Nunes, L. M. Socolovsky, E. De Biasi, J.M.Vargas, *J. Nanosci. Nanotechnol.*, vol. 8, no. 4, pp. 1880–1885, 2008.

[11] K. S. University, *Vis. Quantum Mech. next Gener.*, pp. 1–5, 2001.

[12] X. Batlle and A. Labarta, *J. Phys. D. Appl. Phys.*, vol. 35, no. 6, pp. R15–R42, 2002.

[13] D. B. Fischbach, "*Phys. Rev.*, vol. 123, no. 5, pp. 1613–1614, 1961.

[14] K. B. Tamayo, *Magnetic properties of solids*. 2009.

[15] C. Won *et al.*, *Phys. Rev. B - Condens. Matter Mater. Phys.*, vol. 71, no. 2, pp. 1–5, 2005.

[16] O. G. Shpyrko *et al.*, *Nature*, vol. 447, no. 7140, pp. 68–71, 2007.

[17] J. Xia *et al.*, *J. Electron. Mater.*, vol. 47, no. 11, pp. 6811–6820, 2018.

[18] B. Issa, I. M. Obaidat, B. A. Albiss, and Y. Haik, *Int. J. Mol. Sci.*, vol. 14, no. 11, pp. 21266–21305, Oct. 2013.

[19] A. H. Lu, E. L. Salabas, and F. Schüth, *Angew. Chemie - Int. Ed.*, vol. 46, no. 8, pp. 1222–1244, 2007.

[20] I. Sharifi, H. Shokrollahi, and S. Amiri, *J. Magn. Magn. Mater.*, vol. 324, no. 6, pp. 903–915, 2012.

[21] A. M. Abu-Dief, M. S. M. Abdelbaky, D. Martínez-Blanco, Z. Amghouz, and S. García-Granda, *Mater. Chem. Phys.*, vol. 174, pp. 164–171, 2016.

[22] V. Rudnev, *Handbook of Induction Heating*, 2nd ed. CRC Press, 2002.

[23] D. W. B. Mathieu, "INDUCTION COOKING," 2011.

[24] M. R. Barati, C. Selomulya, K. G. Sandeman, and K. Suzuki, *Appl. Phys. Lett.*, vol. 105, no. 16, 2014.

[25] P. E. Burke, 1986.

[26] W. Rano R. Wells, , Franklin, 1995.

[27] E. Wetzel and B. Fink, *Army Res. Lab.*, vol. 243, no. March, p. 73, 2001.

[28] R. E. Rosensweig, *J. Magn. Magn. Mater.*, vol. 252, no. 1–3 SPEC. ISS., pp. 370–374, 2002.

[29] N. F. Particles, Y. Zhang, and Y. Zhai, *Adv. Induction Microw. Heat. Miner. Org. Mater.*, no.

1, 2011.

[30] A. Okamoto, *2009 IEEE Globecom Work. Gc Work. 2009*, no. 1, pp. 1–6, 2009.

[31] P. Hu *et al.*, *J. Magn. Magn. Mater.*, vol. 322, no. 1, pp. 173–177, 2010.

[32] A. Zapata and G. Herrera, *Ceram. Int.*, vol. 39, no. 7, pp. 7853–7860, 2013.

[33] L. B. de Mello, L. C. Varanda, F. A. Sigoli, and I. O. Mazali, *J. Alloys Compd.*, pp. 698–705, 2019.

[34] E. M. M. Ewais, M. M. Hessien, and A. H. A. El-Geassy, *J. Aust. Ceram. Soc.*, vol. 44, no. 1, pp. 57–62, 2008.

[35] R. S. Yadav, J. Havlica, J. Masilko, J. Tkacz, I. Kuřitka, and J. Vilcakova, *J. Mater. Sci. Mater. Electron.*, vol. 27, no. 6, pp. 5992–6002, 2016.

[36] B. Rezaei, A. Kermanpur, and S. Labbaf, *J. Magn. Magn. Mater.*, vol. 481, no. February, pp. 16–24, 2019.

[37] T. D. Schladt, K. Schneider, H. Schild, and W. Tremel, *Dalt. Trans.*, vol. 40, no. 24, pp. 6315–6343, 2011.

[38] T. Kimijima, K. Kanie, M. Nakaya, and A. Muramatsu, *CrystEngComm*, vol. 16, no. 25, pp. 5591–5597, 2014.

[39] M. Atif, S. K. Hasanain, and M. Nadeem, *Solid State Commun.*, vol. 138, no. 8, pp. 416–421, 2006.

[40] S. A. Seyyed Ebrahimi and S. M. Masoudpanah, *J. Magn. Magn. Mater.*, vol. 357, pp. 77–81, 2014.

[41] P. J. van der Zaag, *Encyclopedia of Materials: Science and Technology*. 2001.

[42] H. L. Andersen, M. Saura-Múzquiz, C. Granados-Miralles, E. Canévet, N. Lock, and M. Christensen, *Nanoscale*, vol. 10, no. 31, pp. 14902–14914, 2018.

[43] P. H. Nam *et al.*, *Phys. B Condens. Matter*, vol. 550, no. May, pp. 428–435, 2018.

[44] D. Limin, H. Zhidong, Z. Yaoming, W. Ze, and Z. Xianyou, *J. Rare Earths*, vol. 24, no. 1 SUPPL. 1, pp. 54–56, 2006.

[45] A. Sutka and G. Mezinskis, *Front. Mater. Sci.*, vol. 6, no. 2, pp. 128–141, 2012.

[46] H. Waqas and A. H. Qureshi, *J. Therm. Anal. Calorim.*, vol. 98, no. 2, pp. 355–360, 2009.

[47] X. Li, R. Sun, B. Luo, A. Zhang, A. Xia, and C. Jin, *J. Mater. Sci. Mater. Electron.*, vol. 28, no. 16, pp. 12268–12272, 2017.

[48] TECHINSTRO, [Online]. Available: https://www.techinstro.com/shop/hydrothermal-autoclave/teflon-lined-hydrothermal-autoclave/. [Accessed: 25-Oct-2019].

[49] N. HAKMEH, no. July 2014, 2015.

[50] Y. C. S. W. H. QI, M. P. WANG, *J. Mater. Sci. Lett. 21*, vol. 21, p. 877– 878 Size, 2002.

[51] B. H. Toby, no. December 2005, pp. 67–70, 2006.

[52] G. K. W. and W. H. HALL, *ACTA Metall.*, vol. 1, 1953.

[53] C. Pereira *et al.*, *Chem. Mater.*, vol. 24, no. 8, pp. 1496–1504, 2012.

[54] H. H. Corporation and P. No, *Microscope*, no. 539, 2002.

[55] H. Bi, S. Li, Y. Zhang, and Y. Du, *J. Magn. Magn. Mater.*, vol. 277, no. 3, pp. 363–367, 2004.

[56] Z. L. Wang, *J. Phys. Chem. B*, vol. 104, no. 6, pp. 1153–1175, 2000.

[57] QuantumDesign, “MPMS3 Product Description.”

[58] M. Mößle *et al.*, *IEEE Trans. Appl. Supercond.*, vol. 15, no. 2 PART I, pp. 757–760, 2005.

[59] K. Martinis, J. M. and Osborne, *Les Houches*, vol. 79, 2004.

[60] J. Effect and S. Electronics, no. October 2005, 2010.

[61] QuantumDesign and M. A. Note, pp. 1–8, 1997.

[62] C. O. Amorim, F. Mohseni, V. S. Amaral, and J. S. Amaral, no. October, pp. 1–3, 2019.

[63] Ferroxcube, pp. 57–60, 2008.

[64] A. N. M. G. Company, p. 7004, 2002.

[65] A. Demir, S. Güner, Y. Bakis, S. Esir, and A. Baykal, *J. Inorg. Organomet. Polym. Mater.*, 2014.

[66] "Zinc oxide (ZnO) crystal structure, lattice parameters," in *II-VI and I-VII Compounds; Semimagnetic Compounds*, Springer, Berlin, Heidelberg, 2005.

[67] S. Mallesh, A. Sunny, M. Vasundhara, and V. Srinivas, *J. Magn. Magn. Mater.*, vol. 418, pp. 112–117, 2016.

[68] R. Gimenes *et al.*, *Ceram. Int.*, vol. 38, no. 1, pp. 741–746, 2012.

[69] S. Wu *et al.*, *J. Magn. Magn. Mater.*, vol. 324, no. 22, pp. 3899–3905, 2012.

[70] Y. Peng, L. Chen, H. Ren, L. Li, J. Yi, and Q. Xia, *J. Mater. Sci. Mater. Electron.*, vol. 27, no. 1, pp. 587–591, 2016.

[71] S. Mallesh and V. Srinivas, *J. Magn. Magn. Mater.*, vol. 475, no. October 2018, pp. 290–303, 2019.

[72] M. A. Ahmed, N. Okasha, and S. I. El-Dek, *Nanotechnology*, vol. 19, no. 6, 2008.

[73] L. Phor and V. Kumar, *J. Mater. Sci. Mater. Electron.*, vol. 30, no. 10, pp. 9322–9333, 2019.

[74] K. Praveena, K. Sadhana, and S. R. Murthy, *Mater. Res. Bull.*, vol. 47, no. 4, pp. 1096–1103, 2012.

[75] C. Rath *et al.*, *J. Appl. Phys.*, vol. 91, no. 3, pp. 2211–2215, 2002.

[76] M. Augustin and T. Balu, *Mater. Today Proc.*, vol. 2, no. 3, pp. 923–927, 2015.

[77] K. R. Lestari, P. Yoo, D. H. Kim, C. Liu, and B. W. Lee, *J. Korean Phys. Soc.*, 2015.

[78] Y. Yang *et al.*, *J. Mater. Chem. C*, vol. 1, no. 16, pp. 2875–2885, 2013.

[79] G. Y. Yimin Xuan, Qiang Li, *J. Magn. Magn. Mater.*, vol. 321, pp. 464–469, 2007.

[80] F. Bødker, M. F. Hansen, and C. B. Koch, *Phys. Rev. B - Condens. Matter Mater. Phys.*, vol. 61, no. 10, pp. 6826–6838, 2000.

[81] E. Brok *et al.*, *J. Phys. D. Appl. Phys.*, vol. 47, no. 36, 2014.

[82] G. F. Goya, vol. 94, no. 5, pp. 3520–3528, 2003.

[83] C. Temperature, *Phys. Rev. Lett.*, vol. 67, no. 25, pp. 3602–3605, 1991.

[84] Y. Su, H. Kang, Y. Wang, J. Li, and G. J. Weng, *Phys. Rev. B*, 2017.

[85] M. H. Mahmoud, A. M. Elshahawy, S. A. Makhlouf, and H. H. Hamdeh, *J. Magn. Magn. Mater.*, vol. 343, pp. 21–26, 2013.

[86] M. Vopsaroiu *et al.*, *J. Appl. Phys.*, vol. 97, no. 10, p. 10N3031-3, 2005.

[87] X. Guo *et al.*, *J. Phys. Chem. C*, vol. 118, no. 51, pp. 30145–30152, 2014.

[88] M. Bałanda and S. M. Dubiel, *J. Alloys Compd.*, vol. 663, pp. 77–81, 2016.

[89] R. Dannert, H. H. Winter, R. Sanctuary, and J. Baller, *Rheol. Acta*, vol. 56, no. 7–8, pp. 615–622, 2017.

# APPENDIX A: AUTOCLAVE TIME

The hydrothermal results presented during the present report belong to samples that went the autoclave procedure for 6 hours. Other samples were synthesized with different autoclave procedure times, 0 and 21 hours, in order to understand the synthesis evolution. Rietveld refinement analysis was employed to determine the spinel-phase percentage.

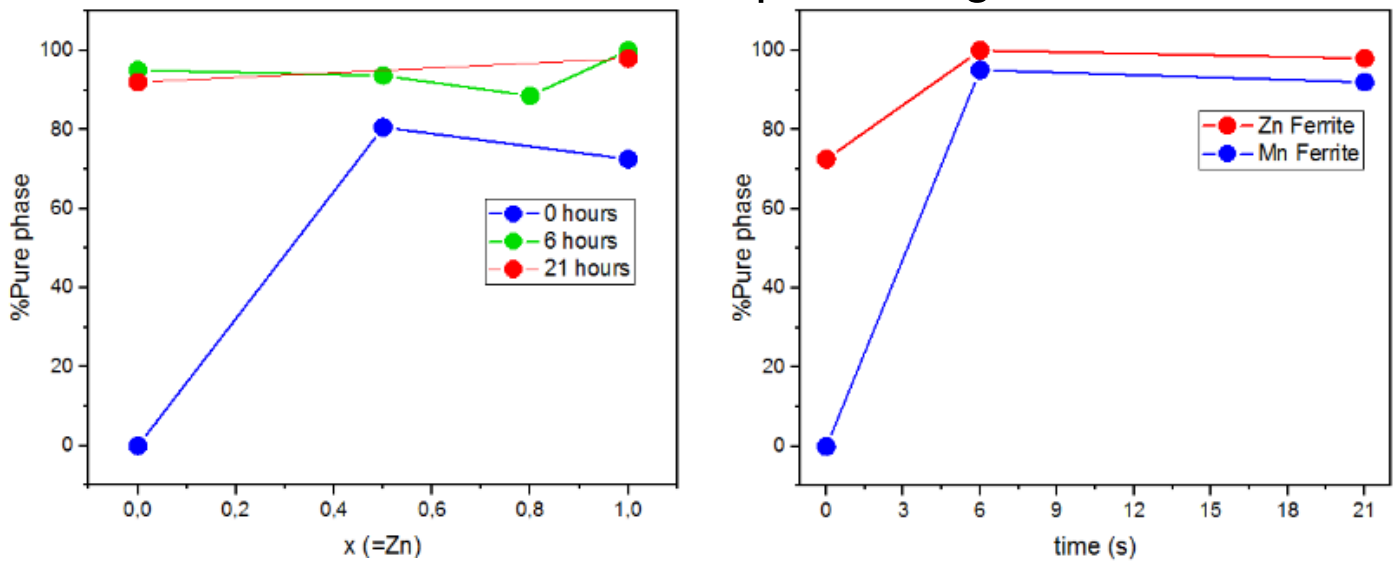

**Figure A.0.1 –** Percentage of pure phase as a function of a) composition and b) autoclave time.

As noticeable from figure A.1, almost pure spinel-phase samples are acquired from both, 6 and 21h. Figure A.1.b) shows the spinel-phase percentage as function of the autoclave time. An increasing of pure-phase percentage from 0 to 6 hours is reported for Mn and Zn ferrites. From 6 to 21h the pure-phase percentage declines.

Samples without the autoclave procedure (0 hours) are very small, which is revealed in XRD diffractogram by the broadened peaks and have low percentage of the spinel-phase. The Rietveld refinement results of these samples, revealed that Zinc ferrite along with a secondary phase of goethite, (FeOOH). The Mn ferrite at 0h revealed the absence of spinel crystal structure, instead it showed Mn-doped goethite ($Mn_{0,18}Fe_{0,82}OOH$) and Mn(III) oxide (MnO3). For higher autoclave times (21h) the pure-phase percentage decreases, comparing with 6h samples. The Rietveld Refinement revealed iron hydroxides of hematite, $Fe_{1,9}O_{2,7}(OH)_{0,3}$, for all samples and goethite for Mn ferrite. These results indicate that autoclave-time plays a major role in single-phase spinel crystal structure formation. Iron Hydroxides are formed before the autoclave procedure and, with temperature and pressure, the metal cations are incorporated in the spinel crystal structure.

The evolution of the properties of Mn and Zn ferrite with autoclave time are present in figure A.2. The 0 hours samples were removed due to high impurity phase.

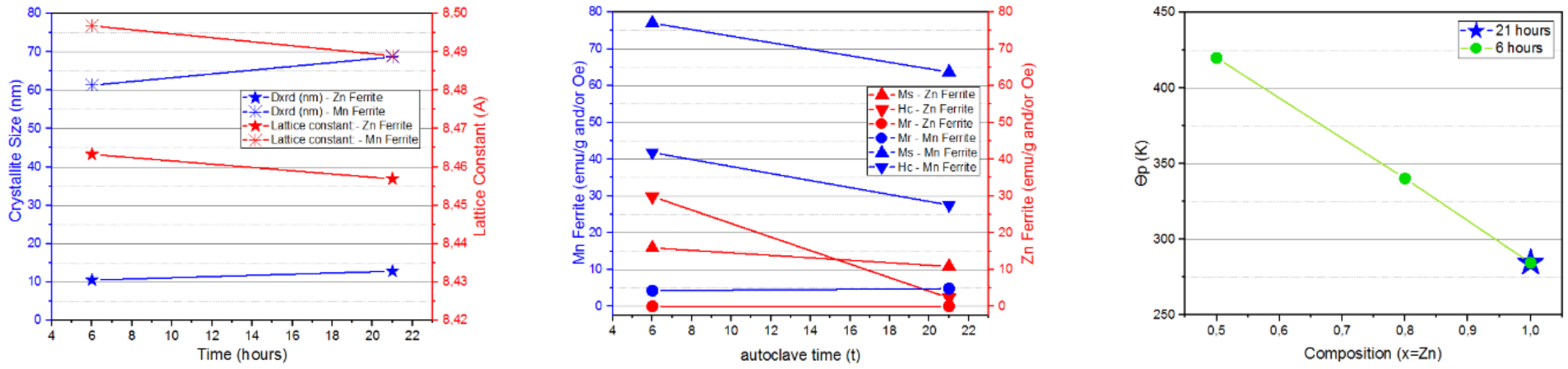

**Figure A.2 –** Mn and Zn ferrite dependency with autoclave time. a) $D_{XRD}(t)$ and a(t). b) $M_S(t)$, $H_C(t)$ and $M_R(t)$ c) $\Theta_P(t)$.

From figure A.2.a it is observable that when increasing the autoclave time, the crystallite size increases and the lattice constant decreases. From figure A.2.b) It is shown the superior magnetic properties of 6 hours samples over 21 hours, even with longer autoclave times. Longer time in autoclave decreases $M_S$, $H_C$ and $M_R$ for both Mn and Zn ferrite. Figure A.2.c) shows that the $\Theta_P$ value of Zn ferrite do not change for longer time in the autoclave procedure.

The presented figure shows minor changes between 6 and 21 hours samples, for this reason they were not included in the thesis body. Its noticeable a slight increase of crystallite size and remnant magnetization and decrease in lattice constant, saturation magnetization and coercivity. The $\Theta_P$ value remains unaltered.

# APPENDIX B: MAGNETIC ORDERING TEMPERATURE

The magnetic ordering temperature, presented in the body text, was determined using the Curie-Weiss law, as detailed in Chapter 4: Experimental Procedure. It was suggested that the temperature derivative of the magnetization, $\frac{\partial M}{\partial T}$, could also be used to calculate the magnetic ordering temperature. Thus, the magnetic ordering temperature would be defined as the temperature at which the FC curve presents a maximum curvature. The Curie temperature was tested with resource to this method, the results are present in figure A.1.

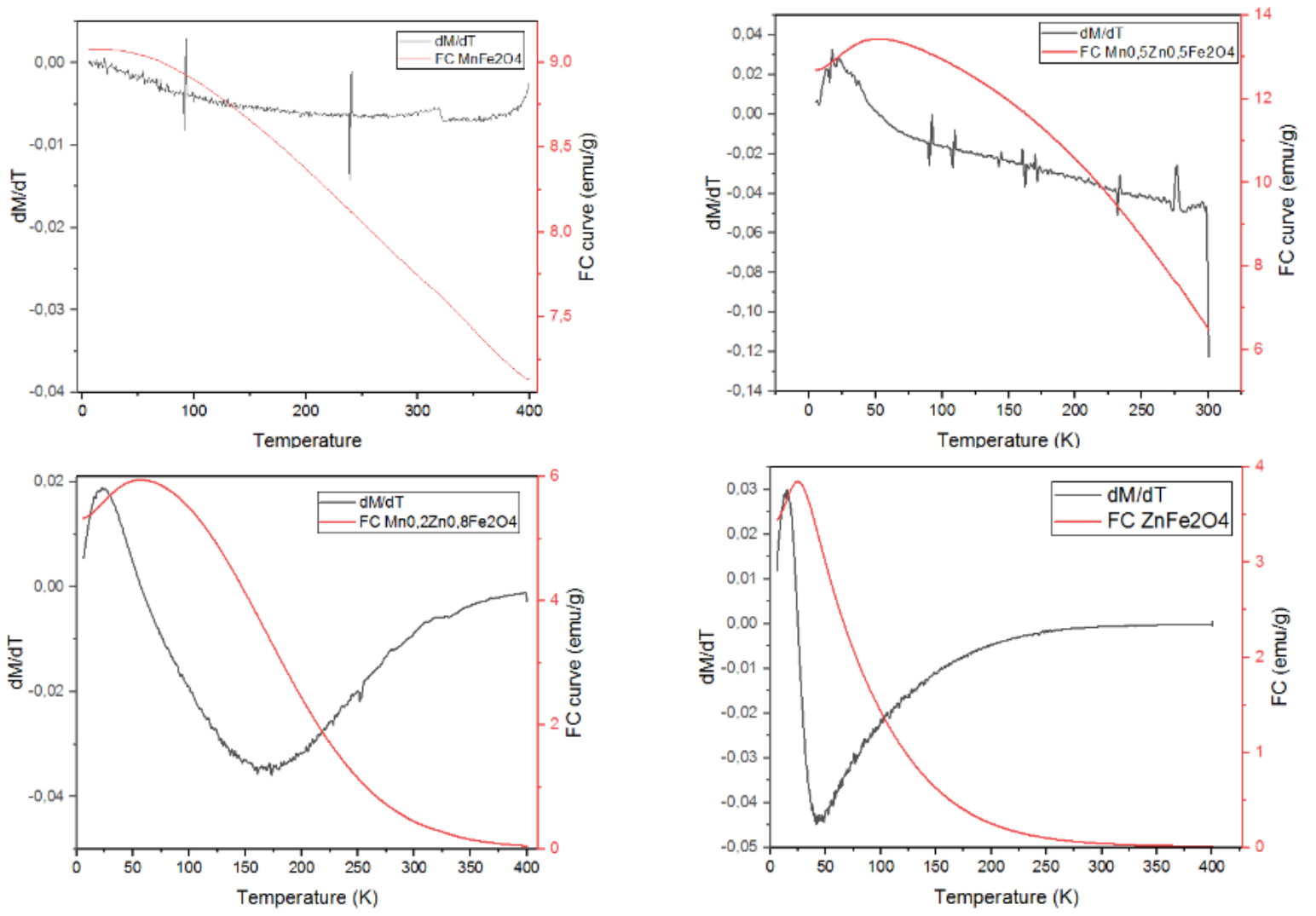

**Figure A.0.1 –** Temperature derivative of the FC curves for a) The red line is the FC measurement and the black line the respective temperature derivative.

Through this method the magnetic ordering temperature would present a lower value of those obtain via the Curie-Weiss law for the paramagnetic state. Furthermore, the magnetic ordering temperatures calculated via $\frac{\partial M}{\partial T}$ are inconsistent with the bibliographic values present in figure 5.19. This method also relies in the range of temperatures where the nanoparticles are under the unblocking procedure, thus, the unblocking of the magnetic domains will influence the maximum of the derivative. For all these reasons, these results were not included in the body text.